\documentclass{article}
\usepackage[dvips]{graphicx}
\usepackage{amsfonts}
\usepackage{amsmath,amsthm}
\usepackage{amssymb}
\usepackage{mathrsfs}
\usepackage{color}
\usepackage{pifont}
\usepackage{epstopdf}
\usepackage{geometry}
\usepackage{tikz}
\usetikzlibrary{shapes.geometric}
\usetikzlibrary{decorations.markings}
\usepgflibrary{decorations.shapes}
\usepgflibrary{shapes.misc}
\usepackage{subfigure}
\usepackage{hyperref}
\definecolor{linkcolour}{rgb}{0,0.2,0.6}
\hypersetup{colorlinks,breaklinks,
            linkcolor=linkcolour,citecolor=linkcolour,
            filecolor=linkcolour, urlcolor=linkcolour}

\newcommand{\nFer}{N} 
\newcommand{\diffBosFer}{M} 
\newcommand{\nRepBulk}{L} 
\newcommand{\nRepLeft}{m} 
\newcommand{\nRepRight}{n} 
\newcommand{\LMon}{\mathcal{L}} 
\newcommand{\RMon}{\mathcal{R}} 
\newcommand{\fug}{\beta} 
\newcommand{\fugL}{\beta_1} 
\newcommand{\fugR}{\beta_2} 
\newcommand{\fugLR}{\beta_{12}} 
\newcommand{\minlr}{\mathfrak{m}} 
\newcommand{\Alg}{\mathcal{A}_{2\nRepBulk,\nRepLeft,
    \nRepRight}} 
\newcommand{\Quot}{\mathcal{Q}_{2\nRepBulk,\nRepLeft,
    \nRepRight}} 
\newcommand{\HLV}{H^V_{\mbox{\scriptsize{left}}}}
\newcommand{\HRV}{H^V_{\mbox{\scriptsize{right}}}}
\newcommand{\HB}{H_{\mbox{\scriptsize{bulk}}}}
\newcommand{\HRVstar}{H^{V^\star}_{\mbox{\scriptsize{right}}}}
\newcommand{\HLVstar}{H^{V^\star}_{\mbox{\scriptsize{left}}}}

\newcommand{\perm}{\mathfrak{S}}
\newcommand{\ket}[1]{\left| #1 \right>}
\newcommand{\smfrac}[2]{\genfrac{}{}{}{1}{#1}{#2}}

\newcommand{\ceff}{c_{\mbox{\scriptsize{eff}}}}
\newcommand{\id}{\mathbf{1}}

\newcommand{\Esl}{E^{\mathfrak{sl}}}
\newcommand{\Psl}{P^{\mathfrak{sl}}}
\newcommand{\Stop}{\Phi_{\mbox{\scriptsize{top}}}}
\newcommand{\ii}{\mathfrak{i}}
\def\pl{\partial}

\newcommand{\gl}[2]{\mathfrak{gl}(#1|#2)}
\newcommand{\ssl}[2]{\mathfrak{sl}(#1|#2)}

\newcommand\slnm{\ensuremath{\text{$\mathfrak{sl}$}(\nFer+\diffBosFer|\nFer)}}

\newcommand\CP{\ensuremath{\text{$\mathbb{CP}$}^{N+M-1|N}}}
\newcommand\Ssn{\ensuremath{S^{2\nFer+2\diffBosFer-1|2\nFer}}}
\newcommand\Un{\ensuremath{\text{U}(\nFer+\diffBosFer\,|\nFer)}}
\newcommand\CPzero{\ensuremath{\text{$\mathbb{CP}$}^{0|1}}}

\newcommand{\tableft}{\begin{tikzpicture} 
\useasboundingbox (0,0) rectangle (1.75,0.45);
\draw (0,0) rectangle (0.35,0.35); 
\draw (0.175,0.175) node {$0$};
\draw (0.35,0) rectangle (1.4,0.35); 
\draw (0.875,0.175) node {$\cdots$};
\draw (1.4,0) rectangle (1.75,0.35); 
\draw (1.575,0.15) node {$\nRepLeft$};
\end{tikzpicture}}
\newcommand{\bleft}{b_{t_1}}
\newcommand{\tableftmone}{
\begin{tikzpicture} 
\useasboundingbox (0,0) rectangle (2.45,0.45);
\draw (0,0) rectangle (0.35,0.35); 
\draw (0.175,0.175) node {$0$};
\draw (0.35,0) rectangle (1.4,0.35); 
\draw (0.875,0.175) node {$\cdots$};
\draw (1.4,0) rectangle (2.45,0.35); 
\draw (1.925,0.15) node {$\nRepLeft-1$};
\end{tikzpicture}}
\newcommand{\bleftmone}{b_{\hat{t}_1}}

\newcommand{\tabright}{
\begin{tikzpicture} 
\useasboundingbox (0,0) rectangle (5.8,0.45);
\draw (0,0) rectangle (2.05,0.35); 
\draw (1.025,0.175) node {$2\nRepBulk+\nRepLeft-1$};
\draw (2.05,0) rectangle (3.1,0.35); 
\draw (2.575,0.175) node {$\cdots$};
\draw (3.1,0) rectangle (5.8,0.35); 
\draw (4.45,0.15) node {$2\nRepBulk+\nRepLeft+\nRepRight-1$};
\end{tikzpicture}}
\newcommand{\bright}{b_{t_2}}
\newcommand{\tabrightmone}{
\begin{tikzpicture}
\useasboundingbox (0,0) rectangle (3.95,0.45);
\draw (0,0) rectangle (1.05,0.35);
\draw (0.525,0.175) node {$\nRepLeft+2$};
\draw (1.05,0) rectangle (2.1,0.35);
\draw (1.575,0.175) node {$\cdots$};
\draw (2.1,0) rectangle (3.95,0.35);
\draw (3.025,0.15) node {$\nRepLeft+\nRepRight+1$};
\end{tikzpicture}}
\newcommand{\brightmone}{b_{\hat{t}_2}}

\DeclareMathOperator{\Det}{Det}
\DeclareMathOperator{\Tr}{Tr}
\DeclareMathOperator{\hook}{hook}
\DeclareMathOperator{\STr}{STr}

\begin{document}
\title{\textbf{Edge states and conformal boundary conditions\\
in super spin chains and super sigma models}}

\author{\\[5mm] Roberto Bondesan$^{1,2}$, Jesper L.\ Jacobsen$^{1,3}$ 
  and Hubert Saleur$^{2,4}$\\
  [3mm] ${}^1$LPTENS, \'Ecole Normale Sup\'erieure, 24 rue Lhomond,
  75231 Paris, France\\
  [3mm] ${}^2$ Institute de Physique Th\'eorique, CEA Saclay,\\ F-91191
  Gif-sur-Yvette, France\\
  [3mm] ${}^3$Universit\'e Pierre et Marie Curie, 4 place Jussieu,
  75252 Paris, France \\
  [3mm] ${}^4$ Physics Department, USC, Los Angeles, CA 90089-0484, USA}

\pagestyle{headings} \date{}

\begin{titlepage}
 \maketitle
 \begin{abstract}
   \noindent 
   The sigma models on projective superspaces $\mathbb{CP}^{N+M-1|N}$
   with topological angle $\theta=\pi\hbox{ mod } 2\pi$ flow to
   non-unitary, logarithmic conformal field theories in the low-energy
   limit. In this paper, we determine the exact spectrum of these
   theories for all open boundary conditions preserving the full
   global symmetry of the model, generalizing recent work on the
   particular case $M=0$ [C.~Candu {\em et al}, JHEP02(2010)015].  In
   the sigma model setting, these boundary conditions are associated
   with complex line bundles, and are labelled by an integer, related
   with the exact value of $\theta$.  Our approach relies on a spin
   chain regularization, where the boundary conditions now correspond
   to the introduction of additional edge states. The exact values of
   the exponents then follow from a lengthy algebraic analysis, a
   reformulation of the spin chain in terms of crossing and
   non-crossing loops (represented as a certain subalgebra of the
   Brauer algebra), and earlier results on the so-called one- and
   two-boundary Temperley Lieb algebras (also known as blob algebras).
   A remarkable result is that the exponents, in general, turn out to
   be irrational. The case $M=1$ has direct applications to the spin
   quantum Hall effect, which will be discussed in a sequel.
 \end{abstract}
\end{titlepage}

\addtolength{\baselineskip}{3pt}

\section{Introduction}
\label{sec:intro}

Two-dimensional sigma models on supergroups and supercosets play a
fundamental role in several areas of theoretical physics, such as
phase transitions in $2+1$ non interacting disordered electronic
systems \cite{Efetov1983,Read1987,Weidenmuller1987} (for reviews see
\cite{Zirnbauer1999,Bocquet2000}), or the AdS/CFT duality
\cite{Witten2004,Mann2004,Mann2005,Aharony2008}. The study of these
models is on the other hand quite difficult, for a variety of reasons
all related to the non unitarity of the target. Progress has thus been
rather slow, but has picked up pace recently, thanks to the use of
mini superspace technology \cite{SCHOMERUS2006}, a better
understanding of algebraic aspects
\cite{Quella2008,Gotz2005,Mitev2008,Candu2010a}, and the introduction
of lattice regularizations
\cite{Candu2009a,Candu2010,Read2001,Read2007a,Read2007,Pearce2006,Pearce2010}
inspired from network models \cite{Read1993,Zirnbauer1997,Kondev1997}.

A particularly interesting question for critical super sigma models
concerns the classification of their conformal boundary
conditions. While for Wess Zumino Witten models the question is well
understood \cite{Creutzig2007,Creutzig2009b}, the situation for other
models---such as the superprojective sigma models at $\theta=\pi$---is
much less under control. In a recent paper \cite{Candu2010}, the sigma
models on
\begin{equation}
  \ensuremath{\text{$\mathbb{CP}$}^{N-1|N}}
  =\ {\rm U}(N|N) / \big( {\rm U}(1) \times {\rm U}(N-1|N) \big)
\end{equation}
at $\theta=\pi$ were considered, and the set of conformal boundary
conditions invariant under the full group ${\rm PU}(\nFer|\nFer) ={\rm
  U}(\nFer|\nFer)/{\rm U}(1)$ classified. These boundary conditions
are parametrized by an integer, and correspond, in string theory
parlance, to volume filling branes equipped with a monopole line
bundle and a connection. In sigma model language, the existence of
these boundary conditions has to do with the fact that, while the bulk
theory only depends on the value of $\theta$ modulo $2\pi$, when there
is a boundary, the exact value of the topological angle matters, and
enters, for instance, the classical equations of motion
\cite{KHMELNITSKII1994,Maslov1993,Xiong1997}.

The sigma model considered in \cite{Candu2010} is peculiar because it
is critical for a large domain of values of the coupling constant
$g_\sigma^2$ including the minisuperspace limit $g_\sigma^2\to 0$. The
solution using harmonic analysis on the target in this limit is a
precious help in finding out the exact spectrum for all values of
$g_\sigma^2$ up to the critical value $g_\sigma^2=1$ (in
normalizations of \cite{Candu2010}) beyond which the sigma model
exhibits different behavior. For other types of superprojective sigma
models, such a line of fixed points is not available, and one must
directly solve the model at finite coupling. The lattice
regularizations of \cite{Read2001} constitute a powerful means of
obtaining such a solution, and we shall take this route in the present
paper.

The main objective of this work is to extend the $\mathbb{CP}^{N-1|N}$
study of \cite{Candu2010} to the considerably more difficult and
richer case of $\mathbb{CP}^{N+M-1|N}$. The use of lattice
regularizations in this context is not only a practical tool, but also
provides a physical intuition for understanding conformal boundary
conditions in terms of edge states.\footnote{We note that the role of
  edge states in sigma models has recently been revisited in
  \cite{Pruisken2008}, although the relation with our and other's work
  is not clear to us.} This in turn has applications to the
description of the transition between plateaus in the spin quantum
Hall effect \cite{Gruzberg1999}, which corresponds to the $M=1$ case
of the general construction. We will discuss the implications of
this observation in a subsequent work \cite{GruzbergObuse}.

Our paper is organized as follows. In section 2, we revisit some of
the main results in \cite{Candu2010} and discuss boundary conditions
in sigma models. We review in detail the easy case of $M=0$, $N=1$
(symplectic fermions) which will serve as a benchmark for the
subsequent developments. In section 3, we discuss briefly the role of
boundaries in mapping spin chains to sigma models, together with the
concept of edge states. In section 4, we discuss the spin chains
relevant for the solution of the boundary superprojective sigma
models, together with their geometrical formulation in terms of loops
and in terms of lattice algebras. Section 5 is the most technical:
this is where we obtain the solution of the loop model based on
earlier results on one and two boundary Temperley-Lieb algebra,
together with exact solutions for
$\mathbb{CP}^{N-1|N}$ and exact diagonalizations
of finite size systems otherwise. In section 6, we discuss in details
the relation between the spectrum of the spin chains and loop model,
while section 7 contains conclusions. Some particularly technical
results are discussed in the appendices.

\subsection{Notations}
\label{sec:notations}

For the reader's convenience we collect  here some notations to be used in
this paper.
\begin{itemize}
\item $\nFer$ and $\nFer+\diffBosFer$ denote respectively the number of 
  bosonic and fermionic coordinates of the superspaces used.
\item $\nRepBulk$ is the number of pairs of alternating
  representations in the bulk of the spin chain.
\item $\nRepLeft$ is the number of additional representations on the
  left boundary of the spin chain.
\item $\nRepRight$ is the number of additional representations on the
  right boundary of the spin chain.
\item $\minlr = {\rm min}(\nRepLeft,\nRepRight)$ is the minimal number of
  uncontractible pairs of boundary representations.
\item $\LMon$ and $\RMon$ are integer (monopole) numbers labelling
  different boundary conditions at the ends of the world-sheet in the
  sigma model formulation.
\item $\fug$ is the fugacity of bulk loops in the geometrical
  formulation of the model.
\item $\fugL$ and $\fugR$ are the fugacities of loops touching
  respectively the left and the right boundary in the
  two-boundary loop model, and $\fugLR$ is the fugacity of loops
  touching both.
\item ${\cal B}_{2L}$ is the Brauer algebra on $2L$ strands.
\item $\Alg$ is the subalgebra of the Brauer algebra involved in the
  formulation of the lattice model.
\item $j$ is the number of non contracted pairs of bulk lines.
\item $k$ is the number of non contracted pairs of boundary lines.
\item $p$ is the minimal model index, corresponding to setting
  $\beta = 2 \cos \left( \frac{\pi}{p+1} \right)$.
\item $h_{r,s}$ is the Kac table of conformal weights.
\end{itemize}

\section{Boundary sigma models}

We first briefly review the definition of the \CP\ models, and then
move on to describe the role of boundaries in the sigma models.

\subsection{Superprojective sigma models at $\theta=\pi$}

Complex projective superspaces \CP\ are built much like their bosonic
cousins \cite{Read2001,Candu2010}.  Begin with superspace
$\mathbb{C}^{\nFer+\diffBosFer|\nFer}$; the $\nFer+\diffBosFer$
complex bosonic coordinates are denoted by $z_a$ and we use $\xi_a$
for the $\nFer$ fermionic directions. Within this complex superspace,
consider the odd (real) dimensional supersphere defined by the
equation
\begin{equation} 
\label{SSconstraint}
  \sum_{a=1}^{\nFer+\diffBosFer} z_a z_a^\ast +
  \sum_{a=1}^\nFer \xi_a \xi_a^\ast = 1 \ \ . 
\end{equation}
The supersphere \Ssn\ carries an action of $U(1)$ by simultaneous
phase rotations of all bosonic and fermionic coordinates,
\begin{equation}\label{phrot}
 z_a \ \longrightarrow \ e^{i \varpi} z_a \ \ \ \ \ , \ \ \ \
   \xi_a \ \longrightarrow\ e^{i\varpi} \xi_a \ \ .
\end{equation}
Note that this transformation indeed leaves the constraint invariant. 
The complex projective superspace \CP\ is the quotient space
$\Ssn\!/U(1)$.
\smallskip

Functions on the supersphere \Ssn\ carry an action of the Lie
supergroup \Un. These transformations include the phase rotations
\eqref{phrot} which act trivially on \CP. Hence, the stabilizer
subalgebra of a point on the projective superspace is given by u(1)
$\times$ u($\nFer+\diffBosFer-1|\nFer$) where the first factor
corresponds to the action \eqref{phrot}. We conclude that
\begin{eqnarray}
  \CP \ = \ \text{\Un} /\left(\,\text{U(1) $\times$
      U($\nFer+\diffBosFer-1|\nFer$)}\right) \ \ .
\end{eqnarray}
Their simplest representative of interest to us is \CPzero\, i.e. the
space with just two real fermionic coordinates. The sigma model with
this target space is equivalent to the theory of two symplectic
fermions, which has been extensively investigated, as for example in
\cite{Kausch1995,Kausch2000}.

The construction of the sigma model on \CP\ closely parallels this
geometric construction. The model involves a field multiplet
$Z_\alpha=Z_\alpha(z,\bar z)$ with $\nFer+\diffBosFer$ bosonic
components $Z_\alpha = z_\alpha, \alpha =1, \dots, \nFer+\diffBosFer$
and $\nFer$ fermionic fields $Z_\alpha = \xi_{\alpha - N-M}, \alpha =
\nFer+\diffBosFer+1, \dots ,2\nFer+\diffBosFer+1$.  To distinguish
between bosons and fermions we introduce from now on a grading
function $|\cdot|$, which is 0 when evaluated on the labels of bosonic
and 1 on the labels of fermionic quantities.  In addition we also need
a non-dynamical U(1) gauge field $a$. With this field content, the
action takes the form (with a summation convention on the index $\alpha$)
\begin{equation} \label{action}
S \ = \ \frac{1}{2g_\sigma^2} \int {\rm d}^2z \, (\pl_\mu - ia_\mu)
Z_\alpha^\dagger (\partial_\mu + i a_\mu)Z_\alpha -
\frac{i\theta}{2\pi} \int {\rm d}^2z \, \epsilon^{\mu\nu} \pl_\mu a_\nu
\end{equation}
and the fields $Z_\alpha$ are subject to the constraint
$Z_\alpha^\dagger Z_\alpha = 1$.\footnote{Note that we eliminated
the radius $\rho$ of the complex projective space in favor of a
coupling $g_\sigma^{-2}$ entering the action in front of the
metric. Equivalently, we can set $g_\sigma^2=1$ and work with a
radius parameter $\rho$ appearing in the modified constraint
$Z_\alpha^\dagger Z_\alpha = 4\rho^2$.} The integration over the
abelian gauge field can be performed explicitly and it leads to
the replacement
\begin{equation}
a_\mu \ = \ \frac{i}{2} \left[ Z_\alpha^\dagger \pl_\mu Z_\alpha -
(\pl_\mu Z_\alpha^\dagger) Z_\alpha\right] \ \ .
\end{equation}
The term multiplied by $\theta$ does not contribute to the equations
of motion for $a_\mu$. As its bosonic counterpart, the \CP\ sigma
model on a closed surface possesses instanton solutions. The
corresponding instanton number is computed by the term---which we will
denote by $Q$ and refer to as the `topological term'---that multiplies the parameter $\theta$. Since $Q$ is
integer-valued, the parameter $\theta = \theta + 2\pi$ can be
considered periodic {\sl as long as the world-sheet has no boundary}.

The target supermanifold being a symmetric superspace, the metric on
the target space is unique up to a constant factor, so $g_\sigma^2$
and $\theta$ are the only coupling constants. The perturbative beta
function is the same as the one for
$\mathbb{CP}^{\diffBosFer-1}$ \cite{Wegner}
\begin{equation}
\frac{{\rm d}g_\sigma^2}{{\rm d}l}=\beta(g_\sigma^2)=\diffBosFer g_\sigma^4+O(g_\sigma^6)
\end{equation}
The beta function for $\theta$ is zero in perturbation theory, and
that for $g_\sigma^2$ is independent of $\theta$.

For $\diffBosFer>0$, the coupling is weak at short length scales, but
flows to strong values at large length scales. For $\theta\neq \pi$
(mod $2\pi$) the \Un\ symmetry is eventually restored, and the theory
is massive. When $\theta=\pi$ ($\mbox{mod } 2\pi$) and $\diffBosFer\leq
2$, the model flows to a non-trivial fixed point. The corresponding
bulk conformal field theory has been studied in
\cite{Read2001,Read2007}, and presents many
fascinating features. The main purpose of this paper is to study its
boundary properties in more details.

For $\diffBosFer=2$ and $\nFer=0$ we recover the usual flow in the
$O(3)$ sigma model; the conformal field theory in that case is the
$SU(2)$ level 1 Wess Zumino model. For $\diffBosFer=0$, the beta
function is in fact exactly zero, and the sigma model exhibits a line
of fixed points
\cite{Read2001,Candu2010,Candu2010b,Zirnbauer1999,Bershadsky1999}.  The case
$\diffBosFer=1$ is particularly interesting, because it is related
with percolation, and with the spin quantum Hall effect
\cite{Gruzberg1999,GruzbergObuse}.

\subsection{The role of the boundaries}

While the model on a compact oriented manifold exhibits properties
that do not depend on the exact value of $\theta$ provided
$\theta=\pi$ ($\mbox{mod } 2\pi$), things are bound to be quite different in
the presence of a boundary, as first noticed in \cite{Xiong1997}. Strictly
speaking, in a finite system with boundaries, there are no well
defined topologically distinct sectors. One does not expect the
behavior of the RG for the bulk properties to change---the standard
analysis should still apply provided the equivalent of the instantons
(configurations such that the `topological term' $Q$ approaches a
constant in the core, far from the boundary) are well localized within
the system. But boundary properties are expected to now depend on the
exact value of $\theta$; in particular, $\theta$ can now be expected
to affect perturbation theory.

To make things more concrete, we restrict to sigma models on the strip
$\Sigma=[0,\pi]\times \mathbb{R}$ or, equivalently by conformal
transformation, the upper half plane $z = x+iy, y > 0$. The boundary
conditions induced by the `topological term' are
\begin{equation} \label{glue0}\begin{split}
(\partial_y + i a_y) Z_\alpha & = \ \ \ {\theta\over\pi}g_\sigma^2
(\partial_x + i a_x) Z_\alpha\  ,\\[2mm]
(\partial_y - i a_y) Z^\dagger_\alpha & =\  - {\theta\over\pi}
g_\sigma^2 (\partial_x - ia_x) Z_\alpha^\dagger
\end{split}
\end{equation}
for $z = \bar z < 0$ and a similar condition along the right half $z =
\bar z > 0$ of the boundary. While in the weak coupling limit the
value of $\theta$ is irrelevant (and we recover purely Neumann
boundary conditions), in general, $\theta$ appears explicitly.

Of course, except in the special case $\diffBosFer=0$
\cite{Read2001}, we will need $\theta=\pi (\hbox{mod }2\pi)$ to have
a conformal field theory in the low energy limit, but we can now
expect properties to depend on the exact value of $\theta$ itself.

It was indeed argued in \cite{Candu2010} that a more general family of \Un\
symmetric boundary conditions can be obtained which are expressed
through the conditions
\begin{equation} \label{glue}\begin{split}
(\partial_y + i a_y) Z_\alpha & = \ \ \Theta_1 g_\sigma^2
(\partial_x + i a_x) Z_\alpha\  ,\\[2mm]
(\partial_y - i a_y) Z^\dagger_\alpha & =\  - \Theta_1
g_\sigma^2 (\partial_x - ia_x) Z_\alpha^\dagger
\end{split}
\end{equation}
for $z = \bar z < 0$ and a similar condition with $\Theta_1$ replaced
by $\Theta_2$, along the right half $z = \bar z > 0$ of the boundary.
The parameters $\Theta_1 = 2\LMon + \theta/\pi$ and $\Theta_2=2\RMon +
\theta/\pi$ involve now integer (monopole) numbers $\LMon,\RMon$.

In the case $\diffBosFer=0$, it was shown in \cite{Candu2010} how, to
every choice of integer $\RMon=\LMon$, is associated a conformally
invariant boundary condition. The case $\RMon\neq \LMon$ then
corresponds to the insertion of a boundary condition changing
operator.

For $0<\diffBosFer\leq 2$, we expect similarly that conformally
invariant boundary conditions at the fixed point theory are induced by
these boundary terms. In other words, we expect that, in the
conformally invariant fixed points of superprojective sigma models
\CP\ at $\theta=\pi$, there is a discrete family of conformal boundary
conditions labelled by the integer $\RMon \in \mathbb{Z}$: these are
the subject of our study.

We note that the only dependence on the exact value of $\theta$
comes from the boundary conditions. A model with a given value of
$\theta,\LMon,\RMon$ is identical with the model with $\theta+2p\pi,
\LMon-p, \RMon-p$, for any $p \in \mathbb{Z}$.

\subsection{The case \CPzero}
\label{sec:cpzero}

To make things more concrete, we note that the simplest case of
\CPzero\ ($\nFer=1,\diffBosFer=0$) is equivalent to symplectic
fermions.  With the bulk action
\begin{equation}
S={1\over 2g_\sigma^2}\int d^2z \partial_\mu\xi\partial_\mu\xi^\dagger
\end{equation}
the boundary conditions are then of the form
\begin{eqnarray}
\partial_y \xi&=&~~\Theta ~g_\sigma^2 ~\partial_x\xi\nonumber\\
\partial_y \xi^\dagger&=&-\Theta~ g_\sigma^2~\partial_x\xi^\dagger
\end{eqnarray}
They can be represented in terms of a glueing automorphism
\begin{eqnarray}
\left(\begin{array}{c}
\partial\xi\\
\partial\xi^\dagger\end{array}\right)=\Omega\left(\begin{array}{c}
\bar{\partial}\xi\\
\bar{\partial}\xi^\dagger\end{array}\right)
\end{eqnarray}
where 
\begin{equation}
  \Omega={1+W(\Theta)\over 1-W(\Theta)},~~ 
  W(\Theta)=ig_\sigma^2\left(\begin{array}{cc} 
      \Theta &0\\0&-\Theta\end{array}\right)
\end{equation}
If we now have two different values of $\Theta$ on the left and right
boundaries, going around the insertion of the boundary condition
changing operator in the complex plane gives rise to a monodromy
expressed by
\begin{equation}
  \Omega_{12}=\Omega_1\Omega_2^{-1}={\kappa +W(\Theta_1-\Theta_2)\over 
    \kappa-W(\Theta_1-\Theta_2)}
\end{equation}
where we have set $\kappa=1+g_\sigma^4\Theta_1\Theta_2$.  Of course,
$\Omega_{12}$ is of the form
\begin{equation}
\Omega_{12}=\left(\begin{array}{cc}
e^{2i\pi\lambda}&0\\
0&e^{-2i\pi\lambda}
\end{array}\right)
\end{equation}
with the twist parameter $\lambda$ given by 
$2\cos2\pi\lambda=\hbox{Tr }\Omega_{12}$:
\begin{equation}
  \cos2\pi\lambda={(1+g_\sigma^4\Theta_1\Theta_2)^2
    -(\Theta_1-\Theta_2)^2g_\sigma^4\over
    (1+g_\sigma^4\Theta_1\Theta_2)^2
    +(\Theta_1-\Theta_2)^2g_\sigma^4}\label{lambdff}
\end{equation}
The twist parameter vanishes when $\LMon=\RMon$. In general, the
ground state of the theory (which has central charge $c=-2$) scales
with the conformal weight
\begin{equation}
h_\lambda^{gr}={1\over 2}\lambda(\lambda-1)
\end{equation}
and the full operator content in this sector is encoded in the
$U(1)\times\hbox{Vir}$ character \cite{Kausch1995}
\begin{equation}
\begin{split}
  d_{\mu,\lambda}&=\hbox{tr}\left(e^{2i\mu J_0}q^{L_0-c/24}\right)
  =\\
  &= e^{-2i\pi\mu\lambda} q^{(1-6\lambda(1-\lambda))/12}\prod_{n=1}^\infty
  \left(1+e^{2i\pi\mu}q^{n+\lambda-1}\right)
  \left(1+e^{-2i\pi\mu}q^{n-\lambda}\right)
\end{split}
\end{equation}
or 
\begin{equation}
\label{eq:charcpzero}
  d_{\mu,\lambda}
  =
  \eta^{-1}e^{-2i\pi\mu\lambda}
  \sum_{m\in Z}e^{2i\pi\mu m}q^{{1\over 2}(m+\lambda-{1\over 2})^2} \,,
\end{equation}
where $q$ denotes the modular parameter and $\eta$ is the Dedekind eta
function.

The problem is thus fully solved in this case. In this paper, we want
to address the considerably more difficult case of $\diffBosFer\neq
0$, in particular the case $\diffBosFer=1$ which is relevant to the
spin quantum Hall effect \cite{Gruzberg1999,GruzbergObuse}.

\section{Boundary conditions  in super spin chains}

\subsection{Super spin chains}
\label{sec:super_spin_chains}

As mentioned in the introduction, our strategy to solve the boundary sigma model is to use a lattice regularization. There is a well known, profound relationship between sigma models and
spin chains, going back to the earliest developments in the $O(3)$
case \cite{Haldane1983}.  In fact, the conformal fixed points in the
bulk \CP\ sigma models at $\theta=\pi$ have been studied in
\cite{Read2001,Read2007} using homogeneous super spin chains
that represent the strong coupling region. The mapping to the sigma
models follows the well known argument in non graded
\cite{Affleck1985,Read1989} as well as graded
\cite{Zirnbauer1994,Gruzberg1999} cases.

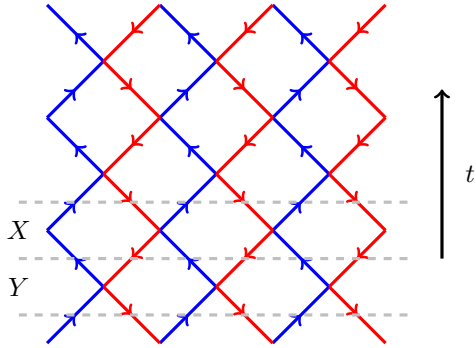
\begin{figure}[htp]
  \centering
\begin{tikzpicture}[very thick,decoration={ markings, mark=at position
    0.5 with {\arrow{>}}}, scale=0.75]
  \foreach \x in {0,2,4} 
  { 
    \foreach \y in {0,2,4} 
    {
      \draw[postaction={decorate},blue] (\x,\y)--(\x+1,\y+1);
      \draw[postaction={decorate},blue] (\x+1,\y+1)--(\x,\y+2); 
    } 
  }
  \foreach \x in {1,3,5} 
  { 
    \foreach \y in {0,2,4} 
    {
      \draw[postaction={decorate},red] (\x+1,\y+2)--(\x,\y+1);
      \draw[postaction={decorate},red] (\x,\y+1)--(\x+1,\y); 
    } 
  }
  \draw[dashed,black!25] (-0.5,0.5)--(6.5,0.5); 
  \node at (-0.5,1) {$Y$}; 
  \draw[dashed,black!25] (-0.5,1.5)--(6.5,1.5); 
  \node at (-0.5,2) {$X$}; 
  \draw[dashed,black!25] (-0.5,2.5)--(6.5,2.5);
  \draw[->] (7,1.5) -- (7,4.5);
  \node at (7.5,3) {$t$};
\end{tikzpicture}
  \caption{The two dimensional lattice corresponding to the
    alternating super spin chain. Edges carrying the fundamental
    representation have up arrows while those with the dual down
    arrows. The transfer matrix is a product of two transfer matrices
    acting on two consecutive layers $T=X Y$, imaginary time flows
    from bottom to top.}
\label{fig:alt_vertex}
\end{figure}

The simplest chain (which will be described in considerably more
details below) is obtained by alternating the fundamental
representation $V$ of \slnm\ and its conjugate $V^*$, and choosing an
antiferromagnetic Heisenberg coupling between nearest neighbours
(since the product $V\otimes V^*$ decomposes generically on two
representations---the identity and the adjoint---this is the most
general nearest neighbour interaction). This coupling in turn can be
conveniently recognized as a representation $\Esl$ of the Temperley-Lieb
algebra acting on $(V\otimes V^*)^{\otimes \nRepBulk}$, and the
Hamiltonian with open (free) boundary conditions can be written as
\begin{equation}
  H = -\sum_{i=1}^{2\nRepBulk-1} \Esl_i
\end{equation}
Detailed expressions in terms of the natural vector basis of $V,V^*$
are given in \cite{Read2001,Read2007a} and we do not reproduce them
for now (but see below). An important point for further study is that
the generators $\Esl_i$ satisfy the Temperley-Lieb relations
\cite{TL1971}
\begin{eqnarray}
  (\Esl_i)^2&=& \diffBosFer \Esl_i\nonumber\\
  \Esl_i\Esl_{i\pm 1}\Esl_i&=&\Esl_i\nonumber\\
  \left[\Esl_i,\Esl_j\right]&=&0, \qquad \mbox{for }|i-j|\geq 2 \, .
\end{eqnarray}
Instead of the Hamiltonian, it is sometimes useful to use a transfer
matrix, the evolution operator of the $1$D chain in discrete imaginary
time. The choice of alternating representations in the spin chain
corresponds then to having oriented edges carrying respectively $V$
and $V^*$ as illustrated in figure \ref{fig:alt_vertex}, making the
potential relationship with network models of quantum localisation
more transparent. The transfer matrix itself is then built as a
product of two diagonal-to-diagonal transfer matrices associated to
two consecutive layers of the lattice, which evolves the states one
unit in time.

For $0<\diffBosFer\leq 2$, the model thus defined flows to the fixed
points discussed in \cite{Read2007}. The Hamiltonian with periodic
boundary conditions has been studied in \cite{Read2001}.  When
$\diffBosFer=0$, recall
\cite{Read2001,Candu2010,Candu2010b,Zirnbauer1999,Bershadsky1999} that
the sigma model admits a line of fixed points parametrized by
$g_\sigma^2$: the model we have just defined corresponds to a
particular value of $g_\sigma^2$ ($=1$). Other values can be obtained
by adding an exactly marginal interaction corresponding to the
exchange of next nearest neighbors degrees of freedom $V$ or $V^*$
(since the products $V\otimes V$ or $V^*\otimes V^*$ decompose
generically on two representations, this is the most generic nearest
neighbor coupling). In the Hamiltonian language, this means
\begin{equation}
H=-\sum_{i=1}^{2\nRepBulk-1} \Esl_i-w\sum_{i=1}^{2\nRepBulk-2}\Psl_{i,i+2} \, ,
\end{equation}
where $\Psl_{i,i+2}$ is the operator permuting states in sites $i$ and
$i+2$.  Note that the $\Psl_{i,i+2}$ terms do not break the \slnm\
symmetry \cite{Candu2010b}, i.e., it does not mix up the $V$ and $V^*$ spaces. A proposal for obtaining the conformal boundary conditions
starting from this spin chain was then made in
\cite{Candu2010}. There, a family of four possible chains was
considered, corresponding to the following spaces (with factors labelled from $0$ to $2L+m+n-1$) and Hamiltonians
\begin{equation}\label{eq:tw_chain_general}
\begin{array}{rlll}
 V^{\otimes \nRepLeft} & \!\!\otimes\,   
 (V\otimes V^\star)^{\otimes \nRepBulk}
 \otimes \ (V^\star)^{\otimes \nRepRight}: 
 \quad\quad   & H^{VV^\star} & = \
 \HLV +
 \HB +
 \HRVstar \\[2mm]
 V^{\otimes \nRepLeft} & \!\!\otimes\,  
 (V\otimes V^\star)^{\otimes \nRepBulk}
 \otimes \ V^{\otimes \nRepRight}: & 
 H^{VV} & =  \ 
 \HLV + 
 \HB +
 \HRV \\[2mm]
 (V^\star)^{\otimes \nRepLeft} & \!\! \otimes \, 
 (V\otimes V^\star)^{\otimes \nRepBulk} \otimes \ 
 (V^\star)^{\otimes \nRepRight}:& 
 H^{V^\star V^\star} & = \ 
 \HLVstar + 
 \HB +
 \HRVstar \\[2mm]
 (V^\star)^{\otimes m} & \!\!\otimes \, (V\otimes V^\star)^{\otimes \nRepBulk}
 \otimes \ V^{\otimes n}:& H^{V^\star V} & =  \ 
 \HLVstar +
 \HB +
 \HRV,
\end{array}
\end{equation}
where the bulk Hamiltonian is
\begin{equation}\label{eq:bulk_ham}
  \HB \ = \ - \sum_{i=m}^{2\nRepBulk+\nRepLeft-2}\Esl_i - w 
  \sum_{i=m}^{2l+m-3}\Psl_{i,i+2}\ ,
\end{equation}
while the boundary Hamiltonians are as follows
\begin{align}\label{eq:b_hams_1}
  \HLV \ =& -u\sum_{i=0}^{\nRepLeft-1} \Psl_{i,i+1} 
  \quad\quad\quad\quad\quad
  \HRVstar \ =\  -v \sum_{i=2\nRepBulk + \nRepLeft - 1}
  ^{2\nRepBulk + \nRepLeft + \nRepRight - 2} \Psl_{i,i+1}\\ \label{eq:b_hams_2}
  \HLVstar \ =&   -u\sum_{i=0}^{\nRepLeft - 2}\Psl_{i,i+1} -  
  w' \Psl_{\nRepLeft-1,\nRepLeft+1} -
  t'\Esl_{\nRepLeft-1}  \\[2mm]
  \HRV \ =  & -t''\Esl_{2\nRepBulk + \nRepLeft - 1} -
  w'' \Psl_{2\nRepBulk + \nRepLeft - 2, 2\nRepBulk + \nRepLeft}-
  v \sum_{i=2\nRepBulk + \nRepLeft}^{2\nRepBulk + \nRepLeft + \nRepRight - 2}
  \Psl_{i,i+1}\ .
 \end{align}
In words, the Temperley-Lieb generators operate only in the bulk spaces,
whereas the permutation operators act on the boundary spaces and on
the first physical space of the {\em same} type as the boundary spaces.

 Numerical evidence was presented \cite{Candu2010} to the effect that these chains
 correspond to the sigma model with monopole boundary conditions
 according to the correspondence
\begin{align}\label{eq:mon_charges_bd_1}
  V^{\otimes \nRepLeft} \otimes
  (V\otimes V^\star)^{\otimes \nRepBulk}
  \otimes \ (V^\star)^{\otimes \nRepRight}:
  & &\LMon\ =\ +\nRepLeft \quad \RMon \ =\  +\nRepRight\, , \\[1mm]
  \label{eq:mon_charges_bd_2}
  V^{\otimes \nRepLeft} \otimes
  (V\otimes V^\star)^{\otimes \nRepBulk}
  \otimes \ V^{\otimes \nRepRight}:
  & &\LMon\ =\ +\nRepLeft \quad \RMon\ =\ -\nRepRight\, , \\[1mm]
  \label{eq:mon_charges_bd_3} 
  (V^\star)^{\otimes \nRepLeft} \otimes 
  (V\otimes V^\star)^{\otimes \nRepBulk} \otimes 
  (V^\star)^{\otimes \nRepRight}:
  & &\LMon\ =\ -\nRepLeft\quad \RMon \ =\  +\nRepRight\, , \\[1mm] 
  \label{eq:mon_charges_bd_4}
  (V^\star)^{\otimes m} \otimes (V\otimes V^\star)^{\otimes \nRepBulk}
  \otimes \ V^{\otimes n}:  
  & &\LMon\ = \ -\nRepLeft\quad \RMon\ =\ -\nRepRight\, .
\end{align}
It was also argued that 
\begin{equation}
\theta=\pi
\end{equation}
and that, for the value $w=0$ to which we restrict here,
\begin{equation}
\diffBosFer=0,~~  g_\sigma^2=1\, .
\end{equation}

\subsection{Mapping on sigma models: the role of boundaries}

We can give a quick heuristic derivation of these results  following
\cite{Ng1994a,Qin1995}. Consider the standard mapping of the spin chain
onto the sigma model. Calling the elementary Berry phase $\Omega(i)$,
the total phase for the usual antiferromagnetic chain of spin $s$
would read
\begin{equation}
  \Stop=s\sum_{i=1}^{2\nRepBulk} (-1)^i\Omega(i)
  \approx 
  {s\over 2} \int_{0}^{2\nRepBulk} 
  {\partial\Omega\over\partial x}dx
  =
  {s\over 2}\left[\Omega(2\nRepBulk)-\Omega(0)+4\pi Q\right]
\end{equation}
where $Q$ is an integer. Let us now suppose we have for instance the
first chain in equation (\ref{eq:mon_charges_bd_1}). Since on the
boundary we have representations of the same kind, the Berry phase on
that side comes unstaggered, so we have (from now on we put $s={1\over
  2}$, which is the right value for our spin chain \cite{Read2001})
\begin{equation}
  \Stop
  =
  -{\nRepLeft \over 2}\Omega(0)
  + 
  {\nRepRight \over 2}\Omega(2\nRepBulk) 
  +
  {1\over 2}\sum_{i=1}^{2\nRepBulk} (-1)^i \Omega(i)
  \approx 
  {\nRepRight + 1/2\over 2}\Omega(2\nRepBulk) 
  -
  {\nRepLeft + 1/ 2\over 2}\Omega(0)
  +\pi Q
\end{equation}
giving immediately rise to the equations (\ref{glue}) with
$\theta=\pi$. Meanwhile, say for the second chain in
(\ref{eq:mon_charges_bd_1}) we have
 \begin{equation}
   \Stop
   =
   -{\nRepLeft \over 2}\Omega(0)
   -
   {\nRepRight \over 2}\Omega(2\nRepBulk)
   +
   {1\over 2}\sum_{i=1}^{2\nRepBulk}(-1)^i \Omega(i)
   \approx 
   {-\nRepRight + 1/2\over 2}\Omega(2\nRepBulk) 
   -
   {\nRepLeft + 1/2\over 2}\Omega(0)
   +\pi Q
\end{equation}
again in agreement with equations (\ref{glue}) this time with
$\LMon=\nRepLeft$ but $\RMon=-\nRepRight$. Note that in this kind of
argument, the exact nature of the couplings on the boundary is
irrelevant. Note also that it does not make a difference whether the
spins on the boundary are actually projected onto the fully symmetric
representation or not: we will discuss this in more details below.

We thus see that adjusting the $\LMon,\RMon$ terms in the formal
boundary action of the sigma model is equivalent to adding extra spins---or
edge states---to the boundaries of the spin chain. As commented
earlier, the model with $\theta=\pi$ and $\LMon=\RMon=p\neq 0$ is
equivalent to the same model with $\theta=\pi+2\pi p$ and
$\LMon=\RMon=0$. Hence shifting $\theta$ by multiples of $2\pi$ leads
to the apparition of extra edge states.  This phenomenon is well known
in the language of QED (see \cite{Coleman1976}), to which the
$\mathbb{CP}^{\diffBosFer-1}$ model is equivalent at large $\diffBosFer$, with
$\diffBosFer$ flavors and a weak gauge coupling $e^2\approx {1\over
  \diffBosFer}$, as reviewed in \cite{Affleck1985}. The topological
term is equivalent to a background electrostatic field $F$ in the
one-dimensional universe, with $F={e\theta\over 2\pi}$. If this field is
too large, it is energetically favorable to produce quark-antiquark
pairs. For one such pair, $l$ being the distance between the quark and
antiquark, the difference in energy between the state with and without
the pair is
$$
\Delta E=l\left[(F\pm e)^2-F^2\right]
$$
While it is not energetically favourable for the vacuum to produce a
pair if $|F|\leq {1\over 2}e$, it becomes so if $|F|>{1\over 2}
e$. Pairs will in fact be produced until $F$ is brought down to a
value $|F_{\mbox{\scriptsize{screened}}}|\leq {1\over 2}e$. So if
$\pi<\theta<3\pi$, one pair is produced, and more generally if
$(2n-1)\pi<\theta<(2n+1)\pi$, $n$ pairs are produced.

This is in fact for $F$ positive. If $F$ is negative, that is if
$\theta<0$, things are quite similar, only one produces this time
antiquark-quark pairs. The picture is thus in agreement with the
previous findings in the case where there are extra representations $V^p$ on the left
and $(V^*)^p$ on the right ($\LMon=\RMon=p$), or $(V^*)^p$ on the left
and $V^p$ on the right ($\LMon=\RMon=-p$), with
$\Theta={\theta\over\pi}+2\RMon=1+2\RMon$. This is once we have
identified the chain with no extra representations as corresponding to
$\theta=\pi$ ($s={1\over 2}$).

Restricting now to the case $\LMon=\RMon$, we observe that another way
to have the same physics would be, instead of taking a spin chain with
$s={1\over 2}$ in the bulk and adding edge states by hand, to take
directly a chain with spin $s={1\over 2}+\LMon={1+2\LMon \over 2}$. In
other words, with a spin $s$ ($s$ half an odd integer) chain in the
bulk, we expect physics of edge states with $\LMon = s-{1\over 2}$, which
corresponds in turn to a spin on the boundary with value
$s'={s-1/2\over 2}$.

In the usual case of the $\mathbb{CP}^1$ model---that is, the $O(3)$
model and the XXX spin chain---all this discussion is somewhat
irrelevant.  Although edge states still present interesting features,
we anticipate that adding extra spins (maybe in a higher dimensional
representation) on the boundary does not give rise to new exponents,
and the spectrum will in the end coincide with that of the XXX chain
with an odd or an even number of sites.  In the case of \CP\ sigma
models however, an infinity of conformal boundary conditions is
available, leading to a much richer behavior.

\section{Spin chains and loop models}

In what follows we will derive results for super spin chains of the type described in section~\ref{sec:super_spin_chains} when
$\diffBosFer$ is not fixed to zero. In particular, we focus now on the case of the chain in equation \eqref{eq:mon_charges_bd_1}, which
we recall:
\begin{equation}        
  \label{eq:space_states}               
  V^{\otimes \nRepLeft} \otimes (V \otimes V^\star)^{\otimes \nRepBulk}\otimes 
  (V^\star)^{\otimes \nRepRight} \, ,
\end{equation}
where as before $V$ and $V^\star$ are two graded vector spaces of even
dimension $N+M$ and odd dimension $N$, taken respectively as the
fundamental and the dual representation of $\slnm$.

\subsection{Hamiltonians and transfer matrices}

We give first the explicit expression of the action of operators on
this spin chain.  Let $e_j$, with $j = 1, \dots, 2\nFer+\diffBosFer$,
be a basis of $V$, and $e^j$, with $j = 1, \dots, 2\nFer+\diffBosFer$,
be the corresponding dual basis of $V^\star$, and let $|j|$ denote the
grade of $e_j$. Then the graded permutation of index $i,i+1$, with $0\le i
\le \nRepLeft-1$ or $\nRepLeft + 2\nRepBulk-1 \le i \le 
\nRepLeft + 2\nRepBulk + \nRepRight - 2$, acts on two
copies of the same representation by interchanging them with a minus
sign if both the elements are odd:
\begin{equation}
  \label{eq:Psli}
  \Psl_{i,i+1} 
  \, e_{j_{0}} \otimes \cdots \otimes e^{j_{\nRepLeft+2\nRepBulk+\nRepRight-1} }
  = (-)^{|j_i||j_{i+1}|} e_{j_0} \otimes \cdots \otimes
  e_{j_{i+1} } \otimes e_{j_{i}} \otimes \cdots \otimes
  e^{j_{\nRepLeft+2\nRepBulk+\nRepRight-1} } \, ,
\end{equation}
The Temperley-Lieb operator of index $i$, $\nRepLeft \le
i \le \nRepLeft+2\nRepBulk-2 $  projects onto the singlet 
in the decomposition of $V\otimes V^\star$ (and $V^\star \otimes V$):
\begin{equation}
  \label{eq:Esli}
  \begin{split}
    \Esl_{i} \, e_{j_{0}} \otimes \cdots \otimes
    e^{j_{\nRepLeft+2\nRepBulk+\nRepRight-1} } =
    (-)^{|j_i||j_{i+1}|} \delta_{i}^{i+1} \sum_{k=1}^{\nFer+\diffBosFer}
    e_{j_0} \otimes \cdots \otimes
    e^{j_{i-1}} \otimes \\
    \otimes e_{k} \otimes e^{k} \otimes e_{j_{i+2}} \otimes \cdot
    \cdots \otimes e^{j_{\nRepLeft+2\nRepBulk+\nRepRight-1} } \, .
\end{split}
\end{equation}

The model we consider is  given by the following
Hamiltonian:
\begin{equation}
  \label{eq:hamiltonian}
  H = - u \sum_{i=0}^{\nRepLeft-1} \Psl_{i,i+1} - 
  \sum_{i=\nRepLeft}^{2\nRepBulk+\nRepLeft-2} \Esl_i -
  v \sum_{i=2\nRepBulk+\nRepLeft-1}^{2\nRepBulk+\nRepLeft+\nRepRight-2} \Psl_{i,i+1}\, .
\end{equation}

\begin{figure}[htp]
  \centering
\begin{tikzpicture}[very thick,decoration={ markings, mark=at position
    0.5 with {\arrow{>}}}, scale=0.75]
  \foreach \x in {-2} { \foreach \y in {0,2,4} {
      \draw[postaction={decorate},blue] (\x,\y)--(\x+1,\y+1);
      \draw[postaction={decorate},blue] (\x+1,\y+1)--(\x,\y+2); } }
  \foreach \x in {-1} { \foreach \y in {0,2,4} {
      \draw[postaction={decorate},blue] (\x+1,\y)--(\x,\y+1);
      \draw[postaction={decorate},blue] (\x,\y+1)--(\x+1,\y+2); } }
  \foreach \x in {0,2,4} { \foreach \y in {0,2,4} {
      \draw[postaction={decorate},blue] (\x,\y)--(\x+1,\y+1);
      \draw[postaction={decorate},blue] (\x+1,\y+1)--(\x,\y+2); } }
  \foreach \x in {1,3,5} { \foreach \y in {0,2,4} {
      \draw[postaction={decorate},red] (\x+1,\y+2)--(\x,\y+1);
      \draw[postaction={decorate},red] (\x,\y+1)--(\x+1,\y); } }
  \foreach \x in {6} { \foreach \y in {0,2,4} {
      \draw[postaction={decorate},red] (\x,\y+2)--(\x+1,\y+1);
      \draw[postaction={decorate},red] (\x+1,\y+1)--(\x,\y); } }
  \foreach \x in {7} { \foreach \y in {0,2,4} {
      \draw[postaction={decorate},red] (\x+1,\y+2)--(\x,\y+1);
      \draw[postaction={decorate},red] (\x,\y+1)--(\x+1,\y); } }
  \draw[dashed,black!25] (-2.5,0.5)--(8.5,0.5);
  \node at (-2.5,1) {$Y$};
  \draw[dashed,black!25] (-2.5,1.5)--(8.5,1.5);
  \node at (-2.5,2) {$X$};
  \draw[dashed,black!25] (-2.5,2.5)--(8.5,2.5);
  \draw[->] (9,1.5) -- (9,4.5);
  \node at (9.5,3) {$t$};
  \begin{scope}[yshift=0.5cm]
  \node at (-1,-1.5) {$m$};
  \node at (3,-1.5) {$2L$};
  \node at (7,-1.5) {$n$};
  \draw[thin,<->] (-2,-1)--(-0.05,-1);
  \draw[thin,<->] (0.05,-1)--(5.95,-1);
  \draw[thin,<->] (6.05,-1)--(8,-1);
\end{scope}
\end{tikzpicture}
  \caption{Vertex model corresponding to the super spin chain with 
    extra edges at the boundaries. Here $\nRepLeft=\nRepRight=2$.}
\label{fig:m_n_vertex}
\end{figure}
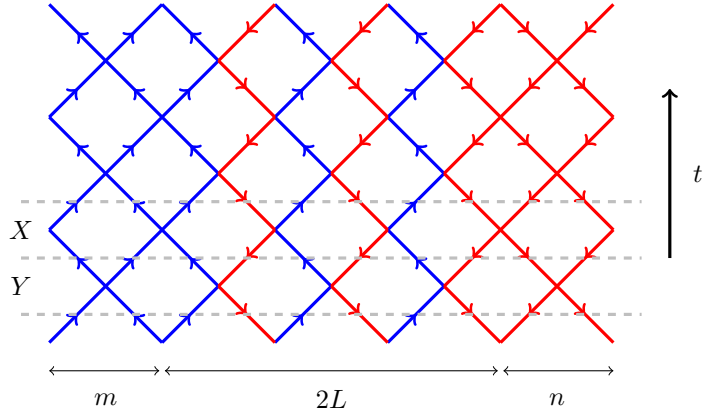

One can naturally associate to this Hamiltonian a vertex model defined
on a strip of length $\nRepLeft+2\nRepBulk+\nRepRight$, see figure
\ref{fig:m_n_vertex}. Like before, the orientations of edges are fixed
by the presence of $V$ and $V^\star$, so there are four types of
vertices. 

The transfer matrix at the critical point will be of the form $T = X
Y$: 
\begin{equation}
  \label{eq:transf_matXY}
  \begin{split}
  X 
  &= 
  \prod_{i=0}^{ \ii^-(\nRepLeft)}
  (1 + u \Psl_{2i+1,2i+2})
  \prod_{i=\ii(\nRepLeft)}
  ^{\nRepBulk-1+\ii^-(\nRepLeft)}
  (1 + \Esl_{2i+1})
  \prod_{i=\nRepBulk+\ii^-(\nRepLeft)}
  ^{\nRepBulk-1+\ii^-(\nRepLeft)+ 
  \ii^+(\nRepRight)}(1 + v \Psl_{2i+1,2i+2}) \\
  Y 
  &= 
  \prod_{i=0}^{\ii^+(\nRepLeft)-1}(1 + u \Psl_{2i,2i+1})
  \prod_{i=\ii^+(\nRepLeft)}
  ^{\nRepBulk-1+\ii(\nRepLeft)}
  (1 + \Esl_{2i}) 
  \prod_{i=\nRepBulk+\ii(\nRepLeft)}
  ^{\nRepBulk -1 +\ii(\nRepLeft)
    + \ii^+(\nRepRight)}
  (1 + v \Psl_{2i,2i+1}) \, ,
  \end{split}
\end{equation}
where $\ii(x)$ and $\ii^\pm(x)$ are the integer part of $x/2$ and
$(x\pm 1)/2$ respectively. The parameters $u,v$ are not necessarily
the same as those in the Hamiltonian, but they conceal the same physics:
they control the strength of the non-trivial part of the boundary interaction.
A more general transfer matrix
could be introduced with the $\Esl_i$ interaction chosen different for
even and odd $i$, as is the case for the purely alternating model
\cite{Read2001}.

\subsection{Formulation within the Brauer algebra}

As mentioned earlier, the generators $\Esl$ obey the defining
relations of the Temperley-Lieb algebra. Adding the $\Psl$ generators
as needed to define the Hamiltonian \eqref{eq:hamiltonian}, the
Temperley-Lieb relations are completed by additional ones, which
define a certain subalgebra of the Brauer algebra that we will denote
$\Alg$.

It is useful at this stage to remind the reader that the full
Brauer algebra on $2\nRepBulk$ sites $\mathcal{B}_{2\nRepBulk}(\fug)$ is
generated by the permutation operators $P_i (:=P_{i,i+1})$, with
$i=0,\dots,2\nRepBulk-2$, and the Temperley-Lieb generators $E_i$,
with $i=0,\dots,2\nRepBulk-2$, and is specified by another parameter $\fug$
which we will call---for reasons that will soon become obvious---the loop fugacity
$\fug$.

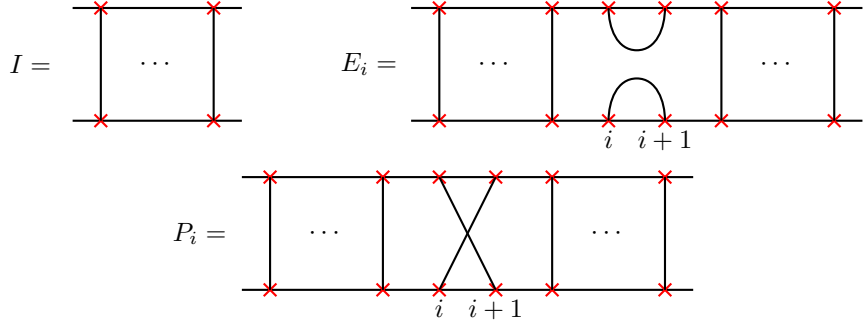
\begin{figure}[htp]
  \centering
\begin{tikzpicture}[thick,scale=0.75,dotdia/.style={cross out, draw,
    solid, red, inner sep=2pt}]
  \begin{scope}[xshift=0cm]
  \node at (-1.25,0) {$I = $};
  \draw (-0.5,-1) -- (2.5,-1);
  \node [dotdia] (a) at (0,-1) {}; 
  \node [dotdia] (b) at (2,-1) {};
  \draw (-0.5,1) -- (2.5,1);
  \node [dotdia] (c) at (0,1) {};
  \node [dotdia] (d) at (2,1) {};
  \node at (1,0) {$\cdots$};
  \draw (0,-1) -- (0,1);
  \draw (2,-1) -- (2,1);
  \end{scope}

  \begin{scope}[xshift=6cm]
  \node at (-1.25,0) {$E_i = $};
  \draw (-0.5,-1) -- (7.5,-1);
  \draw (-0.5,1) -- (7.5,1);
  \node [dotdia] at (0,-1) {};
  \node [dotdia] at (0,1) {};
  \draw (0,-1) -- (0,1);
  \node at (1,0) {$\cdots$};
  \node [dotdia] at (2,-1) {};
  \node [dotdia] at (2,1) {};
  \draw (2,-1) -- (2,1);  
  \node [dotdia] (i) at (3,-1) {};
  \node [below] at (i) {$i$};
  \node [dotdia] at (3,1) {};
  \node [dotdia] (ip) at (4,-1) {};
  \node [below] at (ip) {$i+1$};
  \node [dotdia] at (4,1) {};
  \draw (3,1) to[out=-90,in=180] (3.5,0.25) to[out=0,in=-90] (4,1);
  \draw (3,-1) to[out=90,in=180] (3.5,-0.25) to[out=0,in=90] (4,-1);
  \node [dotdia] at (5,-1) {};
  \node [dotdia] at (5,1) {};
  \draw (5,-1) -- (5,1);  
  \node at (6,0) {$\cdots$};
  \node [dotdia] at (7,-1) {};
  \node [dotdia] at (7,1) {};
  \draw (7,-1) -- (7,1);  
  \end{scope}

  \begin{scope}[yshift=-3cm,xshift=3cm]
  \node at (-1.25,0) {$P_i = $};
  \draw (-0.5,-1) -- (7.5,-1);
  \draw (-0.5,1) -- (7.5,1);
  \node [dotdia] at (0,-1) {};
  \node [dotdia] at (0,1) {};
  \draw (0,-1) -- (0,1);
  \node at (1,0) {$\cdots$};
  \node [dotdia] at (2,-1) {};
  \node [dotdia] at (2,1) {};
  \draw (2,-1) -- (2,1);  
  \node [dotdia] (i) at (3,-1) {};
  \node [below] at (i) {$i$};
  \node [dotdia] at (3,1) {};
  \node [dotdia] (ip) at (4,-1) {};
  \node [below] at (ip) {$i+1$};
  \node [dotdia] at (4,1) {};
  \draw (3,-1) to (4,1);
  \draw (4,-1) to (3,1);
  \node [dotdia] at (5,-1) {};
  \node [dotdia] at (5,1) {};
  \draw (5,-1) -- (5,1);  
  \node at (6,0) {$\cdots$};
  \node [dotdia] at (7,-1) {};
  \node [dotdia] at (7,1) {};
  \draw (7,-1) -- (7,1);  
  \end{scope}  

\end{tikzpicture}
  \caption{Identity, the Temperley-Lieb generator $E_i$ and the
    permutation operator $P_i$, in the diagrammatic representation.}
\label{fig:IdEiPi}
\end{figure}

It is well known that words in $\mathcal{B}_{2\nRepBulk}(\fug)$ can be
represented graphically using diagrams composed by rows of dots
connected in pairs. The generators $I$ (identity), $E_i$ (Temperley-Lieb operator) and $P_i$ (permutation operator) are shown in figure \ref{fig:IdEiPi} 
in this graphical representation.
The Brauer algebra is then the
$\mathbb{C}$-span of diagrams thus obtained.  The product of diagrams
$d_1\cdot d_2$ is defined by placing $d_1$ over $d_2$ and identifying
the bottom dots of $d_1$ with the top dots of $d_2$, replacing every
loop formed by the (complex) weight $\fug$.  For an abstract
introduction to the Brauer algebra, see \cite{Ram1995,Doran1999}, and
for a study in the context of super spin chains, see \cite{Candu2009a}.

The abstract relations defining $\mathcal{B}_{2\nRepBulk}(\fug)$ are
the following:
\begin{align}
 \label{eq:def_brauer1}
 & P_i^2 = 1, \quad E_i^2 = \fug E_i, \quad E_i P_i = P_i E_i = E_i, \\
 \label{eq:def_brauer2}
 & P_i P_{i \pm 1} P_i = P_{i \pm 1} P_i P_{i \pm 1}, \quad
   E_i E_{i \pm 1} E_i = E_i, \\
 \label{eq:def_brauer3}
 & P_i E_{i \pm 1} E_i = P_{i \pm 1} E_i, \quad
   E_i E_{i \pm 1} P_i = E_i P_{i \pm 1} \,.
\end{align}
and for non-adjacent sites, $|i - j| > 1$, all generators commute:
\begin{equation}
 \label{eq:def_brauer4}
 P_i P_j = P_j P_i, \quad E_i E_j = E_j E_i, \quad E_i P_j = P_j E_i \,.
\end{equation}

In the case we wish to study, the full Brauer algebra
$B_{2\nRepBulk}(\fug)$ is not needed, since we restrict to having
permutations on boundary sites only (the Temperley-Lieb generators
still act in the bulk). The corresponding subalgebra will be denoted
$\Alg(\fug)$, and its generators satisfy the relations
\begin{align}
  \label{eq:def_A1}
 & P_i^2 = 1, \quad E_i^2 = \fug E_i, \quad E_i P_i = P_i E_i = E_i, \\ 
  \label{eq:def_A2}
 & P_i P_{i \pm 1} P_i = P_{i \pm 1} P_i P_{i \pm 1}, \quad
   E_i E_{i \pm 1} E_i = E_i, \\
   \label{eq:def_A3}
 & E_i P_{i \pm 1} E_i = E_i, \\ 
   \label{eq:def_A4}
 & P_i P_j = P_j P_i, \quad E_i E_j = E_j E_i, \quad E_i P_j = P_j E_i \,,
\end{align}
where we note that the two relations \eqref{eq:def_brauer3} have been
replaced by the single relation \eqref{eq:def_A3} (which is implied by
the former two within the full Brauer algebra). One can easily verify
these defining relations by expressing the generators as diagrams.

We shall not devote much time to studying the algebra $\Alg(\fug)$
abstractly, as our only purpose is to use it as a tool to diagonalize
our Hamiltonians. Apart from providing a convenient language and
intuition, the algebraic point of view indeed will allow us to
separate the spectrum into different sectors \cite{martin1991potts},
and, within each sector, to considerably reduce the size of the
Hilbert space necessary for numerical determination or verification of
the exponents.

\subsection{Spin chain representation and quotients}
\label{sec:alg_quot}

In the spin chain, a representation of the algebra $\Alg(\fug)$, with
loop fugacity $\beta$ equal to $\STr 1 =\diffBosFer$, is obtained if
we take the action of $P_i$ and $E_i$ given by $\Psl_i$ and $\Esl_i$
(see eqs.~\eqref{eq:Psli} and \eqref{eq:Esli}).  This representation
is however not faithful: in addition to
\eqref{eq:def_A1}--\eqref{eq:def_A4}, the generators acting on the
spin chain satisfy additional relations. These additional relations
define a quotient of $\Alg(\diffBosFer)$, which we now describe.

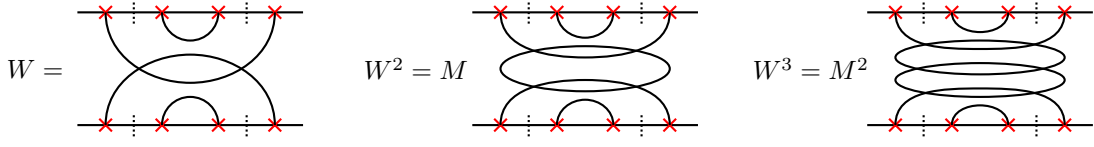
\begin{figure}[htp]
  \centering
\begin{tikzpicture}[thick,scale=0.75,dotdia/.style={cross out, draw,
    solid, red, inner sep=2pt}]
  \begin{scope}[xshift=0cm]
    \node at (-1.25,0) {$W = $};
    \draw (-0.5,-1) -- (3.5,-1);
    \draw (-0.5,1) -- (3.5,1);
    \foreach \x in
    {0,1,2,3}
    {
      \node [dotdia] at (\x,1) {}; 
      \node [dotdia] at (\x,-1) {}; 
    }
    
    \draw (1,-1) to[out=90,in=180]
    (1+0.5,-0.5) to[out=0,in=90] (1+1,-1); 
    \draw (1,1) to[out=-90,in=180]
    (1+0.5,0.5) to[out=0,in=-90] (1+1,1); 
    
    \draw (0,-1) to[out=90,in=180]
    (0+1.5,0.25) to[out=0,in=90] (0+3,-1); 
    \draw (0,1) to[out=-90,in=180]
    (0+1.5,-0.25) to[out=0,in=-90] (0+3,1); 
    \foreach \y in
    {-1,1}
    {
      \draw[densely dotted] (0.5,\y-0.2)--(0.5,\y+0.2); 
      \draw[densely dotted] (2.5,\y-0.2)--(2.5,\y+0.2); 
    }
  \end{scope}

  \begin{scope}[xshift=7cm]
    \node at (-1.5,0) {$W^2 = \diffBosFer$};
    \draw (-0.5,-1) -- (3.5,-1);
    \draw (-0.5,1) -- (3.5,1);
    \foreach \x in
    {0,1,2,3}
    {
      \node [dotdia] at (\x,1) {}; 
      \node [dotdia] at (\x,-1) {}; 
    }
    
    \draw (1,-1) to[out=90,in=180]
    (1+0.5,-0.55) to[out=0,in=90] (1+1,-1); 
    \draw (1,1) to[out=-90,in=180]
    (1+0.5,0.55) to[out=0,in=-90] (1+1,1); 
    
    \draw (0,-1) to[out=90,in=180]
    (0+1.5,-0.2) to[out=0,in=90] (0+3,-1); 
    \draw (0,1) to[out=-90,in=180]
    (0+1.5,0.2) to[out=0,in=-90] (0+3,1); 
    \draw (1.5,0) ellipse (1.5 and 0.4);
    \foreach \y in
    {-1,1}
    {
      \draw[densely dotted] (0.5,\y-0.2)--(0.5,\y+0.2); 
      \draw[densely dotted] (2.5,\y-0.2)--(2.5,\y+0.2); 
    }
  \end{scope}

  \begin{scope}[xshift=14cm]
    \node at (-1.5,0) {$W^3 = \diffBosFer^2$};
    \draw (-0.5,-1) -- (3.5,-1);
    \draw (-0.5,1) -- (3.5,1);
    \foreach \x in
    {0,1,2,3}
    {
      \node [dotdia] at (\x,1) {}; 
      \node [dotdia] at (\x,-1) {}; 
    }
    
    \draw (1,-1) to[out=90,in=180]
    (1+0.5,-0.65) to[out=0,in=90] (1+1,-1); 
    \draw (1,1) to[out=-90,in=180]
    (1+0.5,0.65) to[out=0,in=-90] (1+1,1); 
    
    \draw (0,-1) to[out=90,in=180]
    (0+1.5,-0.35) to[out=0,in=90] (0+3,-1); 
    \draw (0,1) to[out=-90,in=180]
    (0+1.5,0.35) to[out=0,in=-90] (0+3,1);
    \draw (1.5,0.2) ellipse (1.5 and 0.3);
    \draw (1.5,-0.2) ellipse (1.5 and 0.3); 
    \foreach \y in
    {-1,1}
    {
      \draw[densely dotted] (0.5,\y-0.2)--(0.5,\y+0.2); 
      \draw[densely dotted] (2.5,\y-0.2)--(2.5,\y+0.2); 
    }
  \end{scope}

\end{tikzpicture}
  \caption{The diagrams corresponding to the words $W=E_1P_2P_0E_1$,
    $W^2$ and $W^3$ in $\Alg(\diffBosFer)$ with
    $\nRepBulk=\nRepLeft=\nRepRight=1$. 
    Here and in the following dotted lines separate bulk and
  boundary sites.}
\label{fig:w-w2-w3}
\end{figure}

First we note that the algebra $\Alg(\diffBosFer)$ as defined by
\eqref{eq:def_A1}--\eqref{eq:def_A4} is infinite dimensional. We can
understand why by considering the simple system with
$\nRepBulk=\nRepLeft=\nRepRight=1$. If we call $W = E_1P_2P_0E_1$,
then the word $W^p$ (see figure \ref{fig:w-w2-w3}) cannot be reduced
using the defining relations, and increasing the power $p$, the
dimension of the algebra becomes arbitrarily large.

It is also clear that this feature is present only
if both $\nRepLeft$ and $\nRepRight$ are non zero, and the larger
$\nRepLeft$ and $\nRepRight$, the larger the number of words that
cannot be reduced. A similar situation appears also for the
two-boundary Temperley-Lieb algebra \cite{DEGIER2009}, which indeed is
related to our model as will be explained later.

The proper algebraic setting for our model is a quotient of
$\Alg(\diffBosFer)$.  We will now give the expression for the
additional relations present in the spin chain representation. Call
$\minlr = \min(\nRepLeft,\nRepRight)$ and
\begin{equation}
  \label{eq:defs1_quotient}
  E = \prod_{i \in I_1} E_i \prod_{i \in I_2} E_i \, ,
\end{equation}
where $I_1$ ($I_2$) is the set of even (odd) indices of Temperley-Lieb
generators if $\nRepLeft$ is odd or the set of odd (even) indices of
Temperley-Lieb generators if $\nRepLeft$ is even.  To proceed, let us
define the words $W_r$ by the recurrence relations:
\begin{align}
  W_0 &= E \\
  W_r &= E P_{\nRepLeft-1} \cdots P_{\nRepLeft-r}
  P_{\nRepLeft+2\nRepBulk-1} \cdots P_{\nRepLeft+2\nRepBulk+r-2} W_{r-1} \, ,
\end{align}
where $r = 1, \dots, \minlr$. Also define 
\begin{equation}
  \label{eq:gen_kerpspin}
  K_{ir} = P_{\nRepLeft-i} W_r - P_{\nRepLeft+2\nRepBulk+i-2} W_r \, , \quad 
  i = 1,\dots,r \, ,
\end{equation}
for $r=1,\dots,\minlr$, and call the ideal generated by these elements
$I$. We can now  take the quotient $\Quot(\diffBosFer) =
\Alg(\diffBosFer)/I$: we claim that the super spin chain
representation of $\Quot(\diffBosFer)$ is faithful for a sufficiently
large number of states $\diffBosFer$.

The mechanism underlying the relations in the quotient can be
understood from simple examples and easily generalized. Take again the
system $\nRepBulk=\nRepLeft=\nRepRight=1$. Then the additional
relation in the spin chain is $P_0 E_1 P_0 P_2 E_1 = P_2 E_1 P_0 P_2
E_1$. Indeed If we compute the action of $E_1 P_0 P_2 E_1$ on the spin
chain, we have:
\begin{equation}
  \label{eq:e1p0p2e1}
  \begin{split}
    E_1 P_0 P_2 E_1 \, e_{j_{0}} \otimes e_{j_{1}} \otimes e^{j_{2}}
    \otimes e^{j_{3}} = (-)^{|j_1|} \delta_{j_1}^{j_2}
    \delta_{j_0}^{j_3} \sum_{k,r = 1}^{\diffBosFer+\nFer}
    (-)^{|k|(|j_0|+|j_3|)}\, \cdot \\
    \cdot\, e_k \otimes e_r \otimes e^r \otimes e^k \, ,
\end{split}
\end{equation}
that is invariant under the action of $P_0 P_2$.

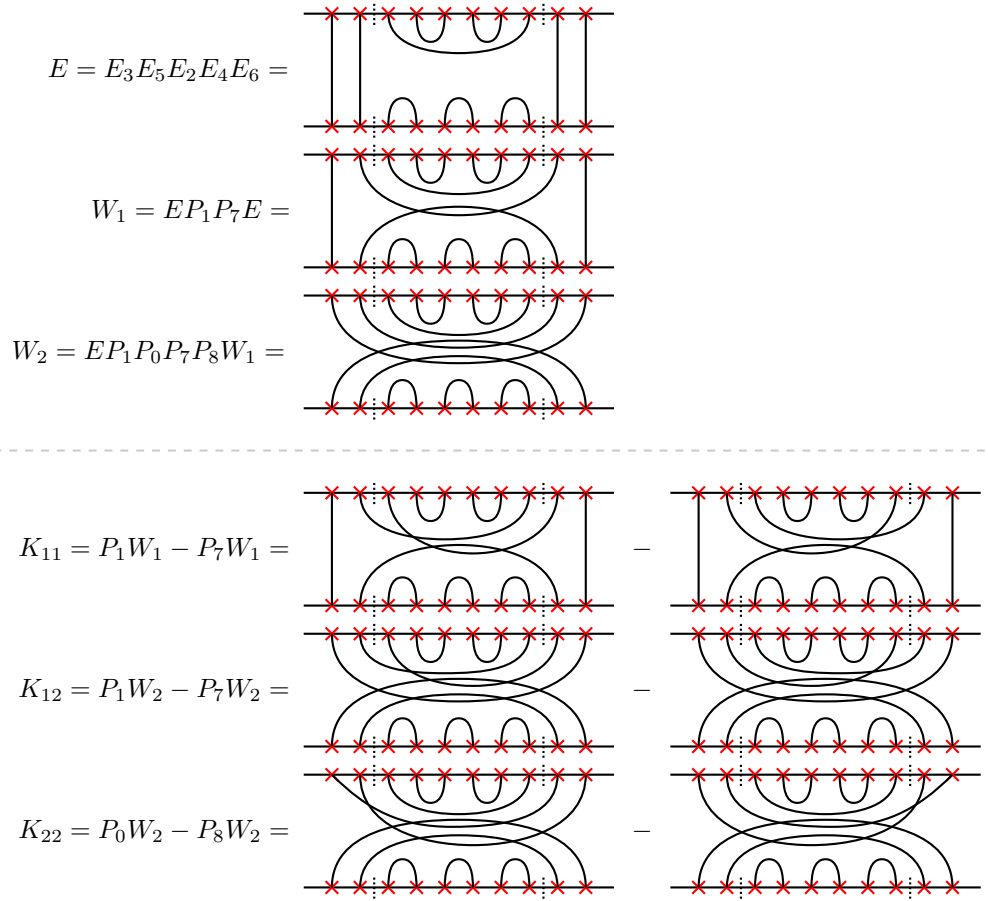
\begin{figure}[htp]
  \centering
\begin{tikzpicture}[thick,scale=0.75,dotdia/.style={cross out, draw,
    solid, red, inner sep=2pt}]
  \begin{scope}[xshift=0cm]
    \node at (-2.9,0) {$E = E_3E_5E_2E_4E_6 =$};
    \draw (-0.5,-1) -- (5,-1); 
    \draw (-0.5,1) -- (5,1); 
    \foreach \x in
    {0,0.5,1,1.5,2,2.5,
      3,3.5,4,4.5}
    {
      \node [dotdia] at (\x,-1) {}; 
      \node [dotdia] at (\x,1) {}; 
    }
    \foreach \x in {0,0.5,4,4.5} 
    { 
      \draw (\x,-1)--(\x,1); 
    } 
    \foreach \x in {1,2,3} 
    { 
      \draw (\x,-1) to[out=90,in=180]
      (\x+0.25,-0.5) to[out=0,in=90] (\x+0.5,-1); 
    }
    \foreach \x in {1.5,2.5} 
    { 
      \draw (\x,1) to[out=-90,in=180]
      (\x+0.25,0.5) to[out=0,in=-90] (\x+0.5,1); 
    }
    \draw (1,1) to[out=-90,in=180]
    (1+1.25,0.3) to[out=0,in=-90] (1+2.5,1); 
    \foreach \y in
    {-1,1}
    {
      \draw[densely dotted] (0.75,\y-0.2)--(0.75,\y+0.2); 
      \draw[densely dotted] (3.75,\y-0.2)--(3.75,\y+0.2); 
    }
  \end{scope}
  \begin{scope}[yshift=-2.5cm]
    \node at (-2.5,0) {$W_1 = EP_1P_7E = $};
    \draw (-0.5,-1) -- (5,-1); 
    \draw (-0.5,1) -- (5,1); 
    \foreach \x in
    {0,0.5,1,1.5,2,2.5,
      3,3.5,4,4.5}
    {
      \node [dotdia] at (\x,-1) {}; 
      \node [dotdia] at (\x,1) {}; 
    }
    \foreach \x in {0,4.5} 
    { 
      \draw (\x,-1)--(\x,1); 
    } 
    \foreach \x in {1,2,3} 
    { 
      \draw (\x,-1) to[out=90,in=180]
      (\x+0.25,-0.5) to[out=0,in=90] (\x+0.5,-1); 
    }
    \foreach \x in {1.5,2.5} 
    { 
      \draw (\x,1) to[out=-90,in=180]
      (\x+0.25,0.5) to[out=0,in=-90] (\x+0.5,1); 
    }
    \draw (1,1) to[out=-90,in=180]
    (1+1.25,0.3) to[out=0,in=-90] (1+2.5,1); 
    \draw (0.5,1) to[out=-90,in=180]
    (0.5+1.75,-0.075) to[out=0,in=-90] (0.5+3.5,1); 
    \draw (0.5,-1) to[out=90,in=180]
    (0.5+1.75,0.075) to[out=0,in=90] (0.5+3.5,-1); 
    \foreach \y in
    {-1,1}
    {
      \draw[densely dotted] (0.75,\y-0.2)--(0.75,\y+0.2); 
      \draw[densely dotted] (3.75,\y-0.2)--(3.75,\y+0.2); 
    }
  \end{scope}
  \begin{scope}[yshift=-5cm]
    \node at (-3.25,0) {$W_2 = EP_1P_0P_7P_8W_1 = $};
    \draw (-0.5,-1) -- (5,-1); 
    \draw (-0.5,1) -- (5,1); 
    \foreach \x in
    {0,0.5,1,1.5,2,2.5,
      3,3.5,4,4.5}
    {
      \node [dotdia] at (\x,-1) {}; 
      \node [dotdia] at (\x,1) {}; 
    }
    \foreach \x in {1,2,3} 
    { 
      \draw (\x,-1) to[out=90,in=180]
      (\x+0.25,-0.5) to[out=0,in=90] (\x+0.5,-1); 
    }
    \foreach \x in {1.5,2.5} 
    { 
      \draw (\x,1) to[out=-90,in=180]
      (\x+0.25,0.5) to[out=0,in=-90] (\x+0.5,1); 
    }
    \draw (1,1) to[out=-90,in=180]
    (1+1.25,0.3) to[out=0,in=-90] (1+2.5,1); 
    \draw (0.5,1) to[out=-90,in=180]
    (0.5+1.75,0.075) to[out=0,in=-90] (0.5+3.5,1); 
    \draw (0,1) to[out=-90,in=180]
    (0+2.25,-0.2) to[out=0,in=-90] (0+4.5,1); 
    \draw (0.5,-1) to[out=90,in=180]
    (0.5+1.75,-0.075) to[out=0,in=90] (0.5+3.5,-1); 
    \draw (0,-1) to[out=90,in=180]
    (0+2.25,0.2) to[out=0,in=90] (0+4.5,-1); 
    \foreach \y in
    {-1,1}
    {
      \draw[densely dotted] (0.75,\y-0.2)--(0.75,\y+0.2); 
      \draw[densely dotted] (3.75,\y-0.2)--(3.75,\y+0.2); 
    }
  \end{scope}

  \draw [dashed,black!20] (-6,-6.75)--(12,-6.75);

  \begin{scope}[yshift=-8.5cm]
    \node at (-3.15,0) {$K_{11} = P_1W_1-P_7W_1 = $};
    \draw (-0.5,-1) -- (5,-1); 
    \draw (-0.5,1) -- (5,1); 
    \foreach \x in
    {0,0.5,1,1.5,2,2.5,
      3,3.5,4,4.5}
    {
      \node [dotdia] at (\x,-1) {}; 
      \node [dotdia] at (\x,1) {}; 
    }
    \foreach \x in {0,4.5} 
    { 
      \draw (\x,-1)--(\x,1); 
    } 
    \foreach \x in {1,2,3} 
    { 
      \draw (\x,-1) to[out=90,in=180]
      (\x+0.25,-0.5) to[out=0,in=90] (\x+0.5,-1); 
    }
    \foreach \x in {1.5,2.5} 
    { 
      \draw (\x,1) to[out=-90,in=180]
      (\x+0.25,0.5) to[out=0,in=-90] (\x+0.5,1); 
    }
    \draw (0.5,1) to[out=-90,in=180]
    (0.5+1.5,0.175) to[out=0,in=-90] (0.5+3,1); 
    \draw (1,1) to[out=-90,in=180]
    (1+1.5,-0.075) to[out=0,in=-90] (1+3,1); 
    \draw (0.5,-1) to[out=90,in=180]
    (0.5+1.75,0.075) to[out=0,in=90] (0.5+3.5,-1); 
 
    \node at (5.5,0) {$-$};
    \foreach \y in
    {-1,1}
    {
      \draw[densely dotted] (0.75,\y-0.2)--(0.75,\y+0.2); 
      \draw[densely dotted] (3.75,\y-0.2)--(3.75,\y+0.2); 
    }

    \begin{scope}[xshift=6.5cm]
      \draw (-0.5,-1) -- (5,-1); 
      \draw (-0.5,1) -- (5,1); 
      \foreach \x in
      {0,0.5,1,1.5,2,2.5,
        3,3.5,4,4.5}
      {
        \node [dotdia] at (\x,-1) {}; 
        \node [dotdia] at (\x,1) {}; 
      }
      \foreach \x in {0,4.5} 
      { 
        \draw (\x,-1)--(\x,1); 
      } 
      \foreach \x in {1,2,3} 
      { 
        \draw (\x,-1) to[out=90,in=180]
        (\x+0.25,-0.5) to[out=0,in=90] (\x+0.5,-1); 
      }
      \foreach \x in {1.5,2.5} 
      { 
        \draw (\x,1) to[out=-90,in=180]
        (\x+0.25,0.5) to[out=0,in=-90] (\x+0.5,1); 
      }
      \draw (1,1) to[out=-90,in=180]
      (1+1.5,0.175) to[out=0,in=-90] (1+3,1); 
      \draw (0.5,1) to[out=-90,in=180]
      (0.5+1.5,-0.075) to[out=0,in=-90] (0.5+3,1); 
      \draw (0.5,-1) to[out=90,in=180]
      (0.5+1.75,0.075) to[out=0,in=90] (0.5+3.5,-1); 
      \foreach \y in
      {-1,1}
      {
        \draw[densely dotted] (0.75,\y-0.2)--(0.75,\y+0.2); 
        \draw[densely dotted] (3.75,\y-0.2)--(3.75,\y+0.2); 
      }
    \end{scope}
  \end{scope}
  \begin{scope}[yshift=-11cm]
    \node at (-3.15,0) {$K_{12} = P_1W_2-P_7W_2 = $};
    \draw (-0.5,-1) -- (5,-1); 
    \draw (-0.5,1) -- (5,1); 
    \foreach \x in
    {0,0.5,1,1.5,2,2.5,
      3,3.5,4,4.5}
    {
      \node [dotdia] at (\x,-1) {}; 
      \node [dotdia] at (\x,1) {}; 
    }
    \foreach \x in {1,2,3} 
    { 
      \draw (\x,-1) to[out=90,in=180]
      (\x+0.25,-0.5) to[out=0,in=90] (\x+0.5,-1); 
    }
    \foreach \x in {1.5,2.5} 
    { 
      \draw (\x,1) to[out=-90,in=180]
      (\x+0.25,0.5) to[out=0,in=-90] (\x+0.5,1); 
    }
    \draw (0.5,-1) to[out=90,in=180]
    (0.5+1.75,-0.075) to[out=0,in=90] (0.5+3.5,-1); 
    \draw (0,-1) to[out=90,in=180]
    (0+2.25,0.2) to[out=0,in=90] (0+4.5,-1); 
    \draw (0.5,1) to[out=-90,in=180]
    (0.5+1.5,0.3) to[out=0,in=-90] (0.5+3,1); 
    \draw (1,1) to[out=-90,in=180]
    (1+1.5,0.075) to[out=0,in=-90] (1+3,1); 
    \draw (0,1) to[out=-90,in=180]
    (0+2.25,-0.2) to[out=0,in=-90] (0+4.5,1); 
    \foreach \y in
    {-1,1}
    {
      \draw[densely dotted] (0.75,\y-0.2)--(0.75,\y+0.2); 
      \draw[densely dotted] (3.75,\y-0.2)--(3.75,\y+0.2); 
    } 

    \node at (5.5,0) {$-$};

    \begin{scope}[xshift=6.5cm]
      \draw (-0.5,-1) -- (5,-1); 
      \draw (-0.5,1) -- (5,1); 
      \foreach \x in
      {0,0.5,1,1.5,2,2.5,
        3,3.5,4,4.5}
      {
        \node [dotdia] at (\x,-1) {}; 
        \node [dotdia] at (\x,1) {}; 
      }
      \foreach \x in {1,2,3} 
      { 
        \draw (\x,-1) to[out=90,in=180]
        (\x+0.25,-0.5) to[out=0,in=90] (\x+0.5,-1); 
      }
      \foreach \x in {1.5,2.5} 
      { 
        \draw (\x,1) to[out=-90,in=180]
        (\x+0.25,0.5) to[out=0,in=-90] (\x+0.5,1); 
      }
      \draw (0.5,-1) to[out=90,in=180]
      (0.5+1.75,-0.075) to[out=0,in=90] (0.5+3.5,-1); 
      \draw (0,-1) to[out=90,in=180]
      (0+2.25,0.2) to[out=0,in=90] (0+4.5,-1); 
      \draw (1,1) to[out=-90,in=180]
      (1+1.5,0.3) to[out=0,in=-90] (1+3,1); 
      \draw (0.5,1) to[out=-90,in=180]
      (0.5+1.5,0.075) to[out=0,in=-90] (0.5+3,1); 
      \draw (0,1) to[out=-90,in=180]
      (0+2.25,-0.2) to[out=0,in=-90] (0+4.5,1); 
      \foreach \y in
      {-1,1}
      {
        \draw[densely dotted] (0.75,\y-0.2)--(0.75,\y+0.2); 
        \draw[densely dotted] (3.75,\y-0.2)--(3.75,\y+0.2); 
      }
    \end{scope}
  \end{scope}
  \begin{scope}[yshift=-13.5cm]
    \node at (-3.15,0) {$K_{22} = P_0W_2-P_8W_2 = $};
    \draw (-0.5,-1) -- (5,-1); 
    \draw (-0.5,1) -- (5,1); 
    \foreach \x in
    {0,0.5,1,1.5,2,2.5,
      3,3.5,4,4.5}
    {
      \node [dotdia] at (\x,-1) {}; 
      \node [dotdia] at (\x,1) {}; 
    }
    \foreach \x in {1,2,3} 
    { 
      \draw (\x,-1) to[out=90,in=180]
      (\x+0.25,-0.5) to[out=0,in=90] (\x+0.5,-1); 
    }
    \foreach \x in {1.5,2.5} 
    { 
      \draw (\x,1) to[out=-90,in=180]
      (\x+0.25,0.5) to[out=0,in=-90] (\x+0.5,1); 
    }
    \draw (0.5,-1) to[out=90,in=180]
    (0.5+1.75,-0.075) to[out=0,in=90] (0.5+3.5,-1); 
    \draw (0,-1) to[out=90,in=180]
    (0+2.25,0.2) to[out=0,in=90] (0+4.5,-1); 
    \draw (1,1) to[out=-90,in=180]
    (1+1.25,0.3) to[out=0,in=-90] (1+2.5,1); 
    \draw (0,1) to[out=-45,in=180]
    (0+2.125,0.075) to[out=0,in=-90] (0+4,1); 
    \draw (0.5,1) to[out=-90,in=180]
    (0.5+2,-0.25) to[out=0,in=-90] (0.5+4,1); 
    \foreach \y in
    {-1,1}
    {
      \draw[densely dotted] (0.75,\y-0.2)--(0.75,\y+0.2); 
      \draw[densely dotted] (3.75,\y-0.2)--(3.75,\y+0.2); 
    } 

    \node at (5.5,0) {$-$};

    \begin{scope}[xshift=6.5cm]
      \draw (-0.5,-1) -- (5,-1); 
      \draw (-0.5,1) -- (5,1); 
      \foreach \x in
      {0,0.5,1,1.5,2,2.5,
        3,3.5,4,4.5}
      {
        \node [dotdia] at (\x,-1) {}; 
        \node [dotdia] at (\x,1) {}; 
      }
      \foreach \x in {1,2,3} 
      { 
        \draw (\x,-1) to[out=90,in=180]
        (\x+0.25,-0.5) to[out=0,in=90] (\x+0.5,-1); 
      }
      \foreach \x in {1.5,2.5} 
      { 
        \draw (\x,1) to[out=-90,in=180]
        (\x+0.25,0.5) to[out=0,in=-90] (\x+0.5,1); 
      }
      \draw (0.5,-1) to[out=90,in=180]
      (0.5+1.75,-0.075) to[out=0,in=90] (0.5+3.5,-1); 
      \draw (0,-1) to[out=90,in=180]
      (0+2.25,0.2) to[out=0,in=90] (0+4.5,-1); 
      \draw (1,1) to[out=-90,in=180]
      (1+1.25,0.3) to[out=0,in=-90] (1+2.5,1); 
    \draw (0.5,1) to[out=-90,in=180]
    (0.5+1.875,0.075) to[out=0,in=-135] (0.5+4,1); 
    \draw (0,1) to[out=-90,in=180]
    (0+2,-0.25) to[out=0,in=-90] (0+4,1); 
    \foreach \y in
    {-1,1}
    {
      \draw[densely dotted] (0.75,\y-0.2)--(0.75,\y+0.2); 
      \draw[densely dotted] (3.75,\y-0.2)--(3.75,\y+0.2); 
    }
  \end{scope}
\end{scope}
  
\end{tikzpicture}
  \caption{$K_{11}=0, K_{12}=0, K_{22}=0$ are the additional relations
    present in the super spin representation for a system with $\nRepBulk=3,
    \nRepLeft = 2, \nRepRight = 2$. }
\label{fig:quotientL3m2mbar2}
\end{figure}

It is useful to represent diagrammatically the relations. In figure
\ref{fig:quotientL3m2mbar2} we have represented them for the case
$\nRepBulk=3, \nRepLeft=\nRepRight=2$. 

Note that the additional relations just  introduced correctly reduce those
words that made the algebra infinite dimensional earlier. 

{}From the operational point of view which will be adopted later for
numerical diagonalization, the quotient tells us that when intertwined
top arcs intersect bottom arcs, their relative intertwining does not matter,
and could be disentangled.

\subsection{Geometrical formulation: the loop model representation}
\label{sec:loop_repr}

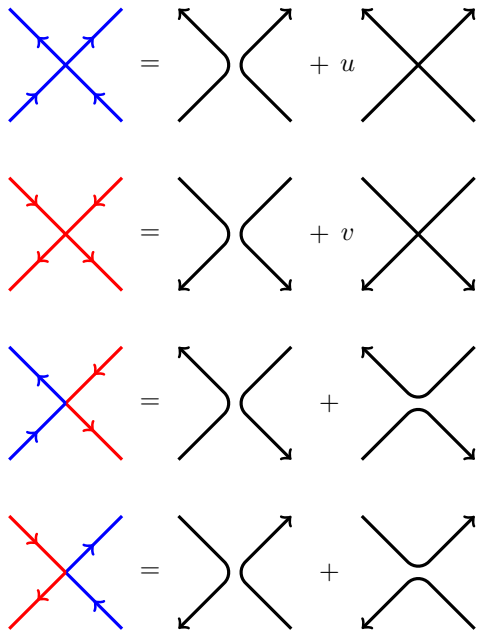
\begin{figure}[htp]
  \centering
\begin{tikzpicture}[very thick,decoration={
    markings,mark=at position 0.5 with {\arrow{>}}}, scale=0.75]
  \draw[postaction={decorate},blue] (0,0)--(1,1);
  \draw[postaction={decorate},blue] (1,1)--(0,2);
  \draw[postaction={decorate},blue] (2,0)--(1,1);
  \draw[postaction={decorate},blue] (1,1)--(2,2);
  \node at (2.5,1) {$=$};
  \draw[->] (3,0) [rounded corners=0.2cm] -- (4,1) -- (3,2);
  \draw[->] (5,0) [rounded corners=0.2cm] -- (4,1) -- (5,2);
  \node at (5.5,1) {$+$};
  \node at (6,1) {$u$};
  \draw[->] (6.25,0) -- (8.25,2);
  \draw[->] (8.25,0) -- (6.25,2);
  \begin{scope}[yshift=-3cm]
    \draw[postaction={decorate},red] (2,2)--(1,1);
    \draw[postaction={decorate},red] (1,1)--(2,0);
    \draw[postaction={decorate},red] (0,2)--(1,1);
    \draw[postaction={decorate},red] (1,1)--(0,0);
    \node at (2.5,1) {$=$};
    \draw[->] (3,2) [rounded corners=0.2cm] -- (4,1) -- (3,0);
    \draw[->] (5,2) [rounded corners=0.2cm] -- (4,1) -- (5,0);
    \node at (5.5,1) {$+$};
    \node at (6,1) {$v$};
    \draw[->] (6.25,2) -- (8.25,0);
    \draw[->] (8.25,2) -- (6.25,0);
  \end{scope}
  \begin{scope}[yshift=-6cm]
    \draw[postaction={decorate},red] (2,2)--(1,1);
    \draw[postaction={decorate},red] (1,1)--(2,0);
    \draw[postaction={decorate},blue] (0,0)--(1,1);
    \draw[postaction={decorate},blue] (1,1)--(0,2);
    \node at (2.5,1) {$=$};
    \draw[->] (3,0) [rounded corners=0.2cm] -- (4,1) -- (3,2);
    \draw[->] (5,2) [rounded corners=0.2cm] -- (4,1) -- (5,0);
    \node at (5.675,1) {$+$};
    \draw[->] (6.25,0) [rounded corners=0.2cm] -- (7.25,1) -- (8.25,0);
    \draw[->] (8.25,2) [rounded corners=0.2cm] -- (7.25,1) -- (6.25,2);
  \end{scope}
  \begin{scope}[yshift=-9cm]
    \draw[postaction={decorate},blue] (2,0)--(1,1);
    \draw[postaction={decorate},blue] (1,1)--(2,2);
    \draw[postaction={decorate},red] (0,2)--(1,1);
    \draw[postaction={decorate},red] (1,1)--(0,0);
    \node at (2.5,1) {$=$};
    \draw[->] (3,2) [rounded corners=0.2cm] -- (4,1) -- (3,0);
    \draw[->] (5,0) [rounded corners=0.2cm] -- (4,1) -- (5,2);
    \node at (5.675,1) {$+$};
    \draw[->] (8.25,0) [rounded corners=0.2cm] -- (7.25,1) -- (6.25,0);
    \draw[->] (6.25,2) [rounded corners=0.2cm] -- (7.25,1) -- (8.25,2);
  \end{scope}
\end{tikzpicture}
  \caption{Vertices of the isotropic vertex model decomposed into the
    oriented loop configurations. $u$ and $v$ are the weights of permutations
    of respectively up and down links.}
\label{fig:vertices}
\end{figure}

The diagram representation of the generators in our algebra naturally
leads to a loop reformulation of the model: we can simply expand the
product (\ref{eq:transf_matXY}) in the vertex model transfer matrix
and translate each term to obtain a sum over configurations of
oriented loops. When viewed along the time direction, {\em bulk}
vertices correspond to the junction of two incoming lines with {\em
  different} orientations.  These can be contracted (by $E_i$) or go
through (by $I$), as shown in the lower two lines of figure
\ref{fig:vertices}. On the other hand, {\em boundary} vertices
correspond to incoming lines having {\em identical} orientations.
These can cross (by $P_i$) or go through (by $I$), as shown in the
upper two lines of figure \ref{fig:vertices}.

There is thus a total of four possible
vertices, which can each be resolved in two ways.

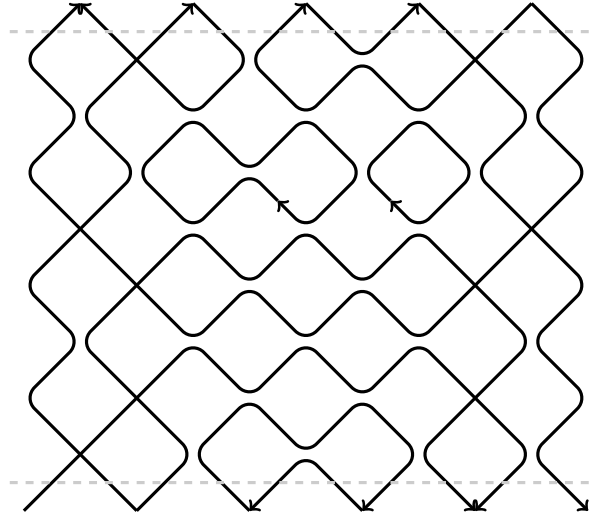
\begin{figure}[htp]
  \centering
\begin{tikzpicture}[very thick, scale=0.75]
  \draw[->,rounded corners=0.2cm]
  (0,0)-- ++(3,3)-- ++(1,-1)-- ++(1,1)-- ++(1,-1)-- ++(1,1)--
  ++(2,-2)-- ++(-1,-1);
  \draw[->,rounded corners=0.2cm]
  (2,0)-- ++(-2,2)-- ++(1,1)-- ++(-1,1)-- ++(2,2)-- ++(-1,1)--
  ++(2,2);
  \draw[->,rounded corners=0.2cm]
  (2,0)-- ++(1,1)-- ++(-2,2)-- ++(2,2)-- ++(1,-1)-- ++(1,1)--
  ++(1,-1)-- ++(1,1)-- ++(2,-2)-- ++(-2,-2)-- ++(1,-1);
  \draw[->,rounded corners=0.2cm]
  (4,0)-- ++(-1,1)-- ++(1,1)-- ++(1,-1)-- ++(1,1)-- ++(1,-1)--
  ++(-1,-1);
  \draw[->,rounded corners=0.2cm]
  (6,0)-- ++(-1,1)-- ++(-1,-1);
  \draw[->,rounded corners=0.2cm]
  (3,9)-- ++(1,-1)-- ++(-1,-1)-- ++(-2,2);
  \draw[->,rounded corners=0.2cm]
  (5,9)-- ++(1,-1)-- ++(1,1);
  \draw[->,rounded corners=0.2cm]
  (7,9)-- ++(2,-2)-- ++(-1,-1)-- ++(2,-2)-- ++(-1,-1)-- ++(1,-1)--
  ++(-1,-1)-- ++(1,-1);
  \draw[->,rounded corners=0.2cm]
  (9,9)-- ++(-2,-2)-- ++(-1,1)-- ++(-1,-1)-- ++(-1,1)-- ++(1,1);
  \draw[->,rounded corners=0.2cm]
  (9,9)-- ++(1,-1)-- ++(-1,-1)-- ++(1,-1)-- ++(-3,-3)-- ++(-1,1)--
  ++(-1,-1)-- ++(-1,1)-- ++(-1,-1)-- ++(-3,3)
  -- ++(1,1)-- ++(-1,1)-- ++(1,1);
  \draw[->,rounded corners=0.2cm]
  (4.5,5.5)-- ++(-0.5,0.5)-- ++(-1,-1)-- ++(-1,1)-- ++(1,1)-- 
  ++(1,-1)-- ++(1,1)-- ++(1,-1)-- ++(-1,-1)-- ++(-0.5,0.5);
  \draw[->,rounded corners=0.2cm]
  (6.5,5.5)-- ++(-0.5,0.5)-- ++(1,1)-- ++(1,-1)-- ++(-1,-1)-- 
  ++(-0.5,0.5);

  \draw[dashed,black!20] (-0.25,0.5)--(10.25,0.5);
  \draw[dashed,black!20] (-0.25,8.5)--(10.25,8.5);

\end{tikzpicture}
  \caption{A configuration of the supersymmetric lattice model in the
    case $\nRepLeft=\nRepRight=2$, $2\nRepBulk=6$. The loop
    orientation follows the presence of the fundamental representation
    of the Lie superalgebra $\slnm$ (links oriented up) or the dual
    representation (links oriented down). Each loop carries a weight $\diffBosFer$.}
\label{fig:model_super_spin}
\end{figure}

Expanding all the vertices into pieces of oriented loops we end up with a
sum over configurations like the one illustrated in figure
\ref{fig:model_super_spin}.  

Note that while the diagrams used earlier to define the Brauer algebra
involve unoriented lines, we have here defined the loop configurations
in terms of oriented lines. The subalgebra $\Alg$ respects this
orientation, in the sense that the associated Hamiltonian
\eqref{eq:hamiltonian} and transfer matrix \eqref{eq:transf_matXY} only
generate diagrams in which each loop has a definite orientation.  We
stress that this orientation is a property of the underlying lattice,
not of the loop itself. In particular, for a loop of a given position
and shape, no sum over orientations is implied, and the states
entering the loop representation do not contain any orientational
information. The weight of the loop $\beta$ is a parameter of the
algebra.

\begin{figure}[htp]
  \centering
\begin{tikzpicture}[thick,scale=1,dotdia/.style={cross out, draw,
    solid, red, inner sep=2pt}]
  \node at (-0.75,0) {$\beta^2$};
  \draw (-0.5,-1) -- (5,-1); 
  \draw (-0.5,1) -- (5,1); 
  \foreach \x in
  {0,0.5,1,1.5,2,2.5,
    3,3.5,4,4.5}
  {
    \node [dotdia] at (\x,-1) {}; 
    \node [dotdia] at (\x,1) {}; 
  }
  \draw (0.5,-1)--(1,1); 
  \draw (4.5,-1)--(3.5,1); 
  \foreach \x in {2} 
  { 
    \draw (\x,-1) to[out=90,in=180]
    (\x+0.25,-0.5) to[out=0,in=90] (\x+0.5,-1); 
  }
  \foreach \x in {2.5} 
  { 
    \draw (\x,1) to[out=-90,in=180]
    (\x+0.25,0.5) to[out=0,in=-90] (\x+0.5,1); 
  }
  \draw (0.5,1) to[out=-90,in=180]
  (0.5+0.5,0.5) to[out=0,in=-90] (0.5+1,1); 
  \draw (2,1) to[out=-90,in=180]
  (2+1,0.25) to[out=0,in=-90] (2+2,1); 
  \draw (0,1) to[out=-90,in=180]
  (0+2.25,-0.125) to[out=0,in=-90] (0+4.5,1); 
  \draw (1.5,-1) to[out=90,in=180]
  (1.5+0.75,-0.3) to[out=0,in=90] (1.5+1.5,-1); 
  \draw (1,-1) to[out=90,in=180]
  (1+1.25,0.125) to[out=0,in=90] (1+2.5,-1); 
  \draw (0,-1) to[out=90,in=180]
  (0+2,-0.2) to[out=0,in=90] (0+4,-1); 
  \foreach \y in
  {-1,1}
  {
    \draw[densely dotted] (0.75,\y-0.2)--(0.75,\y+0.2); 
    \draw[densely dotted] (3.75,\y-0.2)--(3.75,\y+0.2); 
  }

  \node at (5.75,0) {$\equiv$};

  \begin{scope}[xshift=7.25cm]
    \node at (-0.75,0) {$\beta^2$};
    \draw (-0.5,-1) -- (5,-1); 
    \draw (-0.5,1) -- (5,1); 
    \foreach \x in
    {0,0.5,1,1.5,2,2.5,
      3,3.5,4,4.5}
    {
      \node [dotdia] at (\x,-1) {}; 
      \node [dotdia] at (\x,1) {}; 
    }
    \draw (0.5,-1)--(1,1); 
    \draw (4.5,-1)--(3.5,1); 
    \foreach \x in {2} 
    { 
      \draw (\x,-1) to[out=90,in=180]
      (\x+0.25,-0.5) to[out=0,in=90] (\x+0.5,-1); 
    }
    \foreach \x in {2.5} 
    { 
      \draw (\x,1) to[out=-90,in=180]
      (\x+0.25,0.5) to[out=0,in=-90] (\x+0.5,1); 
    }
    \draw (0.5,1) to[out=-90,in=180]
    (0.5+0.5,0.5) to[out=0,in=-90] (0.5+1,1); 
    \draw (2,1) to[out=-90,in=180]
    (2+1,0.25) to[out=0,in=-90] (2+2,1); 
    \draw (0,1) to[out=-90,in=180]
    (0+2.25,-0.125) to[out=0,in=-90] (0+4.5,1); 
    \draw (1.5,-1) to[out=90,in=180]
    (1.5+0.75,-0.3) to[out=0,in=90] (1.5+1.5,-1); 
    \draw (1,-1) to[out=90,in=180]
    (1+1.25,-0.2) to[out=0,in=90] (1+2.5,-1); 
    \draw (0,-1) to[out=90,in=180]
    (0+2,0.125) to[out=0,in=90] (0+4,-1); 
    \foreach \y in
    {-1,1}
    {
      \draw[densely dotted] (0.75,\y-0.2)--(0.75,\y+0.2); 
      \draw[densely dotted] (3.75,\y-0.2)--(3.75,\y+0.2); 
    }
  \end{scope}
  
\end{tikzpicture}
  \caption{The two identified words in the quotient
    $\Quot(\diffBosFer)$ corresponding to figure
    \ref{fig:model_super_spin}.}
\label{fig:quotient_ex}
\end{figure}
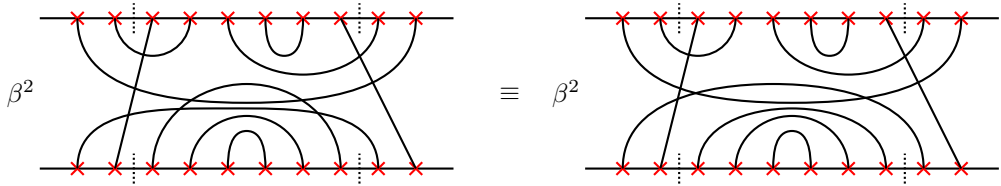

We shall find it convenient to modify the loop model in
such a way that its spectrum is as close as possible to the spectrum
of the spin chain. To this end we obviously set $\beta = M$. Moreover,
we need to go to the quotient
$\Quot(\diffBosFer)$. As a concrete example, the word corresponding to
figure \ref{fig:model_super_spin} in the algebra
$\Alg(\diffBosFer)$, will be reduced as showed in figure
\ref{fig:quotient_ex} using the additional relations of the quotient
discussed above.

These additional rules complete the definition of our loop model, and
we can finally relate the spectrum of the Hamiltonian
\eqref{eq:hamiltonian} in the loop and super spin
representations. This exercise is similar to the better known cases of
the Temperley-Lieb \cite{Read2001} and Brauer \cite{Candu2009a}
algebra.  For the Temperley-Lieb case, the spectrum of the Hamiltonian (on a
given number of sites $2L$) acting on the $\slnm$ chain is the same for every
integer $\nFer$, the only thing that changes are (non zero)
multiplicities, that grow very fast with $\nFer$.  In our case, the
situation is a bit different.  First, we have extensive numerical
evidence that, calling $\Sigma(N)$ the set of eigenvalues of the
$\ssl{\nFer}{\nFer}$ Hamiltonian (up to non zero degeneracies), the
following relations holds:
\begin{equation}
  \label{eq:inclusion_spectra_glNpMN}
  \Sigma(1) \subset \Sigma(2) = \Sigma(3) = \dots = \Sigma(N) = \dots \,
\end{equation}
Moreover, $\Sigma(2)$ is also equal to the set of eigenvalues of the
Hamiltonian in the loop representation: in other words, all spin
chains but the $\ssl{1}{1}$ one have the same spectrum as the geometrical
model. The spectrum of $\ssl{1}{1}$ meanwhile is only a subset; this is
discussed in more details in the appendix \ref{sec:glone_case}.

When we look at $\slnm$, $\diffBosFer>0$, the numerical evidence is
that the spectra are all the same (again, up to non zero
degeneracies). It is very likely that spin chains for a number of
states per site greater or equal to 3 provide faithful representations
of the quotient algebra $\Quot(\diffBosFer)$. This statement is in
line with the findings of \cite{Jones1994}, which should be possible
to generalize in our case.

A final advantage of the loop representation is that we can now study the
geometrical model for arbitrary values of $\beta$, which is now a simple
parameter appearing as the fugacity of loops. Note that, except when
$\beta = M$ is an integer, we do not know of any spin chain representation of the
corresponding algebras, be they $\mathcal{A},\mathcal{Q}$ or
Brauer. This situation has to be contrasted with the ordinary
Temperley-Lieb algebra, where the XXZ spin chain provides such a
representation for all values of the loop fugacity. Clearly the
presence of loop crossings---even if only between boundary lines---makes
the problem considerably more involved. These crossings also make
impossible a Coulomb gas approach to the underlying conformal field theory
(for essentially the same technical reasons).

We remark that the bulk part of the Hamiltonian \eqref{eq:hamiltonian}
does not follow from the usual Yang-Baxter construction for
alternating fundamental and dual representations of $\ssl{
  \nFer+\diffBosFer}{\nFer}$ \cite{Essler2005}.  However, in the bulk,
\eqref{eq:hamiltonian} is solvable exactly, since it belongs to the
Temperley-Lieb algebra, and the realization of this algebra in the
$U_q(\mathfrak{sl}(2))$-invariant XXZ spin chain leads to Bethe ansatz
equations etc (see also \cite{BatchelorYung} for a related situation).
We do not know what the situation for our model with boundary
interactions might be. But it is interesting to observe that in the
particular case of $\mathfrak{sl}(2)$, solvable spin chain
Hamiltonians with boundary impurity spins can be obtained from
solutions of the Yang-Baxter equation \cite{Chen1998}. More work is
needed to clarify the situation.

\section{Critical exponents}
\label{sec:crit_exp}

We now launch into the very technical problem of determining the
critical exponents for the geometrical model. This will allow us later
to reconstruct the spectrum of the super spin chains by the proper
combination of sectors.

Our loop model in the bulk is nothing but the well known dense loop
gas going back to early studies of the Potts model and the dense
$O(n)$ model \cite{Nienhuis1982,DenNijs1979}. It is critical for
$0\leq M\leq 2$.

The central charge of the dense loop gas is
\begin{equation}
 c = 1 - \frac{6}{p(p+1)} \,,
\end{equation}
if we parametrize $\fug = 2 \cos \left( \smfrac{\pi}{p+1} \right)$.
We also recall the Kac formula
\begin{equation}
 h_{r,s} = \frac{( (p+1) r - ps)^2 - 1}{4p(p+1)} \,,
\end{equation}
which is often a convenient means of stating our results on critical
exponents. In the following we are going to diagonalize the
Hamiltonian \eqref{eq:hamiltonian}, although as already stressed, we
believe that the same results can be obtained studying the transfer
matrix \eqref{eq:transf_matXY}.  Conformal invariance relates the
exponents $h$ to the eigenvalues $\mathcal{E}_i$ of the Hamiltonian
acting on a chain of length $L$ via the finite size scaling formula
\cite{Blote1986,Affleck1986}:
\begin{equation}
  \label{eq:c_finite_size}
  \mathcal{E}_i = L f_b + 2 f_s - \frac{v_s \ceff \pi}{24 L} + 
  \mathcal{O}\left(\frac{1}{L^2}\right) \, .
\end{equation}
In this formula $f_b$ and $f_s$ are the bulk and surface free energy,
$\ceff = c - 24 h$ and $v_s = (p+1)\sin\left(\smfrac{\pi}{p+1}\right)$
is the so-called velocity of sound.

We briefly review how critical exponents are related to the 
representation theory of the loop model (see \cite{Richard} for more details).
On the lattice the geometrical transfer matrix and Hamiltonians have a
blockwise lower-triangular structure since the $E_i$
annihilate non-contractible lines (or
strings) in pairs but cannot create any string and $P_i$ does not
alter the number of strings.  The restriction to a sector is done by
setting the action of TL generators on two strings equal to zero.
Further we note that blocks only differing by the configuration of the
bottom row of states are identical.  So for the purpose of computing
the spectrum, we can consider reduced states in which only the top
part of a state on which the Hamiltonian acts is taken.  The critical
exponents we are going to compute, called in the literature watermelon
or $2j$-leg exponents, are the lowest conformal weights in a sector
with fixed number of non-contractible lines.

Before moving on, it is useful to  recall  some results about the
two-boundary loop model ($2$BLM)
\cite{DEGIER2009,Dubail2009} which we will use
later. This model is a dense loop model on a strip in which lines
touching a boundary get marked in different ways, and marked loops
get a weight different from the weight $\beta$ of bulk loops.
These boundary weights are $\fugL$ (resp.\ $\fugR$) for loops touching the
left (resp.\ right) boundary only, and $\fugLR$ for loops touching both boundaries.
Using the parametrization
\begin{equation}
  \label{eq:notation2BLM}
\begin{split}
  &\qquad \quad \fugL = \frac{ \sin \left( (r_1+1) \frac{\pi}{p+1} \right) }
  { \sin \left( r_1 \frac{\pi}{p+1} \right)} \, , \qquad
  \fugR = \frac{ \sin \left( (r_2+1) \frac{\pi}{p+1} \right) }
  { \sin \left( r_2 \frac{\pi}{p+1} \right)} \\
  &\fugLR =\frac{ 
    \sin \left( (r_1+r_2+1-r_{12})\frac{\pi}{2(p+1)}\right) 
    \sin \left( (r_1+r_2+1+r_{12})\frac{\pi}{2(p+1)}\right) 
  }{\sin \left( r_1 \frac{\pi}{p+1} \right)
    \sin \left( r_2 \frac{\pi}{p+1} \right)}  \, ,
\end{split}
\end{equation}
it was found that in the sector with no non-contractible lines,
the conformal weights appearing are
\cite{Dubail2009}
\begin{equation}
  \label{eq:h2BLMj0}
  h_{r_{12}-2n,r_{12}} \, ,
\end{equation}
with $n \in \mathbb{Z}$, while with $2j>0$ non-contractible lines they are
\begin{equation}
  \label{eq:h2BLMj>0}
  h_{\epsilon_1 r_1 + \epsilon_2 r_2 - 1 - 2n, \epsilon_1 r_1 + \epsilon_2 r_2 - 1 + 2j} 
  \, .
\end{equation}
The sign $\epsilon_{1} = \pm 1$ indicate whether the leftmost
non-contractible line is required to touch the left boundary (for
$\epsilon_1 = +1$, referred to as the ``blobbed'' sector), or
forbidden from doing so (for $\epsilon_1 = -1$, referred to as the
``unblobbed'' sector).  The sign $\epsilon_2$ similarly describe the
choice of blobbed/unblobbed sectors at the right boundary, and $n \in
\mathbb{N}$. The parameters $r_1$, $r_2$ and $r_{12}$ are related to
the weight of marked loops through the formulas
\eqref{eq:notation2BLM}.  The case $\fugR=\fug$ and $\fugLR=\fugL$
corresponds to the one-boundary loop model ($1$BLM)
\cite{MartinSaleur}, where the exponents reduce to \cite{Jacobsen2008}
\begin{equation}
 \label{res1btl}
  h_{r_1,r_1+2j \epsilon_1} \, .
\end{equation}

\subsection{One-boundary case}
\label{sec:oneb}

We will start  by investigating the one-boundary problem.  Setting
$\nRepRight=0$, the Hamiltonian we want to study is
\begin{equation}
  \label{eq:Honeb}
  H = - u \sum_{i=0}^{\nRepLeft-1} P_{i,i+1} -
  \sum_{i=\nRepLeft}^{2 \nRepBulk+\nRepLeft-2} E_i \, .
\end{equation}

When computing the spectrum of this operator, one extra representation
theoretical consideration comes into play. Once the number of
non-contractible lines (or ``strings'') at the boundary have been fixed,
(\ref{eq:Honeb}) still acts on them non-trivially by means of permutations.
The Hamiltonian can be block-diagonalized with respect to this action
by a change of basis that fixes
the action of the symmetric group $\perm_{\nRepLeft+j}$ on the
$\nRepLeft+j$ strings according to its irreducible representations,
the so-called Specht modules \cite{Sagan2001}.  Since permutations
occur between boundary lines and the leftmost bulk line (if any), we will actually deal with
$\perm_\nRepLeft$ when $j=0$ or $\perm_{\nRepLeft+1}$ when $j>0$.

We can follow the procedure explained in \cite{Candu2009a} for a  related model
based on the Brauer algebra. Instead of the usual diagrams on which
the Hamiltonian acts, we can consider a tensor product of an
unlabelled diagram, for which states with exchanged strings are
equivalent, and a vector belonging to a basis of a certain Specht
module.  An operator will act on such a state by the multiplication of
diagrams, and by the matrix action in the Specht module of the
permutation keeping track of the reordering of strings.

Then the Hamiltonian \eqref{eq:Honeb} can be put in a block form, each
block indexed by the number of bulk strings and a Young diagram
referring to a Specht module. Accordingly the Hilbert space will be
decomposed in terms of these subspaces.  Numerical diagonalization
shows that the lowest eigenvalue of the Hamiltonian always lies in the
symmetric representation of the symmetric group, where permutations
act as the identity in the Specht module. So for the purpose of computing
the leading critical exponents we can restrict to this sector.

\subsubsection{Exponents and relations with the one-boundary loop model}
\label{sec:rel_oneb}

\begin{figure}[htp]
  \centering
\begingroup
  \makeatletter
  \providecommand\color[2][]{%
    \GenericError{(gnuplot) \space\space\space\@spaces}{%
      Package color not loaded in conjunction with
      terminal option `colourtext'%
    }{See the gnuplot documentation for explanation.%
    }{Either use 'blacktext' in gnuplot or load the package
      color.sty in LaTeX.}%
    \renewcommand\color[2][]{}%
  }%
  \providecommand\includegraphics[2][]{%
    \GenericError{(gnuplot) \space\space\space\@spaces}{%
      Package graphicx or graphics not loaded%
    }{See the gnuplot documentation for explanation.%
    }{The gnuplot epslatex terminal needs graphicx.sty or graphics.sty.}%
    \renewcommand\includegraphics[2][]{}%
  }%
  \providecommand\rotatebox[2]{#2}%
  \@ifundefined{ifGPcolor}{%
    \newif\ifGPcolor
    \GPcolortrue
  }{}%
  \@ifundefined{ifGPblacktext}{%
    \newif\ifGPblacktext
    \GPblacktexttrue
  }{}%
  \let\gplgaddtomacro\g@addto@macro
  \gdef\gplbacktext{}%
  \gdef\gplfronttext{}%
  \makeatother
  \ifGPblacktext
    \def\colorrgb#1{}%
    \def\colorgray#1{}%
  \else
    \ifGPcolor
      \def\colorrgb#1{\color[rgb]{#1}}%
      \def\colorgray#1{\color[gray]{#1}}%
      \expandafter\def\csname LTw\endcsname{\color{white}}%
      \expandafter\def\csname LTb\endcsname{\color{black}}%
      \expandafter\def\csname LTa\endcsname{\color{black}}%
      \expandafter\def\csname LT0\endcsname{\color[rgb]{1,0,0}}%
      \expandafter\def\csname LT1\endcsname{\color[rgb]{0,1,0}}%
      \expandafter\def\csname LT2\endcsname{\color[rgb]{0,0,1}}%
      \expandafter\def\csname LT3\endcsname{\color[rgb]{1,0,1}}%
      \expandafter\def\csname LT4\endcsname{\color[rgb]{0,1,1}}%
      \expandafter\def\csname LT5\endcsname{\color[rgb]{1,1,0}}%
      \expandafter\def\csname LT6\endcsname{\color[rgb]{0,0,0}}%
      \expandafter\def\csname LT7\endcsname{\color[rgb]{1,0.3,0}}%
      \expandafter\def\csname LT8\endcsname{\color[rgb]{0.5,0.5,0.5}}%
    \else
      \def\colorrgb#1{\color{black}}%
      \def\colorgray#1{\color[gray]{#1}}%
      \expandafter\def\csname LTw\endcsname{\color{white}}%
      \expandafter\def\csname LTb\endcsname{\color{black}}%
      \expandafter\def\csname LTa\endcsname{\color{black}}%
      \expandafter\def\csname LT0\endcsname{\color{black}}%
      \expandafter\def\csname LT1\endcsname{\color{black}}%
      \expandafter\def\csname LT2\endcsname{\color{black}}%
      \expandafter\def\csname LT3\endcsname{\color{black}}%
      \expandafter\def\csname LT4\endcsname{\color{black}}%
      \expandafter\def\csname LT5\endcsname{\color{black}}%
      \expandafter\def\csname LT6\endcsname{\color{black}}%
      \expandafter\def\csname LT7\endcsname{\color{black}}%
      \expandafter\def\csname LT8\endcsname{\color{black}}%
    \fi
  \fi
  \setlength{\unitlength}{0.0500bp}%
  \begin{picture}(6120.00,4284.00)%
    \gplgaddtomacro\gplbacktext{%
      \csname LTb\endcsname%
      \put(1474,704){\makebox(0,0)[r]{\strut{}$-0.2$}}%
      \put(1474,1028){\makebox(0,0)[r]{\strut{}$0$}}%
      \put(1474,1353){\makebox(0,0)[r]{\strut{}$0.2$}}%
      \put(1474,1677){\makebox(0,0)[r]{\strut{}$0.4$}}%
      \put(1474,2002){\makebox(0,0)[r]{\strut{}$0.6$}}%
      \put(1474,2326){\makebox(0,0)[r]{\strut{}$0.8$}}%
      \put(1474,2651){\makebox(0,0)[r]{\strut{}$1$}}%
      \put(1474,2975){\makebox(0,0)[r]{\strut{}$1.2$}}%
      \put(1474,3300){\makebox(0,0)[r]{\strut{}$1.4$}}%
      \put(1474,3624){\makebox(0,0)[r]{\strut{}$1.6$}}%
      \put(1606,484){\makebox(0,0){\strut{}$0$}}%
      \put(2024,484){\makebox(0,0){\strut{}$1$}}%
      \put(2443,484){\makebox(0,0){\strut{}$2$}}%
      \put(2861,484){\makebox(0,0){\strut{}$3$}}%
      \put(3280,484){\makebox(0,0){\strut{}$4$}}%
      \put(3698,484){\makebox(0,0){\strut{}$5$}}%
      \put(4116,484){\makebox(0,0){\strut{}$6$}}%
      \put(4535,484){\makebox(0,0){\strut{}$7$}}%
      \put(4953,484){\makebox(0,0){\strut{}$8$}}%
      \put(5372,484){\makebox(0,0){\strut{}$9$}}%
      \put(5790,484){\makebox(0,0){\strut{}$10$}}%
      \put(440,2164){\rotatebox{90}{\makebox(0,0){\strut{}$h$}}}%
      \put(3698,154){\makebox(0,0){\strut{}$u$}}%
      \put(3698,3954){\makebox(0,0){\strut{}One-boundary $\beta=0$}}%
    }%
    \gplgaddtomacro\gplfronttext{%
      \csname LTb\endcsname%
      \put(4935,2703){\makebox(0,0)[r]{\strut{}$m=1, j=0$}}%
      \csname LTb\endcsname%
      \put(4935,2483){\makebox(0,0)[r]{\strut{}$m=1, j=1$}}%
      \csname LTb\endcsname%
      \put(4935,2263){\makebox(0,0)[r]{\strut{}$m=1, j=2$}}%
      \csname LTb\endcsname%
      \put(4935,2043){\makebox(0,0)[r]{\strut{}$m=2, j=0$}}%
      \csname LTb\endcsname%
      \put(4935,1823){\makebox(0,0)[r]{\strut{}$m=2, j=1$}}%
      \csname LTb\endcsname%
      \put(4935,1603){\makebox(0,0)[r]{\strut{}$m=2, j=2$}}%
    }%
    \gplbacktext
    \put(0,0){\includegraphics{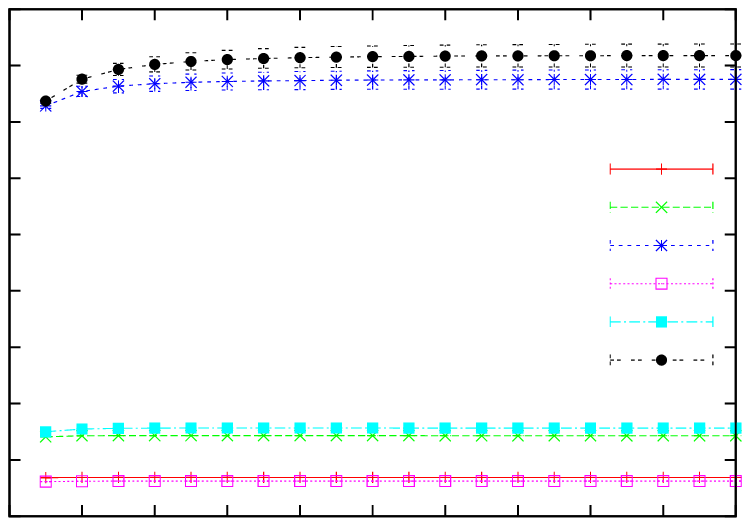}}%
    \gplfronttext
  \end{picture}%
\endgroup
  \caption{Independence of critical exponents $h$ on the coupling $u$
    for $j=0,1,2$ and $\nRepLeft=1,2$ ($\nRepRight=0$) in the model 
    with $\fug=0$. The values are computed by finite-size
    scaling corrections of
    equation \eqref{eq:c_finite_size} with inclusion of a term of order 
    $1/\nRepBulk^2$ 
    , for systems of sizes $\nRepBulk=5$ to $\nRepBulk=9$.
  }
\label{fig:h_m1to2_mbar0_K0to2_Q0_diff_u_L9}
\end{figure}

Diagonalizing numerically the Hamiltonian at different values of
$\fug$ and $\nRepLeft$, we have been able to conjecture the analytical
form of the $(\nRepLeft+2j)$-leg critical exponents
$h^{\nRepLeft,0}(j)$. Using the parametrization of the $1$BLM, if we
define $\fugL$ as
\begin{equation}
  \label{eq:n1_m}
  \fugL = \frac{\nRepLeft+\fug}{\nRepLeft+1} \, ,
\end{equation}
then the exponents are
\begin{equation}
  \label{eq:2j-leg}
  h^{\nRepLeft,0}(j) = h_{r_1(\nRepLeft),r_1(\nRepLeft)+2j} \, ,
\end{equation}
where $r_1(\nRepLeft)$ is given explicitly by
\begin{equation}
  \label{eq:r1m}
  r_1(\nRepLeft) = \frac{1 + p}{\pi} \arctan 
  \left( 
    \smfrac{ (1+\nRepLeft) \sin \left( \frac{\pi}{p+1} \right) }
    {\nRepLeft-(\nRepLeft-1) \cos \left( \frac{\pi}{p+1} \right)} 
  \right)  \, .
\end{equation}
Our numerical findings give strong evidence that this result for the
exponents does {\em not} depend on the coupling $u$ as long as we take
it positive (see figure
\ref{fig:h_m1to2_mbar0_K0to2_Q0_diff_u_L9}). Note also that when
$\fug=1$ (whence $\beta_1 = r_1(m) = 1$ for all $m$), the exponents
are trivial, equal to those of the free boundary case.\footnote{If we
  think of the case $\nRepLeft=1$, one intuitive argument supporting
  this triviality goes as follows. Everytime a loop crosses the
  boundary line, we can think of this configuration as a loop with
  weight $1$ and a straight line times the coupling $-u$:
  \[
  \begin{tikzpicture}[scale=0.25]
    \draw[rounded corners=0.2cm]
    (0,-1.5)-- ++(0,0.5)-- ++(2,2)-- ++(1,-1)-- ++(-1,-1)-- ++(-2,2)--
    ++(0,0.5);
    \node at (5,0) {$=-u$};
  \begin{scope}[xshift=7cm]
    \draw[rounded corners=0.2cm] (0,-1.5)-- ++(0,0.5)-- ++(1,1)--
    ++(-1,1)-- ++(0,0.5); 
    \draw[rounded corners=0.2cm]
    (2.5,-0.5)-- ++(0.5,0.5)-- ++(-1,1)-- ++(-1,-1)-- ++(1,-1)--
    ++(0.5,0.5);
  \end{scope}
  \end{tikzpicture}
 \]
  For general $\nRepLeft$ and $u$ positive we have found the following relation
  $\lambda^0_\nRepLeft = -u + \lambda^0_{\nRepLeft-1}$,
  $\lambda^0_\nRepLeft$ being the lowest eigenvalues of the
  Hamiltonian with $\nRepLeft$ boundary lines, which implies the same
  scaling in $1/\nRepBulk$ of the two eigenvalues. 
}

In view of this result, which is formally identical to (\ref{res1btl})
for a particular choice of the parameter $r_1(m)$, it is natural to
search for a relation between our model and the one-boundary
Temperley-Lieb algebra. We recall that the latter is defined
\cite{MartinSaleur,Jacobsen2008} by endowing the Temperley-Lieb
algebra with an extra blob operator $b$ acting on the leftmost strand,
which we will label $m$ for making easier the connection with our
model. By definition, $b$ is idempotent and satisfies the relation
\begin{equation}
  \label{eq:ebe}
  E_m b E_m = \fugL E_m \, ,
\end{equation}
whose geometrical interpretation is that marked loops get a weight $\fugL$.

For the particular case $m=1$ this relationship is manifest. Namely, setting
$P_0 = 2 b - 1$ and using $b^2 = b$ we get $(P_0)^2 = 1$ as required.
So for a particular choice of the parameter $u$ the boundary interaction
is proportional to $b$ indeed. The corresponding weight of boundary
loops follows from
\begin{equation}
 E_1 \frac{1+P_0}{2} E_1 = \frac{1+\beta}{2} E_1 \,,
\end{equation}
where we have used \eqref{eq:def_A3}. This shows that
$\beta_1 = \frac{1+\beta}{2}$ as in (\ref{eq:n1_m}). The exponents
(\ref{eq:2j-leg}) then follow by invoking the result (\ref{res1btl})
for the one-boundary loop model \cite{Jacobsen2008}.

For higher values $m > 1$ there is no such mapping to the one-boundary
loop model. We can nevertheless derive the results
(\ref{eq:n1_m})--(\ref{eq:r1m}) for an appropriate modification of the
Hamiltonian (\ref{eq:Honeb}). Our argument 
that (\ref{eq:n1_m})--(\ref{eq:r1m}) are correct also for the original
Hamiltonian (\ref{eq:Honeb}) is then based on universality: we present
numerical evidence that the original and modified Hamiltonians have
the same critical exponents.

\begin{figure}[htp]
  \centering
  \subfigure[]{
    \includegraphics[scale=0.7]{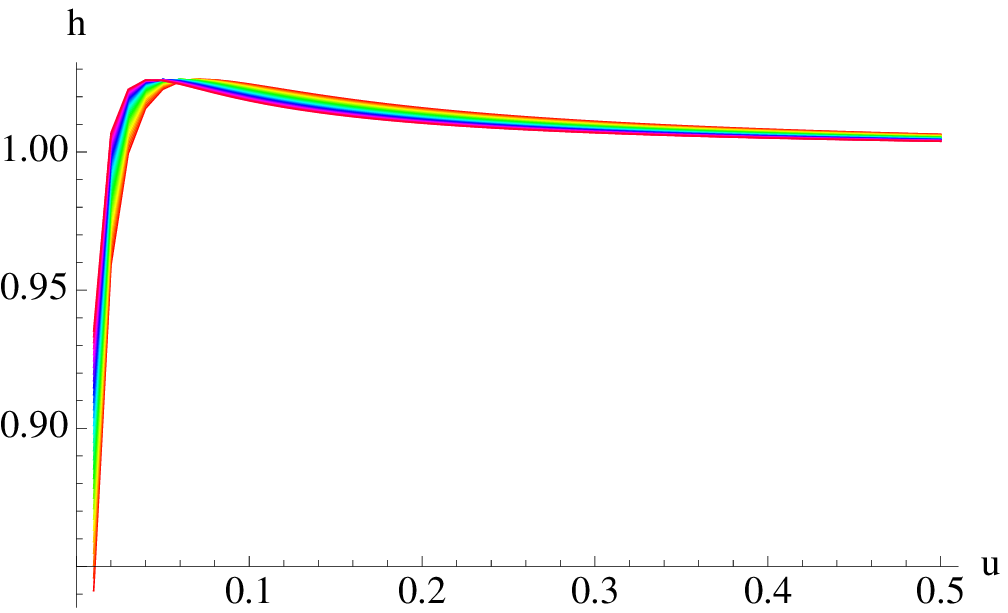}}
  \subfigure[]{
    \includegraphics[scale=0.7]{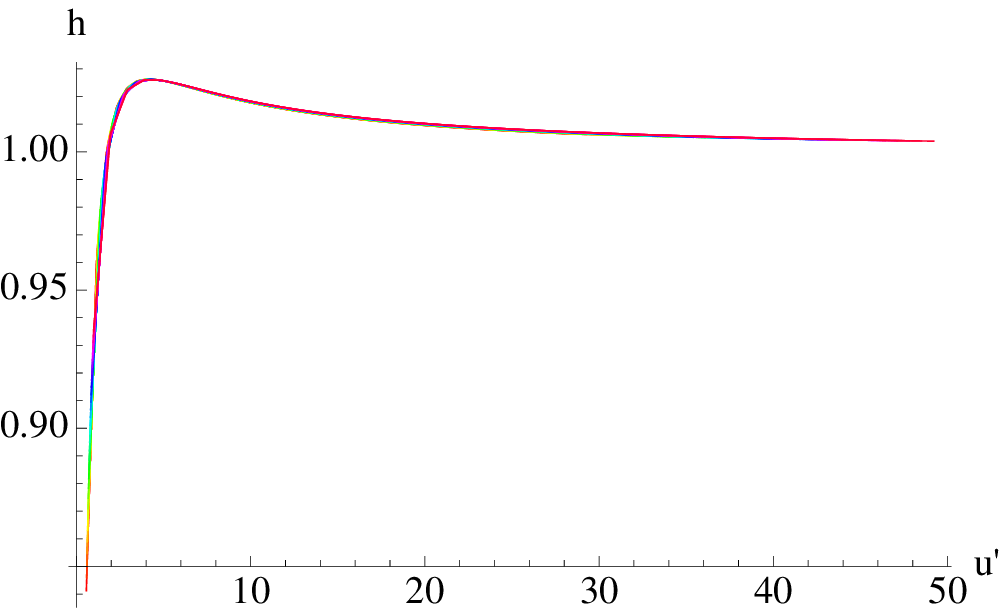}}
  \caption{$h$ vs $u$ in the $\ssl{1}{1}$ chain for sizes of the bulk
    length $\nRepBulk=45$ to $\nRepBulk=75$ (here in different colors)
    and $\nRepLeft=1,\nRepRight=0$. In figure (a) the variable $u$ is
    not rescaled.  In figure (b) we rescale $u$ as $u'=L^{1 -
      h_{r_1(1),r_1(1)}} u$, and we observe a perfect collapse of all
    curves.}
  \label{fig:scaling_h_gl11}
\end{figure}

As a warmup, we first discuss our numerical observation that within the
space of Hamiltonians of the form (\ref{eq:Honeb}), the exponents
(\ref{eq:2j-leg}) are independent of the boundary coupling $u$.
This should be interpreted in terms of boundary
renormalization group flow. The situation is akin to that of the
Ising model with a boundary field, where the fixed boundary condition
is a stable fixed point, and the free boundary condition an unstable
one \cite{Affleck1991}.  At a given non-zero value of the boundary field, there
is a crossover, and the system flows to the stable fixed point for a
sufficiently large system.  In our case, the boundary perturbation to
the Lagrangian is $u\phi_{r_1(\nRepLeft),r_1(\nRepLeft)}$, and it
drives the system from the $\nRepLeft=0$ to the $\nRepLeft > 0$
boundary condition. Since the dimension of the field
$\phi_{r_1(\nRepLeft),r_1(\nRepLeft)}$ is
$h_{r_1(\nRepLeft),r_1(\nRepLeft)}$, by dimensional arguments the
relevant length scale in this boundary flow is given by
$L^{1-h_{r_1(\nRepLeft),r_1(\nRepLeft)}}$. We have explicitly checked
that curves describing the critical exponents as a function of $u$,
collapse after appropriate rescaling in the $\ssl{1}{1}$ case, where we
can access large sizes of the system, see figure
\ref{fig:scaling_h_gl11}.


Extending these ideas of universality, we now construct a modified
Hamiltonian $\widetilde{H}$ for the lattice model with edge states,
which---unlike the original Hamiltonian $H$
of equation (\ref{eq:Honeb})---is in precise correspondence with the
one-boundary loop model. The key step is to construct an idempotent
blob operator $b$ satisfying (\ref{eq:ebe}).

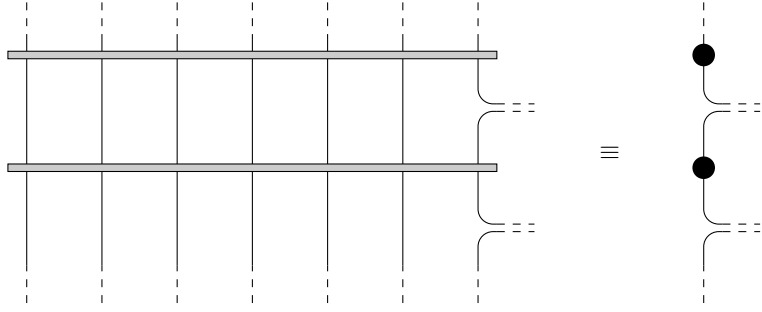
\begin{figure}[htp]
  \centering
\begin{tikzpicture}[scale=1]

  \foreach \x in {0,1,2,3,4,5}
  {
    \draw (\x,0)--(\x,3);
  }
  \foreach \x in {0,1,2,3,4,5,6,9}
  {
    \draw[dashed] (\x,-0.5)--(\x,0);
    \draw[dashed] (\x,3.5)--(\x,3);
  }

  \foreach \y in {0.45,0.55,2.05,2.15}
  {
    \draw[dashed] (6.25,\y)--(6.75,\y);
    \draw[dashed] (9.25,\y)--(9.75,\y);
  }

  \node at (7.75,1.5) {$\equiv$};

  \foreach \x in {6,9}
  {
    \draw[rounded corners=0.2cm]
    (\x,3)-- ++(0,-0.85)-- ++(0.25,0);
    \draw[rounded corners=0.2cm]
    (\x+0.25,2.05)-- ++(-0.25,0)-- ++(0,-1.5)-- ++(0.25,0);
    \draw[rounded corners=0.2cm]
    (\x+0.25,0.45)-- ++(-0.25,0)-- ++(0,-0.45);
  }

  \draw[fill=black!20] (-0.25,1.25) rectangle (6.25,1.35);
  \draw[fill=black!20] (-0.25,2.75) rectangle (6.25,2.85);
  \fill (9,1.3) circle (0.15);
  \fill (9,2.8) circle (0.15);

\end{tikzpicture}
  \caption{Symmetrization (here represented by horizontal gray bars)
    over the $\nRepLeft+1$ leftmost strands can be interpreted as
    blobbing the first bulk strand.}
\label{fig:b_symmetrizer}
\end{figure}

In the Hilbert space of our spin chain, the idempotent words of the
boundary algebra are given by the Young symmetrizers of the symmetric
group $\perm_{\nRepLeft+1}$. In appendix \ref{sec:YS_blob} we show
that indeed every symmetrizer satisfies a relation such as
\eqref{eq:ebe}. The value of $\beta_1$ (\ref{eq:n1_m}) we are
interested in is obtained if we take the fully symmetric
representation of the symmetric group (see figure
\ref{fig:b_symmetrizer} for a graphical representation of the relation
between the symmetrizer and the blob operator).

Define therefore the  standard Young
tableau\footnote{In what follows we will adopt the convention of labelling the
  boxes of a Young diagram of size $\nRepLeft+1$ with the elements of
  the set $\{0,\dots,\nRepLeft\}$ when related to the left boundary
  action of the permutation group} $t_1 = \tableft$, and the
associated Young symmetrizer $\bleft=\smfrac{1}{(\nRepLeft+1)!}
\sum_{\sigma \in \perm_{\nRepLeft+1}} \sigma$. It then turns out that 
$\bleft$
satisfies:
\begin{equation}
  \label{eq:EbzeromE}
  E_\nRepLeft \bleft E_\nRepLeft = 
  \frac{\nRepLeft+\fug}{\nRepLeft+1} E_\nRepLeft \bleftmone \, ,
\end{equation}
where $\hat{t}_1 = \tableftmone$. To prove formula \eqref{eq:EbzeromE} we can use the following simple
identity
\begin{equation}
  \label{eq:bzerom_bzerom-1}
  \bleft = \frac{1}{\nRepLeft+1} \bleftmone (1 + (0,\nRepLeft) + 
  (1,\nRepLeft) + \dots + (\nRepLeft-1,\nRepLeft)) \, ,
\end{equation}
$(i,j)$ denoting the transposition of $i$ and $j$, and the relations
of the algebra.  

Relation \eqref{eq:ebe} now follows
with our value of $\fugL$ \eqref{eq:n1_m} if we replace $E_i$ by
$\widetilde{E}_i = E_i \bleftmone$, an operation which does not affect
relations of the algebra $\Alg$. In other words, the following 
modified Hamiltonian
\begin{equation}
  \label{eq:Honeballperm}
   \widetilde{H} = 
   - u\, \frac{1}{(m+1)!} \sum_{\sigma \in \perm_{m+1}} \sigma 
   - \sum_{i=m}^{2L+m-2} E_i\, ,
\end{equation}
has precisely the form of a one-boundary loop model (the first term
being $-u b$) and by (\ref{eq:EbzeromE}) the weight of a loop marked
by $b$ is exactly (\ref{eq:n1_m}). It follows from (\ref{res1btl}) that
the $2j$-leg exponents are indeed given by (\ref{eq:2j-leg}).

Obviously the boundary terms in (\ref{eq:Honeb}) and (\ref{eq:Honeballperm})
are different. However, just as the spectrum of (\ref{eq:Honeb}) was shown
numerically to be independent of $u$, it is a natural hypothesis that
the critical exponents of $H$ and $\widetilde{H}$ are identical.
In table \ref{tab:h_m2} we present strong numerical support for this
universality hypothesis.
\begin{table}[h!c]
  \centering
  \begin{tabular}{|c|c|c|c|c|}
    \hline
    $\fug$ & $j$ & $h$ & $\tilde{h}$ & $h_{2,0}(j)$ \\
    \hline
    $0$ & $0$ & $-0.07606\pm 0.00010$ & $-0.07615 \pm 0.00015$ 
    & $-0.076068$ \\
    \hline
    $0$ & $1$ & $0.1099\pm 0.0004$ & $0.1090 \pm 0.0006$ & $0.111099$\\
    \hline
    $0$ & $2$ & $1.339\pm 0.010$ & $1.321 \pm 0.006$ & $1.298266$\\
    \hline
    $1$ & $0$ & $0$ & $0$ & $0$ \\
    \hline
    $1$ & $1$ & $0.33478\pm 0.00009$ & $0.33478 \pm 0.00009$ & $1/3$ \\ 
    \hline
    $1$ & $2$ & $2.099\pm 0.016$ & $2.099 \pm 0.016$ & $2$ \\
    \hline
    $\sqrt{2}$ & $0$ & $0.01374\pm 0.00016$ & $0.01302 \pm 0.00021$ 
    & $0.014542$ \\ 
    \hline
    $\sqrt{2}$ & $1$ & $0.4384\pm 0.0011$ & $0.4432 \pm 0.0007$ 
    & $0.438772$ \\
    \hline
    $\sqrt{2}$ & $2$ & $2.4256\pm 0.0013$ & $2.4410 \pm 0.0009$ 
    & $2.363001$ \\
    \hline 
\end{tabular}
\caption{Comparison of critical exponents for the model described by
  \eqref{eq:Honeb}, $h$, and for the one described by \eqref{eq:Honeballperm},
  $\tilde{h}$, listed for different values of $\fug$ and $j$, 
  in a system with $\nRepLeft=2,\nRepRight=0$.  
  The values are obtained by fitting eigenvalues for bulk sizes
  $\nRepBulk=6\to 10$, using formula \eqref{eq:c_finite_size}
  with a term of order $1/\nRepBulk^2$,
  and although these numbers should be equal when $\nRepBulk\to\infty$,
  already with these relatively small sizes we find good
  agreement between the values of the exponents in the model and 
  the exact value predicted in the continuum limit $h_{2,0}(j)$,
  given in the last column.}
  \label{tab:h_m2}
\end{table}

We remark that for having the universal behaviour described above, it
is crucial that the signs in front of the permutation generators are
all negative. In this case, adding to the Hamiltonian \eqref{eq:Honeb}
other terms of the symmetric group than the generators, is a redundant
perturbation when they have the same negative coefficients.  We expect
that taking different couplings $u_i$ for each $P_{i,i+1}$ in the
boundary Hamiltonian, will in general drive the model to different universality
classes according to the relative signs (see also appendix
\ref{sec:YS_blob}). In particular, it appears an appealing possibility
that each choice of Young symmetrizer in the above construction
might lead to a different universality class, but we have not investigated
this issue in sufficient detail.

Finally we have numerical evidence that the universality class of $H$
and $\widetilde{H}$
is more general, and that, in particular, the following
Hamiltonian also has the same continuum limit:
\begin{equation}
  \label{eq:Hproj}
  \widetilde{\widetilde{H}}= \bleftmone H \bleftmone \, ,
\end{equation}
corresponding to the case where the $\nRepLeft$ strands on the
boundary are fully symmetrized.  From an algebraic point of view this
version of the model is equivalent to replacing the $\nRepLeft$ copies
of the fundamental representation of the superalgebra in the spin
chain formulation with a higher-dimensional one.

\subsection{Two-boundary case}
\label{sec:twob}

We now move to the two-boundary problem, which is obviously much richer.

\subsubsection{New sectors}
\label{sec:newsec}

\begin{figure}[htp]
  \centering
\begin{tikzpicture}[thick,scale=0.75,dotdia/.style={cross out, draw,
    solid, red, inner sep=2pt}]

  \draw (-0.5,0) -- (5,0);
  \foreach \x in
  {0,0.5,1,1.5,2,2.5,
    3,3.5,4,4.5}
  {
    \node [dotdia] at (\x,0) {}; 
  }
  \foreach \x in {1,2,3} 
  { 
    \draw (\x,0) to[out=-90,in=180]
    (\x+0.25,-0.5) to[out=0,in=-90] (\x+0.5,0); 
  }
  \draw (0.5,0) to[out=-90,in=180]
  (0.5+1.75,-1) to[out=0,in=-90] (0.5+3.5,0); 
  \draw (0,0) to[out=-90,in=180]
  (0+2.25,-1.25) to[out=0,in=-90] (0+4.5,0); 
  \foreach \y in
  {0}
  {
    \draw[densely dotted] (0.75,\y-0.2)--(0.75,\y+0.2); 
    \draw[densely dotted] (3.75,\y-0.2)--(3.75,\y+0.2); 
  }

  \begin{scope}[xshift=6cm]
    \draw (-0.5,0) -- (5,0);
    \foreach \x in
    {0,0.5,1,1.5,2,2.5,
      3,3.5,4,4.5}
    {
      \node [dotdia] at (\x,0) {}; 
    }
    \foreach \x in {0,4.5} 
    { 
      \draw (\x,0)--(\x,-1.5); 
    } 
    \foreach \x in {2,3} 
    { 
      \draw (\x,0) to[out=-90,in=180]
      (\x+0.25,-0.5) to[out=0,in=-90] (\x+0.5,0); 
    }
    \draw (0.5,0) to[out=-90,in=180]
    (0.5+0.5,-0.75) to[out=0,in=-90] (0.5+1,0); 
    \draw (1,0) to[out=-90,in=180]
    (1+1.5,-1.5) to[out=0,in=-90] (1+3,0); 
    \foreach \y in
    {0}
    {
      \draw[densely dotted] (0.75,\y-0.2)--(0.75,\y+0.2); 
      \draw[densely dotted] (3.75,\y-0.2)--(3.75,\y+0.2); 
    }
  \end{scope}

  \begin{scope}[xshift=12cm]
    \draw (-0.5,0) -- (5,0);
    \foreach \x in
    {0,0.5,1,1.5,2,2.5,
      3,3.5,4,4.5}
    {
      \node [dotdia] at (\x,0) {}; 
    }
    \foreach \x in {0,0.5,1,1.5,2,2.5,3.5,4.5} 
    { 
      \draw (\x,0)--(\x,-1.5); 
    } 
    \draw (3,0) to[out=-90,in=180]
    (3+0.5,-0.75) to[out=0,in=-90] (3+1,0); 
    \foreach \y in
    {0}
    {
      \draw[densely dotted] (0.75,\y-0.2)--(0.75,\y+0.2); 
      \draw[densely dotted] (3.75,\y-0.2)--(3.75,\y+0.2); 
    }
  \end{scope}

\end{tikzpicture}
  \caption{Different sectors in a system with $\nRepLeft=\nRepRight=2$
  and $\nRepBulk=3$. From left to right, $(j=0,k=0)$, $(j=0,k=1)$,
  $(j=2,k=2)$.}
\label{fig:example_sectors}
\end{figure}
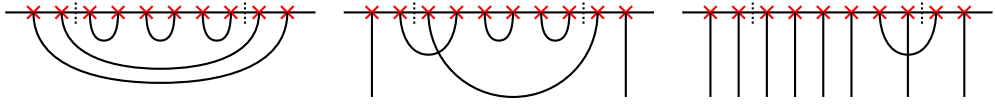

The quotient given by the action of the Hamiltonian on the super spin
chain discussed previously in section \ref{sec:alg_quot} allows
contraction of boundary lines, introducing a new quantum number in the
problem: $k$, the number of couples of left-right boundary lines not
contracted. So we will restrict to $(j,k)$ sectors with $j$ and $k$
fixed and the action of TL generators on a pair of strings (non
contractible lines) set to zero. See figure \ref{fig:example_sectors}
for examples of states in different sectors.

\begin{figure}[htp]
  \centering
  \subfigure[]{
    \label{fig:secKtoK-1j0}
\begin{tikzpicture}[thick,scale=0.75,dotdia/.style={cross out, draw,
    solid, red, inner sep=2pt}]

  \draw (-0.5,0) -- (8,0);
  \foreach \x in
  {0,1.5,2,2.5,3,4.5,5,5.5,6,7.5}
  {
    \node [dotdia] at (\x,0) {}; 
    \node [dotdia] at (\x,2) {}; 
  }
  \draw (-0.5,2) -- (8,2);
  \foreach \x in {0,1.5,6,7.5} 
  { 
    \draw (\x,0)--(\x,2); 
  } 
  \foreach \x in {0,1.5,2,5.5,6,7.5} 
  { 
    \draw (\x,0)--(\x,-1); 
  } 
  \draw (2.5,0)--(2,2); 
  \draw (5,0)--(5.5,2); 
  \foreach \x in {2.5,4.5} 
  { 
    \draw (\x,0) to[out=-90,in=180]
    (\x+0.25,-0.5) to[out=0,in=-90] (\x+0.5,0); 
    \draw (\x,2) to[out=-90,in=180]
    (\x+0.25,2-0.5) to[out=0,in=-90] (\x+0.5,2); 
  }
  \draw (3,0) to[out=-90,in=180](3+0.25,0.5);
  \draw (4.25,0.5) to[out=0,in=-90](4.25+0.25,0);

  \draw (2,0) to[out=90,in=180]
  (2+1.75,1) to[out=0,in=90] (2+3.5,0); 
  \node at (0.75,-0.5) {$\cdots$};
  \node at (0.75,1) {$\cdots$};
  \node at (6+0.75,-0.5) {$\cdots$};
  \node at (6+0.75,1) {$\cdots$};
  \node at (3.75,0.25) {$\cdots$};
  \node at (3.75,2-0.25) {$\cdots$};
  \node at (3.75,-0.25) {$\cdots$};
  \node at (1,-1.75) {$m$};
  \node at (6.5,-1.75) {$n$};
  \draw[thin,<->] (0,-1.5)--(2,-1.5);
  \draw[thin,<->] (5.5,-1.5)--(7.5,-1.5);
  \foreach \y in
  {0,2}
  {
    \draw[densely dotted] (2.25,\y-0.2)--(2.25,\y+0.2); 
    \draw[densely dotted] (5.25,\y-0.2)--(5.25,\y+0.2); 
  }

  \node at (8.5,0) {$=$};

  \begin{scope}[xshift=9.5cm]
  \draw (-0.5,0) -- (8,0);
  \foreach \x in
  {0,1.5,2,2.5,3,4.5,5,5.5,6,7.5}
  {
    \node [dotdia] at (\x,0) {}; 
  }
  \foreach \x in {0,1.5,6,7.5} 
  { 
    \draw (\x,0)--(\x,-1); 
  } 
  \foreach \x in {2.5,4.5} 
  { 
    \draw (\x,0) to[out=-90,in=180]
    (\x+0.25,-0.5) to[out=0,in=-90] (\x+0.5,0); 
  }
  \draw (2,0) to[out=-90,in=180]
  (2+1.75,-1) to[out=0,in=-90] (2+3.5,0); 
  \node at (0.75,-0.5) {$\cdots$};
  \node at (6+0.75,-0.5) {$\cdots$};
  \node at (3.75,-0.25) {$\cdots$};    
  \node at (0.75,-1.75) {$m-1$};
  \node at (6.75,-1.75) {$n-1$};
  \draw[thin,<->] (0,-1.5)--(1.5,-1.5);
  \draw[thin,<->] (6,-1.5)--(7.5,-1.5);
  \foreach \y in
  {0}
  {
    \draw[densely dotted] (2.25,\y-0.2)--(2.25,\y+0.2); 
    \draw[densely dotted] (5.25,\y-0.2)--(5.25,\y+0.2); 
  }
  \end{scope}

\end{tikzpicture}}
  \\
  \subfigure[]{
    \label{fig:secKtoK-1j1}
\begin{tikzpicture}[thick,scale=0.75,dotdia/.style={cross out, draw,
    solid, red, inner sep=2pt}]

  \draw (-0.5,0) -- (9,0);
  \foreach \x in
  {0,1.5,2,2.5,3,3.5,4,5.5,6,6.5,7,8.5}
  {
    \node [dotdia] at (\x,0) {}; 
    \node [dotdia] at (\x,2) {}; 
  }
  \draw (-0.5,2) -- (9,2);
  \foreach \x in {0,1.5,2.5,7,8.5} 
  { 
    \draw (\x,0)--(\x,2); 
  } 
  \foreach \x in {0,1.5,2,2.5,3,6.5,7,8.5} 
  { 
    \draw (\x,0)--(\x,-1); 
  } 
  \draw (6,0)--(3,2); 
  \foreach \x in {3.5,5.5} 
  { 
    \draw (\x,0) to[out=-90,in=180]
    (\x+0.25,-0.5) to[out=0,in=-90] (\x+0.5,0); 
  }
  \foreach \x in {5.5} 
  { 
    \draw (\x,2) to[out=-90,in=180]
    (\x+0.25,2-0.5) to[out=0,in=-90] (\x+0.5,2); 
  }
  \draw (3.5,2) to[out=-90,in=180]
  (3.5+0.25,2-0.35) to[out=0,in=-90] (3.5+0.5,2);
  \foreach \x in {3} 
  { 
    \draw (\x,0) to[out=90,in=180]
    (\x+0.25,0.5) to[out=0,in=90] (\x+0.5,0); 
  }

  \draw (4,0) to[out=-90,in=180](4+0.25,0.5);
  \draw (5.25,0.35) to[out=0,in=-90](5.25+0.25,0);
   
  \draw (2,0) to[out=90,in=180]
  (2+2.25,0.9) to[out=0,in=90] (2+4.5,0); 
  \draw (2,2) to[out=-90,in=180]
  (2+2.25,1.1) to[out=0,in=-90] (2+4.5,2); 
  
  \node at (0.75,-0.5) {$\cdots$};
  \node at (0.75,1) {$\cdots$};
  \node at (7+0.75,-0.5) {$\cdots$};
  \node at (7+0.75,1) {$\cdots$};
  \node at (4.75,0.25) {$\cdots$};
  \node at (4.75,2-0.25) {$\cdots$};
  \node at (4.75,-0.25) {$\cdots$};
  \node at (1,-1.75) {$m$};
  \node at (7.5,-1.75) {$n$};
  \draw[thin,<->] (0,-1.5)--(2,-1.5);
  \draw[thin,<->] (6.5,-1.5)--(8.5,-1.5);
  \foreach \y in
  {0,2}
  {
    \draw[densely dotted] (2.25,\y-0.2)--(2.25,\y+0.2); 
    \draw[densely dotted] (6.25,\y-0.2)--(6.25,\y+0.2); 
  }

  \node at (9.5,0) {$=$};

  \begin{scope}[xshift=10.5cm]
  \draw (-0.5,0) -- (8,0);
  \foreach \x in
  {0,1.5,2,2.5,3,4.5,5,5.5,6,7.5}
  {
    \node [dotdia] at (\x,0) {}; 
  }
  \foreach \x in {0,1.5,2.5,3,6,7.5} 
  { 
    \draw (\x,0)--(\x,-1); 
  } 
  \foreach \x in {4.5} 
  { 
    \draw (\x,0) to[out=-90,in=180]
    (\x+0.25,-0.5) to[out=0,in=-90] (\x+0.5,0); 
  }
  \draw (2,0) to[out=-90,in=180]
  (2+1.75,-1) to[out=0,in=-90] (2+3.5,0); 
  \node at (0.75,-0.5) {$\cdots$};
  \node at (6+0.75,-0.5) {$\cdots$};
  \node at (3.75,-0.25) {$\cdots$};    
  \node at (0.75,-1.75) {$m-1$};
  \node at (6.75,-1.75) {$n-1$};
  \draw[thin,<->] (0,-1.5)--(1.5,-1.5);
  \draw[thin,<->] (6,-1.5)--(7.5,-1.5);
  \foreach \y in
  {0}
  {
    \draw[densely dotted] (2.25,\y-0.2)--(2.25,\y+0.2); 
    \draw[densely dotted] (5.25,\y-0.2)--(5.25,\y+0.2); 
  }
  \end{scope}

\end{tikzpicture}}
  \\
  \subfigure[]{
    \label{fig:secKj>1}
\begin{tikzpicture}[thick,scale=0.75,dotdia/.style={cross out, draw,
    solid, red, inner sep=2pt}]

  \draw (-0.5,0) -- (10,0);
  \foreach \x in
  {0,1.5,2,2.5,3,3.5,4,4.5,5,6.5,7,7.5,8,9.5}
  {
    \node [dotdia] at (\x,0) {}; 
  }
  \foreach \x in {0,1.5,2.5,3,3.5,4,8,9.5} 
  { 
    \draw (\x,0)--(\x,-1); 
  } 
  \foreach \x in {4.5,6.5} 
  { 
    \draw (\x,0) to[out=-90,in=180]
    (\x+0.25,-0.5) to[out=0,in=-90] (\x+0.5,0); 
  }
  \foreach \x in {4} 
  { 
    \draw (\x,0) to[out=90,in=180]
    (\x+0.25,0.5) to[out=0,in=90] (\x+0.5,0); 
  }

  \draw (5,0) to[out=-90,in=180](5+0.25,0.5);
  \draw (6.25,0.5) to[out=0,in=-90](6.25+0.25,0);

  \draw (2.5,0)--(2,1.5);
  \draw (7,0)--(7.5,1.5);
 
  \draw[dashed] (2,0)--(2,-1);
  \draw[dashed] (7.5,0)--(7.5,-1);
  \draw[dashed] (2,0) to[out=90,in=180](2+0.5,1.5);
  \draw[dashed] (7.5,0) to[out=90,in=0](7.5-3.5,1.5); 
  \draw (3,0) to[out=90,in=180] (3.5,1.5);
  \draw (3.5,0) to[out=90,in=0](3,1.5);   
  \draw [red] (3.25,1.45) circle (0.5);
  \node at (3.25,2.25) {NO!};

  \node at (0.75,-0.5) {$\cdots$};
  \node at (8+0.75,-0.5) {$\cdots$};
  \node at (5.75,-0.25) {$\cdots$};
  \node at (5.75,0.25) {$\cdots$};
  \node at (1,-1.75) {$m$};
  \node at (8.5,-1.75) {$n$};
  \draw[thin,<->] (0,-1.5)--(2,-1.5);
  \draw[thin,<->] (7.5,-1.5)--(9.5,-1.5);
  \foreach \y in
  {0}
  {
    \draw[densely dotted] (2.25,\y-0.2)--(2.25,\y+0.2); 
    \draw[densely dotted] (7.25,\y-0.2)--(7.25,\y+0.2); 
  }

\end{tikzpicture}
}
  \caption{(a) Starting from the reference (reduced) state on top left
    at $j=0$ with $k=\minlr$, we pass to the state with $k=\minlr-1$ on
    the right upon acting with the word juxtaposed above.  (b)
    Similarly, when $j=1$ multiplication by an element of the algebra
    allows passing from $k=\minlr$ to $k=\minlr-1$.
    (c) When $j=2$, we have to exchange bulk strings in order that
    the two boundary strings (the dashed lines) can be permuted with
    the bulk strands, and come on adjacent sites.  This is
    forbidden by the interactions in our model. The situation is
    similar with higher values of $j$.}
  \label{fig:secKtoK-1}
\end{figure}
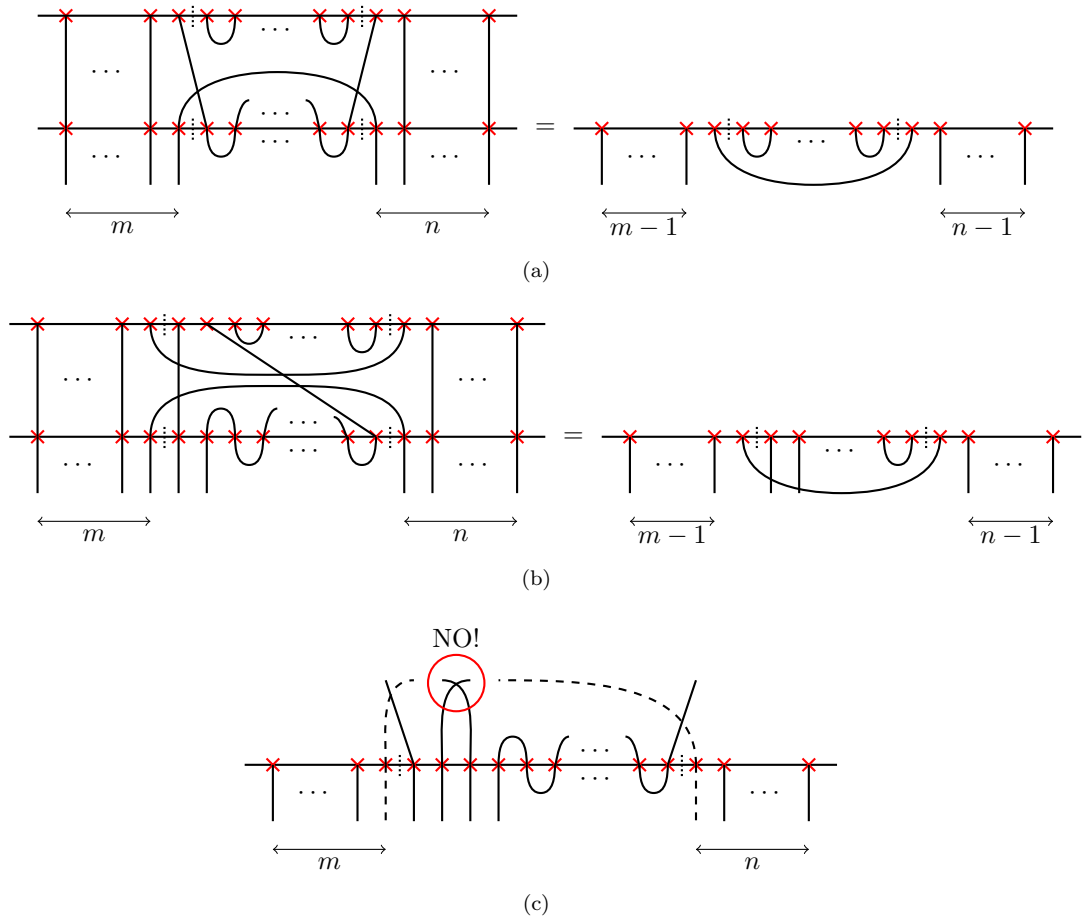

A sector of given $k$ appears with multiplicity
\begin{equation}
  \label{eq:multsectork}
  \binom{\nRepRight}{\minlr - k} \binom{\nRepLeft}{\minlr - k} \, .
\end{equation}
where recall that $\minlr = \min(\nRepLeft,\nRepRight)$. The
possible values of $k$ for $j=0,1$ are $k = 0,1,\dots,\minlr$.  Indeed
when $j=0$ or $j=1$ we can always go from a configuration with $k>0$
contracted boundary lines to another with $k-1$ contracted boundary
lines by acting with elements of the algebra. In figure
\ref{fig:secKtoK-1}, it is shown how we can go from the sector
$k=\minlr$ to $k=\minlr-1$, and it is clear that previous exchange of
boundary lines allows contracting up to $\minlr$ lines.  For $j>1$ meanwhile,
 there is only the sector $k=\minlr$. In this case the boundary
lines cannot come on adjacent sites since lines in the bulk cannot
cross, as illustrated for $j=2$.

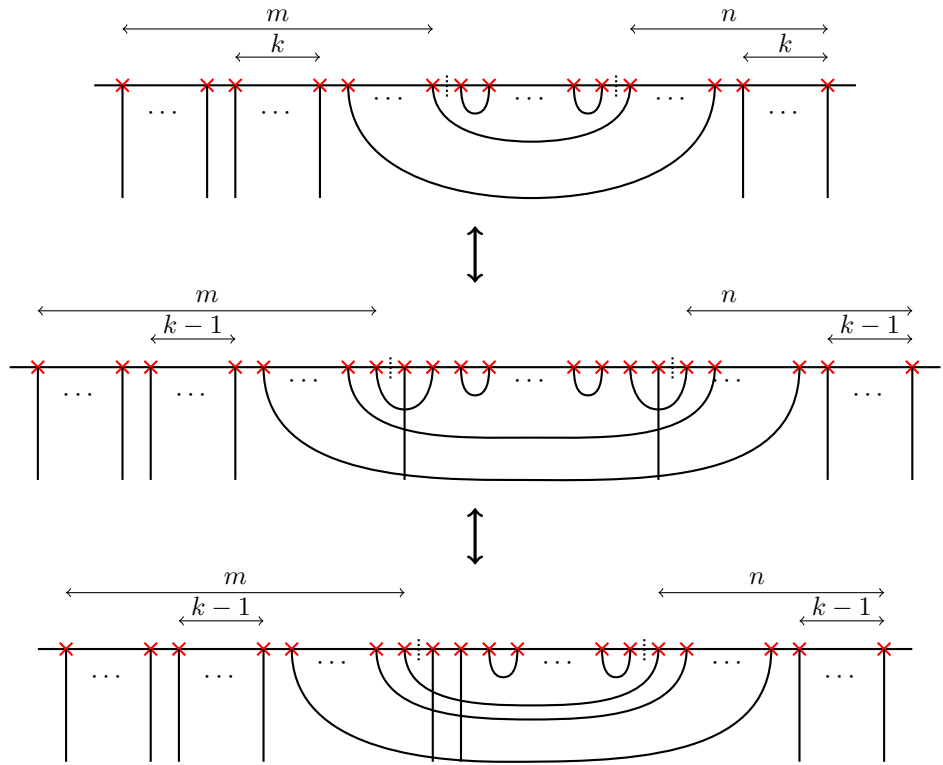
\begin{figure}[htp]
  \centering
\begin{tikzpicture}[thick,scale=0.75,dotdia/.style={cross out, draw,
    solid, red, inner sep=2pt}]

  \draw (-0.5,0) -- (13,0);
  \foreach \x in
  {0,1.5,2,3.5,4,5.5,6,6.5,8,8.5,9,10.5,11,12.5}
  {
    \node [dotdia] at (\x,0) {}; 
  }
  \foreach \x in {0,1.5,2,3.5,11,12.5} 
  { 
    \draw (\x,0)--(\x,-2); 
  } 
  \foreach \x in {6,8} 
  { 
    \draw (\x,0) to[out=-90,in=180]
    (\x+0.25,-0.5) to[out=0,in=-90] (\x+0.5,0); 
  }
  \draw (5.5,0) to[out=-90,in=180]
  (5.5+1.75,-1) to[out=0,in=-90] (5.5+3.5,0); 
  \draw (4,0) to[out=-90,in=180]
  (4+3.25,-2) to[out=0,in=-90] (4+6.5,0); 

  \node at (0.75,-0.5) {$\cdots$};
  \node at (11+0.75,-0.5) {$\cdots$};
  \node at (2+0.75,-0.5) {$\cdots$};
  \node at (4.75,-0.25) {$\cdots$};    
  \node at (9.75,-0.25) {$\cdots$};    
  \node at (7.25,-0.25) {$\cdots$};    
  \node at (2.75,0.75) {$k$};
  \node at (11.75,0.75) {$k$};
  \draw[thin,<->] (2,0.5)--(3.5,0.5);
  \draw[thin,<->] (11,0.5)--(12.5,0.5);
  \node at (2.75,1.25) {$m$};
  \node at (10.75,1.25) {$n$};
  \draw[thin,<->] (0,1)--(5.5,1);
  \draw[thin,<->] (9,1)--(12.5,1);
  \foreach \y in
  {0}
  {
    \draw[densely dotted] (5.75,\y-0.2)--(5.75,\y+0.2); 
    \draw[densely dotted] (8.75,\y-0.2)--(8.75,\y+0.2); 
  }

  \draw[<->,very thick] (6.25,-2.5)--(6.25,-3.5);

  \begin{scope}[yshift=-5cm,xshift=-1.5cm]
    \draw (-0.5,0) -- (16,0);
    \foreach \x in
    {0,1.5,2,3.5,4,5.5,6,6.5,7,7.5,8,9.5,10,10.5,11,11.5,
      12,13.5,14,15.5}
    {
      \node [dotdia] at (\x,0) {}; 
    }
    \foreach \x in {0,1.5,2,3.5,6.5,11,14,15.5} 
    { 
      \draw (\x,0)--(\x,-2); 
    } 
    \foreach \x in {7.5,9.5} 
    { 
      \draw (\x,0) to[out=-90,in=180]
      (\x+0.25,-0.5) to[out=0,in=-90] (\x+0.5,0); 
    }
    \foreach \x in {6,10.5} 
    { 
      \draw (\x,0) to[out=-90,in=180]
      (\x+0.5,-0.75) to[out=0,in=-90] (\x+1,0); 
    }

    \draw (5.5,0) to[out=-90,in=180]
    (5.5+3.25,-1.25) to[out=0,in=-90] (5.5+6.5,0); 
    \draw (4,0) to[out=-90,in=180]
    (4+4.25,-2) to[out=0,in=-90] (4+9.5,0); 
    
    \node at (0.75,-0.5) {$\cdots$};
    \node at (2+0.75,-0.5) {$\cdots$};
    \node at (4.75,-0.25) {$\cdots$};    
    \node at (8.75,-0.25) {$\cdots$};    
    \node at (12.25,-0.25) {$\cdots$};
    \node at (14+0.75,-0.5) {$\cdots$};    
    \node at (2.75,0.75) {$k-1$};
    \node at (14.75,0.75) {$k-1$};
    \draw[thin,<->] (2,0.5)--(3.5,0.5);
    \draw[thin,<->] (14,0.5)--(15.5,0.5);
    \node at (3,1.25) {$m$};
    \node at (12.25,1.25) {$n$};
    \draw[thin,<->] (0,1)--(6,1);
    \draw[thin,<->] (11.5,1)--(15.5,1);
    \draw[<->,very thick] (7.75,-2.5)--(7.75,-3.5);
    \foreach \y in
    {0}
    {
      \draw[densely dotted] (6.25,\y-0.2)--(6.25,\y+0.2); 
      \draw[densely dotted] (11.25,\y-0.2)--(11.25,\y+0.2); 
    }

  \end{scope}

  \begin{scope}[yshift=-10cm,xshift=-1cm]
    \draw (-0.5,0) -- (15,0);
    \foreach \x in
    {0,1.5,2,3.5,4,5.5,6,6.5,7,7.5,8,9.5,10,10.5,11,12.5,13,14.5}
    {
      \node [dotdia] at (\x,0) {}; 
    }
    \foreach \x in {0,1.5,2,3.5,6.5,7,13,14.5} 
    { 
      \draw (\x,0)--(\x,-2); 
    } 
    \foreach \x in {7.5,9.5} 
    { 
      \draw (\x,0) to[out=-90,in=180]
      (\x+0.25,-0.5) to[out=0,in=-90] (\x+0.5,0); 
    }

    \draw (6,0) to[out=-90,in=180]
    (6+2.25,-1) to[out=0,in=-90] (6+4.5,0); 

    \draw (5.5,0) to[out=-90,in=180]
    (5.5+2.75,-1.25) to[out=0,in=-90] (5.5+5.5,0); 
    \draw (4,0) to[out=-90,in=180]
    (4+3.75,-2) to[out=0,in=-90] (4+8.5,0); 
    
    \node at (0.75,-0.5) {$\cdots$};
    \node at (2+0.75,-0.5) {$\cdots$};
    \node at (4.75,-0.25) {$\cdots$};    
    \node at (8.75,-0.25) {$\cdots$};    
    \node at (11.75,-0.25) {$\cdots$};
    \node at (13+0.75,-0.5) {$\cdots$};    
    \node at (2.75,0.75) {$k-1$};
    \node at (13.75,0.75) {$k-1$};
    \draw[thin,<->] (2,0.5)--(3.5,0.5);
    \draw[thin,<->] (13,0.5)--(14.5,0.5);
    \node at (3,1.25) {$m$};
    \node at (12.25,1.25) {$n$};
    \draw[thin,<->] (0,1)--(6,1);
    \draw[thin,<->] (10.5,1)--(14.5,1);
    \foreach \y in
    {0}
    {
      \draw[densely dotted] (6.25,\y-0.2)--(6.25,\y+0.2); 
      \draw[densely dotted] (10.25,\y-0.2)--(10.25,\y+0.2); 
    }
  \end{scope}

\end{tikzpicture}
  \caption{From the reference state of the sector $(j=0,k)$ on top, we
    can reach the state of the sector $(j=1,k-1)$ on bottom, by the
    two steps depicted. First, for each boundary we pass one of $k$
    lines into the bulk by applying permutations (middle), and then we
    order the strings using TL generators (we have assumed
    $\nRepLeft\ge \nRepRight$).  Going from bottom to top works in
    the same manner, and this procedure can be done for every states
    in the two sectors, which are then equivalent.}
\label{fig:j0Ktoj1K-1}
\end{figure}

Note that with this definition, the sectors $(j=0,k)$ and
$(j=1,k-1)$ are actually the same, and that a state can be regarded as
belonging to one or the other sector (see figure
\ref{fig:j0Ktoj1K-1}).

Therefore it remains to study the sectors $(j=0, k = 0,1, \dots,
\minlr)$ and $(j,k= \minlr)$ when $j>0$. In each such sector the total
(bulk plus boundary) number of non-contractible lines is different and
correspondingly the value of the exponents will depend on $j+k$.
Calling $\ell = |\nRepLeft-\nRepRight|$, we will refer to the lowest
exponent in a sector $(j,k)$ as the $(2j+2k+\ell)$-leg exponent,
$h^{\nRepLeft,\nRepRight}(j+k)$.  Note anyway that if we would like to
look at the lowest exponent when we fix only $j$, this will be present
in the sector with the minimum value of $k$ possible.

\subsubsection{Relations with two-boundary loop models and solution}
\label{sec:rel_2blm}

In the analysis of the one-boundary problem we have found that our
model is in the same universality class as a $1$BLM with weight of
marked loops given by \eqref{eq:n1_m}. This weight can be obtained by
taking as blob operator the Young symmetrizer of the fully symmetric
representation of the symmetric group.  This identification now turns
out to be very useful for solving the two-boundary problem which we
recall is described by
\begin{equation}
  \label{eq:hamiltonian_loop}
  H = - u \sum_{i=0}^{\nRepLeft-1} P_{i,i+1} - 
  \sum_{i=\nRepLeft}^{2\nRepBulk+\nRepLeft-2} E_i -
  v \sum_{i=2\nRepBulk+\nRepLeft-1}^{2\nRepBulk+\nRepLeft+\nRepRight-2} P_{i,i+1}\, .
\end{equation}
Indeed, by naive analogy,  we expect  the following Hamiltonian to be in the
same universality class as our starting problem \eqref{eq:hamiltonian_loop}:
\begin{align}
  \tilde{H} 
  &= 
  - u\, \frac{1}{(\nRepLeft+1)!}  
  \sum_{\sigma \in \perm^{\mbox{\tiny{left}}}} \sigma -
  \sum_{i=\nRepLeft}^{2\nRepBulk+\nRepLeft-2} E_i
  -v\, \frac{1}{(\nRepRight+1)!} 
  \sum_{\sigma \in \perm^{\mbox{\tiny{right}}}} \sigma \\
   \label{eq:Hallperm}
   &= -u \, \bleft 
   - \sum_{i=\nRepLeft}^{2\nRepBulk+\nRepLeft-2} E_i
   -v \, \bright
\end{align}
where we have introduced the symmetric group of the $m+1$ leftmost
strands $\perm^{\mbox{\scriptsize{left}}} =
\perm_{\{0,\dots,m\}}$, the symmetric group of the $n+1$ rightmost
strands $\perm^{\mbox{\scriptsize{right}}} =
\perm_{\{2\nRepBulk+\nRepLeft-1,\dots,2\nRepBulk+\nRepLeft+\nRepRight-1\}}$,
as well as the standard Young tableaux
$t_1 = \tableft$ and $t_2 = \tabright$. As before $b_t$ stands
for the Young symmetrizer related to the tableau $t$. 

\begin{figure}[htp]
  \centering
\begingroup
  \makeatletter
  \providecommand\color[2][]{%
    \GenericError{(gnuplot) \space\space\space\@spaces}{%
      Package color not loaded in conjunction with
      terminal option `colourtext'%
    }{See the gnuplot documentation for explanation.%
    }{Either use 'blacktext' in gnuplot or load the package
      color.sty in LaTeX.}%
    \renewcommand\color[2][]{}%
  }%
  \providecommand\includegraphics[2][]{%
    \GenericError{(gnuplot) \space\space\space\@spaces}{%
      Package graphicx or graphics not loaded%
    }{See the gnuplot documentation for explanation.%
    }{The gnuplot epslatex terminal needs graphicx.sty or graphics.sty.}%
    \renewcommand\includegraphics[2][]{}%
  }%
  \providecommand\rotatebox[2]{#2}%
  \@ifundefined{ifGPcolor}{%
    \newif\ifGPcolor
    \GPcolortrue
  }{}%
  \@ifundefined{ifGPblacktext}{%
    \newif\ifGPblacktext
    \GPblacktexttrue
  }{}%
  \let\gplgaddtomacro\g@addto@macro
  \gdef\gplbacktext{}%
  \gdef\gplfronttext{}%
  \makeatother
  \ifGPblacktext
    \def\colorrgb#1{}%
    \def\colorgray#1{}%
  \else
    \ifGPcolor
      \def\colorrgb#1{\color[rgb]{#1}}%
      \def\colorgray#1{\color[gray]{#1}}%
      \expandafter\def\csname LTw\endcsname{\color{white}}%
      \expandafter\def\csname LTb\endcsname{\color{black}}%
      \expandafter\def\csname LTa\endcsname{\color{black}}%
      \expandafter\def\csname LT0\endcsname{\color[rgb]{1,0,0}}%
      \expandafter\def\csname LT1\endcsname{\color[rgb]{0,1,0}}%
      \expandafter\def\csname LT2\endcsname{\color[rgb]{0,0,1}}%
      \expandafter\def\csname LT3\endcsname{\color[rgb]{1,0,1}}%
      \expandafter\def\csname LT4\endcsname{\color[rgb]{0,1,1}}%
      \expandafter\def\csname LT5\endcsname{\color[rgb]{1,1,0}}%
      \expandafter\def\csname LT6\endcsname{\color[rgb]{0,0,0}}%
      \expandafter\def\csname LT7\endcsname{\color[rgb]{1,0.3,0}}%
      \expandafter\def\csname LT8\endcsname{\color[rgb]{0.5,0.5,0.5}}%
    \else
      \def\colorrgb#1{\color{black}}%
      \def\colorgray#1{\color[gray]{#1}}%
      \expandafter\def\csname LTw\endcsname{\color{white}}%
      \expandafter\def\csname LTb\endcsname{\color{black}}%
      \expandafter\def\csname LTa\endcsname{\color{black}}%
      \expandafter\def\csname LT0\endcsname{\color{black}}%
      \expandafter\def\csname LT1\endcsname{\color{black}}%
      \expandafter\def\csname LT2\endcsname{\color{black}}%
      \expandafter\def\csname LT3\endcsname{\color{black}}%
      \expandafter\def\csname LT4\endcsname{\color{black}}%
      \expandafter\def\csname LT5\endcsname{\color{black}}%
      \expandafter\def\csname LT6\endcsname{\color{black}}%
      \expandafter\def\csname LT7\endcsname{\color{black}}%
      \expandafter\def\csname LT8\endcsname{\color{black}}%
    \fi
  \fi
  \setlength{\unitlength}{0.0500bp}%
  \begin{picture}(6120.00,4284.00)%
    \gplgaddtomacro\gplbacktext{%
      \csname LTb\endcsname%
      \put(1606,704){\makebox(0,0)[r]{\strut{}$-0.05$}}%
      \put(1606,1069){\makebox(0,0)[r]{\strut{}$0$}}%
      \put(1606,1434){\makebox(0,0)[r]{\strut{}$0.05$}}%
      \put(1606,1799){\makebox(0,0)[r]{\strut{}$0.1$}}%
      \put(1606,2164){\makebox(0,0)[r]{\strut{}$0.15$}}%
      \put(1606,2529){\makebox(0,0)[r]{\strut{}$0.2$}}%
      \put(1606,2894){\makebox(0,0)[r]{\strut{}$0.25$}}%
      \put(1606,3259){\makebox(0,0)[r]{\strut{}$0.3$}}%
      \put(1606,3624){\makebox(0,0)[r]{\strut{}$0.35$}}%
      \put(1738,484){\makebox(0,0){\strut{}$0$}}%
      \put(2143,484){\makebox(0,0){\strut{}$1$}}%
      \put(2548,484){\makebox(0,0){\strut{}$2$}}%
      \put(2954,484){\makebox(0,0){\strut{}$3$}}%
      \put(3359,484){\makebox(0,0){\strut{}$4$}}%
      \put(3764,484){\makebox(0,0){\strut{}$5$}}%
      \put(4169,484){\makebox(0,0){\strut{}$6$}}%
      \put(4574,484){\makebox(0,0){\strut{}$7$}}%
      \put(4980,484){\makebox(0,0){\strut{}$8$}}%
      \put(5385,484){\makebox(0,0){\strut{}$9$}}%
      \put(5790,484){\makebox(0,0){\strut{}$10$}}%
      \put(440,2164){\rotatebox{90}{\makebox(0,0){\strut{}$h$}}}%
      \put(3764,154){\makebox(0,0){\strut{}$u=v$}}%
      \put(3764,3954){\makebox(0,0){\strut{}Two-boundaries $\beta=1$}}%
    }%
    \gplgaddtomacro\gplfronttext{%
      \csname LTb\endcsname%
      \put(4935,2930){\makebox(0,0)[r]{\strut{}$m=n=1, j=0,k=0$}}%
      \csname LTb\endcsname%
      \put(4935,2710){\makebox(0,0)[r]{\strut{}$m=n=1, j=0,k=1$}}%
      \csname LTb\endcsname%
      \put(4935,2490){\makebox(0,0)[r]{\strut{}$m=n=1, j=1,k=1$}}%
      \csname LTb\endcsname%
      \put(4935,2270){\makebox(0,0)[r]{\strut{}$m=2,n=1, j=0,k=0$}}%
      \csname LTb\endcsname%
      \put(4935,2050){\makebox(0,0)[r]{\strut{}$m=2,n=1, j=0,k=1$}}%
      \csname LTb\endcsname%
      \put(4935,1830){\makebox(0,0)[r]{\strut{}$m=2,n=1, j=1,k=1$}}%
    }%
    \gplbacktext
    \put(0,0){\includegraphics{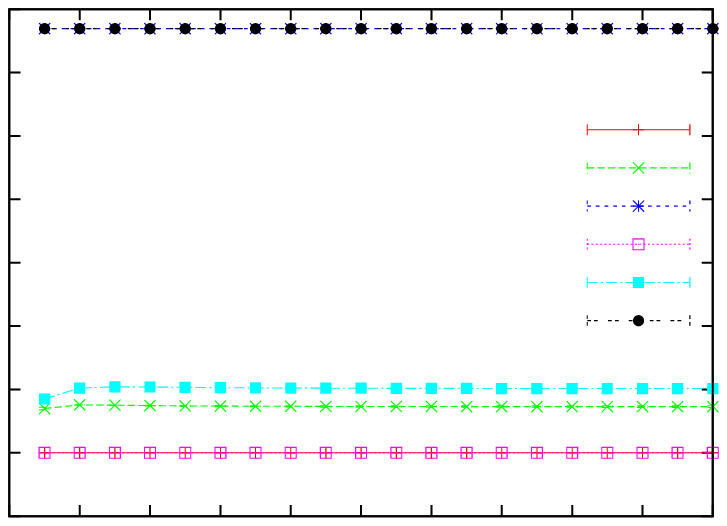}}%
    \gplfronttext
  \end{picture}%
\endgroup
  \caption{Independence of the first few critical exponents on $u=v$
    for $\nRepLeft=\nRepRight=1$ and $\nRepLeft=2,\nRepRight=1$ when
    $\fug=1$.  The values are computed by finite size scaling fitting
    equation \eqref{eq:c_finite_size} with a term of order
    $1/\nRepBulk^2$, for systems of sizes $\nRepBulk=4$ to
    $\nRepBulk=8$.}
  \label{fig:indep_u_v}
\end{figure}

We have checked numerically the hypothesis that adding these additional
permutations to our former Hamiltonian $H$ is a redundant perturbation
also in this two-boundary case.
See table \ref{tab:h_m2_mbar2} for a sample of evidence in
the case $\nRepLeft=\nRepRight=2$.
As in the one-boundary case, we claim that the results are unchanged
as long as $u$ and $v$ are taken positive, as we have verified this
numerically (see figure \ref{fig:indep_u_v}).
\begin{table}[h!c]
  \centering
  \begin{tabular}{|c|c|c|c|c|c|}
    \hline
    $\fug$ & $j$ & $k$ & $h$ & $\tilde{h}$ & $h_{2,2}(j+k)$ \\
    \hline
    $0$ & $0$ & $0$ & $-0.0026\pm 0.0010$ & $-0.0041 \pm 0.0013$ 
    & $0$ \\
    \hline
    $0$ & $0$ & $1$ & $-0.0026\pm 0.0010$ & $-0.0041 \pm 0.0013$ 
    & $0$ \\
    \hline
    $0$ & $0$ & $2$ & $0.0504\pm 0.0015$ & $0.0479 \pm 0.0021$ 
    & $0.054734$ \\
    \hline
    $0$ & $1$ & $2$ & $0.247\pm 0.003$ & $ 0.238 \pm 0.005$ 
    & $0.257230$ \\
    \hline
    $0$ & $2$ & $2$ & $1.670\pm 0.007$ & $1.601 \pm 0.006$ 
    & $1.631564$ \\
    \hline
    $1$ & $0$ & $0$ & $0$ & $0$ & $0$ \\
    \hline
    $1$ & $0$ & $1$ & $0.0152\pm 0.0002$ & $0.0147 \pm 0.0004$ 
    & $0.015906$ \\
    \hline
    $1$ & $0$ & $2$ & $0.0700\pm 0.0012$ & $0.0678\pm 0.0018$ 
    & $0.073136$ \\
    \hline
    $1$ & $1$ & $2$ & $0.3346\pm 0.0003$ & $0.3346\pm 0.0003$ & $1/3$\\
    \hline
    $1$ & $2$ & $2$ & $2.118 \pm 0.008$ & $2.118\pm 0.008$ & $2$ \\
    \hline
\end{tabular}
\caption{Comparison of critical exponents for the model described by
  \eqref{eq:hamiltonian_loop}, $h$, and for the one described by 
  \eqref{eq:Hallperm}, $\tilde{h}$, listed for different values of $\fug$,
  $j$ and $k$, in a system with $\nRepLeft=2,\nRepRight=2$.  
  The values are obtained by fitting eigenvalues for bulk sizes
  $\nRepBulk=4\to 8$, and although these numbers should be equal when 
  $\nRepBulk\to\infty$, already with these relatively small sizes we have 
  good agreement between these two values and with 
  the exact value predicted, given in the last column.}
  \label{tab:h_m2_mbar2}
\end{table}

The advantage of studying $\widetilde{H}$ is that it belongs to the
two-boundary Temperley-Lieb algebra, the two-boundary analogue of the
blob algebra, and for that problem we know how to compute the
exponents \cite{Dubail2009}.  In the language of the two-boundary
loop model ($2$BLM), loops
marked on the left boundary have weight \eqref{eq:n1_m} and those
marked on the right boundary
\begin{equation}
  \label{eq:n2_mbar}
  \fugR = \frac{\nRepRight+\fug}{\nRepRight+1} \, .
\end{equation}
All that remains to be done is to compute $\fugLR$, the value of loops
marked by both blob operators, as expressed in the following algebraic
relation (we can consider for simplicity the situation with $L=1$; in
the general case replace $E_m$ by $E$ defined in (\ref{eq:defs1_quotient})):
\begin{equation}
  \label{eq:eb1b2e}
  E_m b_1 b_2 E_m = \fugLR E_m \, ,
\end{equation}
where $b_1$ and $b_2$ act respectively on the first and second strand labelled
$m$ and $m+1$.

So we fix $\nRepBulk=1$, and consider the $j=0$ sector only.  Define
$\hat{t}_1 = \tableftmone$ , $\hat{t}_2 = \tabrightmone$.  If we
replace respectively $b_1$ and $b_2$ in equation \eqref{eq:eb1b2e}
with our operators $\bleft$ and $\bright$, and $E_0$ with
$\tilde{E}_m= E_m \bleftmone \brightmone$, we would like that the
following equation holds for a certain value of $\fugLR$:
\begin{equation}
  \label{eq:e_bleft_bright_e}
  \tilde{E}_\nRepLeft \bleft \bright \tilde{E}_\nRepLeft 
  = 
  \fugLR \tilde{E}_\nRepLeft \, .
\end{equation}
This equation can however only be satisfied and meaningful in our
model, if we consider it for a fixed value of $k$, as we will explain
below. The sector $(j=0,k)$ of the Hamiltonian \eqref{eq:Hallperm}
will then be described for different $k$'s by different quotients
fixing $\fugLR$, which we will now call $\fugLR^k$ to indicate  the
$k$-dependence. Taking the quotient as described in section
\ref{sec:alg_quot}, we have already implicitly chosen particular
values of $\fugLR^k$, which we are going to compute.  We remark that
the representation of the two-boundary Temperley-Lieb algebra given by
\eqref{eq:Hallperm} is not faithful in general, and the structure of
the sectors cannot be directly given from this mapping.  However the
leading eigenvalues will be present and we explain now how to
compute them in the different sectors.

\begin{figure}[htp]
  \centering
\begin{tikzpicture}[thick,scale=0.75,dotdia/.style={cross out, draw,
    solid, red, inner sep=2pt}]
  \begin{scope}[xshift=0cm]
    \node at (-1.25,0) {$W = $};
    \draw (-0.5,-1) -- (5.5,-1);
    \draw (-0.5,1) -- (5.5,1);
    \foreach \x in
    {0,1,2,3,4,5}
    {
      \node [dotdia] at (\x,1) {}; 
      \node [dotdia] at (\x,-1) {}; 
    }
    \foreach \x in {0,5} 
    { 
      \draw (\x,-1)--(\x,1); 
    } 
    
    \draw (2,-1) to[out=90,in=180]
    (2+0.5,-0.5) to[out=0,in=90] (2+1,-1); 
    \draw (2,1) to[out=-90,in=180]
    (2+0.5,0.5) to[out=0,in=-90] (2+1,1); 
    
    \draw (1,-1) to[out=90,in=180]
    (1+1.5,0.25) to[out=0,in=90] (1+3,-1); 
    \draw (1,1) to[out=-90,in=180]
    (1+1.5,-0.25) to[out=0,in=-90] (1+3,1); 

    \foreach \x in
    {-0.15,3.85}
    {
      \draw[fill=black!20] (\x,0.75) rectangle (\x+1.3,0.8); 
      \draw[fill=black!20] (\x,-0.75) rectangle (\x+1.3,-0.8); 
    }
    \foreach \y in
    {-1,1}
    {
      \draw[densely dotted] (1.5,\y-0.2)--(1.5,\y+0.2); 
      \draw[densely dotted] (3.5,\y-0.2)--(3.5,\y+0.2); 
    }
  \end{scope}

  \begin{scope}[xshift=10cm]
    \node at (-1.25,0) {$\tilde{E}_2 = $};
    \draw (-0.5,-1) -- (5.5,-1);
    \draw (-0.5,1) -- (5.5,1);
    \foreach \x in
    {0,1,2,3,4,5}
    {
      \node [dotdia] at (\x,1) {}; 
      \node [dotdia] at (\x,-1) {}; 
    }
    \foreach \x in {0,1,4,5} 
    { 
      \draw (\x,-1)--(\x,1); 
    } 

    \draw (2,-1) to[out=90,in=180]
    (2+0.5,-0.5) to[out=0,in=90] (2+1,-1); 
    \draw (2,1) to[out=-90,in=180]
    (2+0.5,0.5) to[out=0,in=-90] (2+1,1); 
    
    \foreach \x in
    {-0.15,3.85}
    {
      \draw[fill=black!20] (\x,0.75) rectangle (\x+1.3,0.8); 
      \draw[fill=black!20] (\x,-0.75) rectangle (\x+1.3,-0.8); 
    }
    \foreach \y in
    {-1,1}
    {
      \draw[densely dotted] (1.5,\y-0.2)--(1.5,\y+0.2); 
      \draw[densely dotted] (3.5,\y-0.2)--(3.5,\y+0.2); 
    }
  \end{scope}

\end{tikzpicture}
  \caption{The words $W$ and $\tilde{E}_2$ in the diagram
    representation.  As before, the gray bars indicate
    symmetrization over the lines and $W$ is obtained using the
    quotient adopted in section \ref{sec:loop_repr}.}
\label{fig:w-e2}
\end{figure}
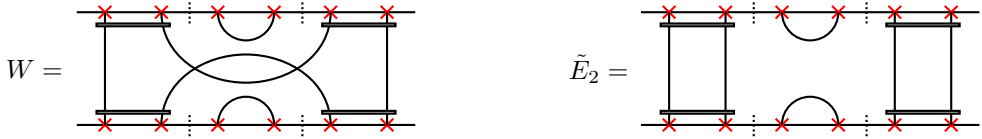

Before showing the general procedure, we will give an example in the
case $\nRepLeft=\nRepRight=2$, where we can understand how the
mechanism works.  First it is convenient to express $\bleft$ and
$\bright$ using the following identities:
\begin{align}
  \bleft &= \frac{1}{3} \bleftmone (1 + (0,2) + (1,2)) 
  = \frac{1}{3} \bleftmone (1 + P_1 (1 + P_0)) \\
  \bright &= \frac{1}{3} (1 + (3,4) + (3,5)) \brightmone 
  = \frac{1}{3} (1 + (1 + P_4)P_3) \brightmone \, .
\end{align}
Then, using the commutation relations of the algebra, we have:
\begin{align}
  \tilde{E}_2 \bleft \bright \tilde{E}_2 &= 
  \frac{1}{9} E_2 \bleftmone (1 + P_1 (1 + P_0))
  (1 + (1 + P_4)P_3) \brightmone
  E_2 = \nonumber \\
  \label{eq:ebbe_m2mbar2}
  &= \frac{1}{9} \left((\fug + 4)\tilde{E}_2 
    + 4 \bleftmone \brightmone E_2 P_1 P_3 E_2 \bleftmone \brightmone
  \right) \, .
\end{align}
The first piece of equation \eqref{eq:ebbe_m2mbar2} is already in the
desired form, while the second piece is not proportional to
$\tilde{E}_2$.  As anticipated, we can only achieve this
by fixing a sector $k$, and computing the action of the operator
$W:=\bleftmone \brightmone E_2 P_1 P_3 E_2 \bleftmone \brightmone$ on 
the states belonging to the sector. The graphical representation of
$W$ and $\tilde{E}_2$ is depicted in figure \ref{fig:w-e2}.

Let us first consider the case $k=2$, when no contractions of boundary
lines is allowed. It is clear from figure \ref{fig:w-e2} that $W$ acts
as zero on every state in this sector, since it diminishes the number
of strings from $2$ to $1$.  So, we have found that every state of the
sector $k=2$ is in the kernel of the operator
\begin{equation}
  \label{eq:R_m2_mbar2}
  R := \left(\tilde{E}_2 \bleft \bright - \fugLR^k\right) \tilde{E}_2 \, ,
\end{equation}
with 
\begin{equation}
  \label{eq:n122_m2_mbar2}
  \fugLR^2 = \frac{\fug+4}{9} \, .
\end{equation}

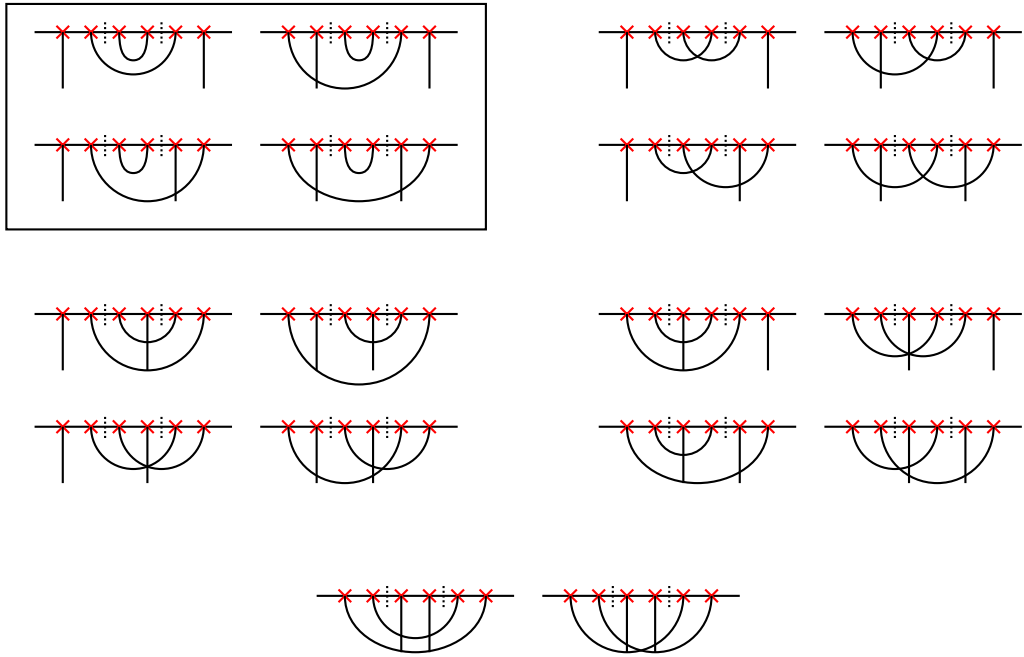
\begin{figure}[htp]
  \centering
\begin{tikzpicture}[thick,scale=0.75,dotdia/.style={cross out, draw,
    solid, red, inner sep=2pt}]
  \draw (-1,0.5) rectangle (7.5,-3.5);
  \draw (-0.5,0) -- (3,0);
  \foreach \x in
  {0,0.5,1,1.5,2,2.5}
  {
    \node [dotdia] at (\x,0) {}; 
  }
  \foreach \x in {0,2.5}
  { 
    \draw (\x,0)--(\x,-1); 
  } 
  \foreach \x in {1}
  { 
    \draw (\x,0) to[out=-90,in=180]
    (\x+0.25,-0.5) to[out=0,in=-90] (\x+0.5,0); 
  }
  \draw (0.5,0) to[out=-90,in=180]
  (0.5+0.75,-0.75) to[out=0,in=-90] (0.5+1.5,0); 
  \foreach \y in
  {0}
  {
    \draw[densely dotted] (0.75,\y-0.2)--(0.75,\y+0.2); 
    \draw[densely dotted] (1.75,\y-0.2)--(1.75,\y+0.2); 
  }
  \begin{scope}[xshift=4cm]
    \draw (-0.5,0) -- (3,0);
    \foreach \x in
    {0,0.5,1,1.5,2,2.5}
    {
      \node [dotdia] at (\x,0) {}; 
    }
    \foreach \x in {0.5,2.5}
    { 
      \draw (\x,0)--(\x,-1); 
    } 
    \foreach \x in {1}
    { 
      \draw (\x,0) to[out=-90,in=180]
      (\x+0.25,-0.5) to[out=0,in=-90] (\x+0.5,0); 
    }
    \draw (0,0) to[out=-90,in=180]
    (0+1,-1) to[out=0,in=-90] (0+2,0); 
    \foreach \y in
    {0}
    {
      \draw[densely dotted] (0.75,\y-0.2)--(0.75,\y+0.2); 
      \draw[densely dotted] (1.75,\y-0.2)--(1.75,\y+0.2); 
    }
  \end{scope}
  \begin{scope}[yshift=-2cm]
    \draw (-0.5,0) -- (3,0);
    \foreach \x in
    {0,0.5,1,1.5,2,2.5}
    {
      \node [dotdia] at (\x,0) {}; 
    }
    \foreach \x in {0,2}
    { 
      \draw (\x,0)--(\x,-1); 
    } 
    \foreach \x in {1}
    { 
      \draw (\x,0) to[out=-90,in=180]
      (\x+0.25,-0.5) to[out=0,in=-90] (\x+0.5,0); 
    }
    \draw (0.5,0) to[out=-90,in=180]
    (0.5+1,-1) to[out=0,in=-90] (0.5+2,0); 
    \foreach \y in
    {0}
    {
      \draw[densely dotted] (0.75,\y-0.2)--(0.75,\y+0.2); 
      \draw[densely dotted] (1.75,\y-0.2)--(1.75,\y+0.2); 
    }
  \end{scope}
  \begin{scope}[xshift=4cm,yshift=-2cm]
    \draw (-0.5,0) -- (3,0);
    \foreach \x in
    {0,0.5,1,1.5,2,2.5}
    {
      \node [dotdia] at (\x,0) {}; 
    }
    \foreach \x in {0.5,2}
    { 
      \draw (\x,0)--(\x,-1); 
    } 
    \foreach \x in {1}
    { 
      \draw (\x,0) to[out=-90,in=180]
      (\x+0.25,-0.5) to[out=0,in=-90] (\x+0.5,0); 
    }
    \draw (0,0) to[out=-90,in=180]
    (0+1.25,-1) to[out=0,in=-90] (0+2.5,0); 
    \foreach \y in
    {0}
    {
      \draw[densely dotted] (0.75,\y-0.2)--(0.75,\y+0.2); 
      \draw[densely dotted] (1.75,\y-0.2)--(1.75,\y+0.2); 
    }
  \end{scope}

  \begin{scope}[xshift=10cm]
  \draw (-0.5,0) -- (3,0);
  \foreach \x in
  {0,0.5,1,1.5,2,2.5}
  {
    \node [dotdia] at (\x,0) {}; 
  }
  \foreach \x in {0,2.5}
  { 
    \draw (\x,0)--(\x,-1); 
  } 
  \foreach \x in {0.5,1}
  { 
    \draw (\x,0) to[out=-90,in=180]
    (\x+0.5,-0.5) to[out=0,in=-90] (\x+1,0); 
  }
    \foreach \y in
    {0}
    {
      \draw[densely dotted] (0.75,\y-0.2)--(0.75,\y+0.2); 
      \draw[densely dotted] (1.75,\y-0.2)--(1.75,\y+0.2); 
    }

  \begin{scope}[xshift=4cm]
    \draw (-0.5,0) -- (3,0);
    \foreach \x in
    {0,0.5,1,1.5,2,2.5}
    {
      \node [dotdia] at (\x,0) {}; 
    }
    \foreach \x in {0.5,2.5}
    { 
      \draw (\x,0)--(\x,-1); 
    } 
    \draw (0,0) to[out=-90,in=180]
    (0+0.75,-0.75) to[out=0,in=-90] (0+1.5,0); 
    \draw (1,0) to[out=-90,in=180]
    (1+0.5,-0.5) to[out=0,in=-90] (1+1,0); 
    \foreach \y in
    {0}
    {
      \draw[densely dotted] (0.75,\y-0.2)--(0.75,\y+0.2); 
      \draw[densely dotted] (1.75,\y-0.2)--(1.75,\y+0.2); 
    }

  \end{scope}
  \begin{scope}[yshift=-2cm]
    \draw (-0.5,0) -- (3,0);
    \foreach \x in
    {0,0.5,1,1.5,2,2.5}
    {
      \node [dotdia] at (\x,0) {}; 
    }
    \foreach \x in {0,2}
    { 
      \draw (\x,0)--(\x,-1); 
    } 
    \draw (0.5,0) to[out=-90,in=180]
    (0.5+0.5,-0.5) to[out=0,in=-90] (0.5+1,0); 
    \draw (1,0) to[out=-90,in=180]
    (1+0.75,-0.75) to[out=0,in=-90] (1+1.5,0); 
    \foreach \y in
    {0}
    {
      \draw[densely dotted] (0.75,\y-0.2)--(0.75,\y+0.2); 
      \draw[densely dotted] (1.75,\y-0.2)--(1.75,\y+0.2); 
    }
  \end{scope}
  \begin{scope}[xshift=4cm,yshift=-2cm]
    \draw (-0.5,0) -- (3,0);
    \foreach \x in
    {0,0.5,1,1.5,2,2.5}
    {
      \node [dotdia] at (\x,0) {}; 
    }
    \foreach \x in {0.5,2}
    { 
      \draw (\x,0)--(\x,-1); 
    } 
    \draw (0,0) to[out=-90,in=180]
    (0+0.75,-0.75) to[out=0,in=-90] (0+1.5,0); 
    \draw (1,0) to[out=-90,in=180]
    (1+0.75,-0.75) to[out=0,in=-90] (1+1.5,0); 
    \foreach \y in
    {0}
    {
      \draw[densely dotted] (0.75,\y-0.2)--(0.75,\y+0.2); 
      \draw[densely dotted] (1.75,\y-0.2)--(1.75,\y+0.2); 
    }
  \end{scope}
  \end{scope}

  \begin{scope}[yshift=-5cm]
  \draw (-0.5,0) -- (3,0);
  \foreach \x in
  {0,0.5,1,1.5,2,2.5}
  {
    \node [dotdia] at (\x,0) {}; 
  }
  \foreach \x in {0,1.5}
  { 
    \draw (\x,0)--(\x,-1); 
  } 
  \draw (0.5,0) to[out=-90,in=180]
  (0.5+1,-1) to[out=0,in=-90] (0.5+2,0); 
  \draw (1,0) to[out=-90,in=180]
  (1+0.5,-0.5) to[out=0,in=-90] (1+1,0); 
    \foreach \y in
    {0}
    {
      \draw[densely dotted] (0.75,\y-0.2)--(0.75,\y+0.2); 
      \draw[densely dotted] (1.75,\y-0.2)--(1.75,\y+0.2); 
    }

  \begin{scope}[xshift=4cm]
    \draw (-0.5,0) -- (3,0);
    \foreach \x in
    {0,0.5,1,1.5,2,2.5}
    {
      \node [dotdia] at (\x,0) {}; 
    }
    \foreach \x in {0.5,1.5}
    { 
      \draw (\x,0)--(\x,-1); 
    } 
    \draw (0,0) to[out=-90,in=180]
    (0+1.25,-1.25) to[out=0,in=-90] (0+2.5,0); 
    \draw (1,0) to[out=-90,in=180]
    (1+0.5,-0.5) to[out=0,in=-90] (1+1,0); 
    \foreach \y in
    {0}
    {
      \draw[densely dotted] (0.75,\y-0.2)--(0.75,\y+0.2); 
      \draw[densely dotted] (1.75,\y-0.2)--(1.75,\y+0.2); 
    }

  \end{scope}
  \begin{scope}[yshift=-2cm]
    \draw (-0.5,0) -- (3,0);
    \foreach \x in
    {0,0.5,1,1.5,2,2.5}
    {
      \node [dotdia] at (\x,0) {}; 
    }
    \foreach \x in {0,1.5}
    { 
      \draw (\x,0)--(\x,-1); 
    } 
    \draw (0.5,0) to[out=-90,in=180]
    (0.5+0.75,-0.75) to[out=0,in=-90] (0.5+1.5,0); 
    \draw (1,0) to[out=-90,in=180]
    (1+0.75,-0.75) to[out=0,in=-90] (1+1.5,0); 
    \foreach \y in
    {0}
    {
      \draw[densely dotted] (0.75,\y-0.2)--(0.75,\y+0.2); 
      \draw[densely dotted] (1.75,\y-0.2)--(1.75,\y+0.2); 
    }

  \end{scope}
  \begin{scope}[xshift=4cm,yshift=-2cm]
    \draw (-0.5,0) -- (3,0);
    \foreach \x in
    {0,0.5,1,1.5,2,2.5}
    {
      \node [dotdia] at (\x,0) {}; 
    }
    \foreach \x in {0.5,1.5}
    { 
      \draw (\x,0)--(\x,-1); 
    } 
    \draw (0,0) to[out=-90,in=180]
    (0+1,-1) to[out=0,in=-90] (0+2,0); 
    \draw (1,0) to[out=-90,in=180]
    (1+0.75,-0.75) to[out=0,in=-90] (1+1.5,0); 
    \foreach \y in
    {0}
    {
      \draw[densely dotted] (0.75,\y-0.2)--(0.75,\y+0.2); 
      \draw[densely dotted] (1.75,\y-0.2)--(1.75,\y+0.2); 
    }
  \end{scope}
  \end{scope}

  \begin{scope}[yshift=-5cm,xshift=10cm]
  \draw (-0.5,0) -- (3,0);
  \foreach \x in
  {0,0.5,1,1.5,2,2.5}
  {
    \node [dotdia] at (\x,0) {}; 
  }
  \foreach \x in {1,2.5}
  { 
    \draw (\x,0)--(\x,-1); 
  } 
  \draw (0,0) to[out=-90,in=180]
  (0+1,-1) to[out=0,in=-90] (0+2,0); 
  \draw (0.5,0) to[out=-90,in=180]
  (0.5+0.5,-0.5) to[out=0,in=-90] (0.5+1,0); 
    \foreach \y in
    {0}
    {
      \draw[densely dotted] (0.75,\y-0.2)--(0.75,\y+0.2); 
      \draw[densely dotted] (1.75,\y-0.2)--(1.75,\y+0.2); 
    }
  \begin{scope}[xshift=4cm]
    \draw (-0.5,0) -- (3,0);
    \foreach \x in
    {0,0.5,1,1.5,2,2.5}
    {
      \node [dotdia] at (\x,0) {}; 
    }
    \foreach \x in {1,2.5}
    { 
      \draw (\x,0)--(\x,-1); 
    } 
    \draw (0,0) to[out=-90,in=180]
    (0+0.75,-0.75) to[out=0,in=-90] (0+1.5,0); 
    \draw (0.5,0) to[out=-90,in=180]
    (0.5+0.75,-0.75) to[out=0,in=-90] (0.5+1.5,0); 
    \foreach \y in
    {0}
    {
      \draw[densely dotted] (0.75,\y-0.2)--(0.75,\y+0.2); 
      \draw[densely dotted] (1.75,\y-0.2)--(1.75,\y+0.2); 
    }

  \end{scope}
  \begin{scope}[yshift=-2cm]
    \draw (-0.5,0) -- (3,0);
    \foreach \x in
    {0,0.5,1,1.5,2,2.5}
    {
      \node [dotdia] at (\x,0) {}; 
    }
    \foreach \x in {1,2}
    { 
      \draw (\x,0)--(\x,-1); 
    } 
    \draw (0,0) to[out=-90,in=180]
    (0+1.25,-1) to[out=0,in=-90] (0+2.5,0); 
    \draw (0.5,0) to[out=-90,in=180]
    (0.5+0.5,-0.5) to[out=0,in=-90] (0.5+1,0); 
    \foreach \y in
    {0}
    {
      \draw[densely dotted] (0.75,\y-0.2)--(0.75,\y+0.2); 
      \draw[densely dotted] (1.75,\y-0.2)--(1.75,\y+0.2); 
    }

  \end{scope}
  \begin{scope}[xshift=4cm,yshift=-2cm]
    \draw (-0.5,0) -- (3,0);
    \foreach \x in
    {0,0.5,1,1.5,2,2.5}
    {
      \node [dotdia] at (\x,0) {}; 
    }
    \foreach \x in {1,2}
    { 
      \draw (\x,0)--(\x,-1); 
    } 
    \draw (0,0) to[out=-90,in=180]
    (0+0.75,-0.75) to[out=0,in=-90] (0+1.5,0); 
    \draw (0.5,0) to[out=-90,in=180]
    (0.5+1,-1) to[out=0,in=-90] (0.5+2,0); 
    \foreach \y in
    {0}
    {
      \draw[densely dotted] (0.75,\y-0.2)--(0.75,\y+0.2); 
      \draw[densely dotted] (1.75,\y-0.2)--(1.75,\y+0.2); 
    }

  \end{scope}
  \end{scope}

  \begin{scope}[yshift=-10cm,xshift=5cm]
  \draw (-0.5,0) -- (3,0);
  \foreach \x in
  {0,0.5,1,1.5,2,2.5}
  {
    \node [dotdia] at (\x,0) {}; 
  }
  \foreach \x in {1,1.5}
  { 
    \draw (\x,0)--(\x,-1); 
  } 
  \draw (0,0) to[out=-90,in=180]
  (0+1.25,-1) to[out=0,in=-90] (0+2.5,0); 
  \draw (0.5,0) to[out=-90,in=180]
  (0.5+0.75,-0.75) to[out=0,in=-90] (0.5+1.5,0); 
    \foreach \y in
    {0}
    {
      \draw[densely dotted] (0.75,\y-0.2)--(0.75,\y+0.2); 
      \draw[densely dotted] (1.75,\y-0.2)--(1.75,\y+0.2); 
    }

  \begin{scope}[xshift=4cm]
    \draw (-0.5,0) -- (3,0);
    \foreach \x in
    {0,0.5,1,1.5,2,2.5}
    {
      \node [dotdia] at (\x,0) {}; 
    }
    \foreach \x in {1,1.5}
    { 
      \draw (\x,0)--(\x,-1); 
    } 
    \draw (0,0) to[out=-90,in=180]
    (0+1,-1) to[out=0,in=-90] (0+2,0); 
    \draw (0.5,0) to[out=-90,in=180]
    (0.5+1,-1) to[out=0,in=-90] (0.5+2,0); 
    \foreach \y in
    {0}
    {
      \draw[densely dotted] (0.75,\y-0.2)--(0.75,\y+0.2); 
      \draw[densely dotted] (1.75,\y-0.2)--(1.75,\y+0.2); 
    }

  \end{scope}
  \end{scope}
  
\end{tikzpicture}
  \caption{The $18$ reduced states of the sector $k=1$, for a system
    with $\nRepLeft=\nRepRight=2, L=1$. Those related by the action of
    $\bleftmone \brightmone$ are grouped together.}
\label{fig:states_k1_m2_mbar2}
\end{figure}

We now move to $k=1$. In this sector we have $18$ states. They are
drawn in figure \ref{fig:states_k1_m2_mbar2}, where we have grouped
those related by the action of $\bleftmone \brightmone =
(1+P_0+P_1+P_0P_1)/4$.
Acting with $\tilde{E}_2$ on an arbitrary state will always produce
the sum of the four states contained in the box in figure
\ref{fig:states_k1_m2_mbar2} times a constant which can eventually be
zero, as when $E_2$ acts on two strings.  The role of $\tilde{E}_2$ is
indeed to project onto the state in which the boundary strands are
fully symmetrized and the bulk ones are contracted. This remark is of
crucial importance since it tells us that the action of $R$ is zero on
every state, with the same value of $\fugLR^k$ then characterizing
this sector. We can evaluate $R$ on a reference state, say the
one on the top left in figure \ref{fig:states_k1_m2_mbar2} to obtain
the value of $\fugLR^k$. $R$ on this state is zero if:
\begin{equation}
  \label{eq:n121_m2_mbar2}
  \fugLR^1 = \frac{2\fug+6}{9} \, .
\end{equation}

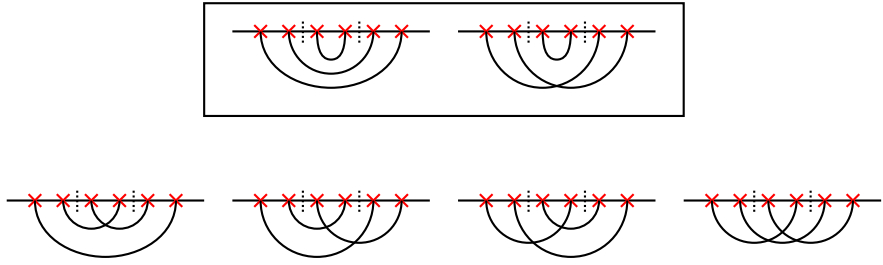
\begin{figure}[htp]
  \centering
\begin{tikzpicture}[thick,scale=0.75,dotdia/.style={cross out, draw,
    solid, red, inner sep=2pt}]
  \draw (-1,0.5) rectangle (7.5,-1.5);
  \draw (-0.5,0) -- (3,0);
  \foreach \x in
  {0,0.5,1,1.5,2,2.5}
  {
    \node [dotdia] at (\x,0) {}; 
  }
  \foreach \x in {1}
  { 
    \draw (\x,0) to[out=-90,in=180]
    (\x+0.25,-0.5) to[out=0,in=-90] (\x+0.5,0); 
  }
  \draw (0.5,0) to[out=-90,in=180]
  (0.5+0.75,-0.75) to[out=0,in=-90] (0.5+1.5,0); 
  \draw (0,0) to[out=-90,in=180]
  (0+1.25,-1) to[out=0,in=-90] (0+2.5,0); 
    \foreach \y in
    {0}
    {
      \draw[densely dotted] (0.75,\y-0.2)--(0.75,\y+0.2); 
      \draw[densely dotted] (1.75,\y-0.2)--(1.75,\y+0.2); 
    }

  \begin{scope}[xshift=4cm]
  \draw (-0.5,0) -- (3,0);
  \foreach \x in
  {0,0.5,1,1.5,2,2.5}
  {
    \node [dotdia] at (\x,0) {}; 
  }
  \foreach \x in {1}
  { 
    \draw (\x,0) to[out=-90,in=180]
    (\x+0.25,-0.5) to[out=0,in=-90] (\x+0.5,0); 
  }
  \draw (0.5,0) to[out=-90,in=180]
  (0.5+1,-1) to[out=0,in=-90] (0.5+2,0); 
  \draw (0,0) to[out=-90,in=180]
  (0+1,-1) to[out=0,in=-90] (0+2,0); 
    \foreach \y in
    {0}
    {
      \draw[densely dotted] (0.75,\y-0.2)--(0.75,\y+0.2); 
      \draw[densely dotted] (1.75,\y-0.2)--(1.75,\y+0.2); 
    }

  \end{scope}

  \begin{scope}[yshift=-3cm,xshift=-4cm]
  \draw (-0.5,0) -- (3,0);
  \foreach \x in
  {0,0.5,1,1.5,2,2.5}
  {
    \node [dotdia] at (\x,0) {}; 
  }
  \draw (0.5,0) to[out=-90,in=180]
  (0.5+0.5,-0.5) to[out=0,in=-90] (0.5+1,0); 
  \draw (1,0) to[out=-90,in=180]
  (1+0.5,-0.5) to[out=0,in=-90] (1+1,0); 
  \draw (0,0) to[out=-90,in=180]
  (0+1.25,-1) to[out=0,in=-90] (0+2.5,0); 
    \foreach \y in
    {0}
    {
      \draw[densely dotted] (0.75,\y-0.2)--(0.75,\y+0.2); 
      \draw[densely dotted] (1.75,\y-0.2)--(1.75,\y+0.2); 
    }

  \begin{scope}[xshift=4cm]
  \draw (-0.5,0) -- (3,0);
  \foreach \x in
  {0,0.5,1,1.5,2,2.5}
  {
    \node [dotdia] at (\x,0) {}; 
  }
  \draw (0.5,0) to[out=-90,in=180]
  (0.5+0.5,-0.5) to[out=0,in=-90] (0.5+1,0); 
  \draw (1,0) to[out=-90,in=180]
  (1+0.75,-0.75) to[out=0,in=-90] (1+1.5,0); 
  \draw (0,0) to[out=-90,in=180]
  (0+1,-1) to[out=0,in=-90] (0+2,0); 
    \foreach \y in
    {0}
    {
      \draw[densely dotted] (0.75,\y-0.2)--(0.75,\y+0.2); 
      \draw[densely dotted] (1.75,\y-0.2)--(1.75,\y+0.2); 
    }

  \end{scope}

  \begin{scope}[xshift=8cm]
  \draw (-0.5,0) -- (3,0);
  \foreach \x in
  {0,0.5,1,1.5,2,2.5}
  {
    \node [dotdia] at (\x,0) {}; 
  }
  \draw (0.5,0) to[out=-90,in=180]
  (0.5+1,-1) to[out=0,in=-90] (0.5+2,0); 
  \draw (1,0) to[out=-90,in=180]
  (1+0.5,-0.5) to[out=0,in=-90] (1+1,0); 
  \draw (0,0) to[out=-90,in=180]
  (0+0.75,-0.75) to[out=0,in=-90] (0+1.5,0); 
    \foreach \y in
    {0}
    {
      \draw[densely dotted] (0.75,\y-0.2)--(0.75,\y+0.2); 
      \draw[densely dotted] (1.75,\y-0.2)--(1.75,\y+0.2); 
    }

  \end{scope}

  \begin{scope}[xshift=12cm]
  \draw (-0.5,0) -- (3,0);
  \foreach \x in
  {0,0.5,1,1.5,2,2.5}
  {
    \node [dotdia] at (\x,0) {}; 
  }
  \foreach \x in
  {0,0.5,1}
  {
    \draw (\x,0) to[out=-90,in=180]
    (\x+0.75,-0.75) to[out=0,in=-90] (\x+1.5,0); 
  }
    \foreach \y in
    {0}
    {
      \draw[densely dotted] (0.75,\y-0.2)--(0.75,\y+0.2); 
      \draw[densely dotted] (1.75,\y-0.2)--(1.75,\y+0.2); 
    }

  \end{scope}
  \end{scope}
  
\end{tikzpicture}
  \caption{The $6$ reduced states of the sector $k=0$, for a system
    with $\nRepLeft=\nRepRight=2, \nRepBulk=1$. States on top and on
    bottom grouped together are related by the action of $\bleftmone
    \brightmone$.}
\label{fig:states_k0_m2_mbar2}
\end{figure}

The last case we have to consider is $k=0$, when no strings are left.
In this sector we have $6$ states depicted in figure
\ref{fig:states_k0_m2_mbar2}, where again we have grouped those
connected by the action of $\bleftmone \brightmone$.
As in the previous case, we realize that $\tilde{E}_2$ acting on every
such states produces the sum of the two states contained in the box in
figure \ref{fig:states_k0_m2_mbar2} times a constant.  Then,
evaluating $R$ on a given state gives
\begin{equation}
  \label{eq:n120_m2_mbar2}
  \fugLR^0 = \frac{\fug+2}{3} \, .
\end{equation}

Having understood this simple example, we can consider the general
case with $\nRepLeft$ and $\nRepRight$ arbitrary.  We first express
$\bleft$ and $\bright$ in terms of $\bleftmone$ and $\brightmone$:
\begin{align}
  \bleft &= \frac{1}{\nRepLeft+1} \bleftmone 
  (1 + (0,\nRepLeft) + \dots + (\nRepLeft-1,\nRepLeft)) 
  \nonumber \\
  &= \frac{1}{\nRepLeft+1} \bleftmone 
  \left(1 + P_{\nRepLeft-1} \left(\sum_{i=0}^{\nRepLeft-1}
  \prod_{j=0}^{i-1} P_{\nRepLeft-2-j}\right)\right) \\
  \bright &= \frac{1}{\nRepRight+1} 
  (1 +  (\nRepLeft+1,\nRepLeft+2) + \dots + 
  (\nRepLeft+1,\nRepLeft+\nRepRight+1))
  \brightmone \nonumber \\
  &= \frac{1}{\nRepRight+1} 
  \left(1 + \left(\sum_{i=0}^{\nRepRight-1}
      \prod_{j=0}^{i-1} P_{\nRepLeft+1+i-j} \right) 
    P_{\nRepLeft+1} \right) \brightmone \, .
\end{align}
Then we arrive at the generalization of formula \eqref{eq:ebbe_m2mbar2}:
\begin{align}
  \tilde{E}_\nRepLeft \bleft \bright \tilde{E}_\nRepLeft
  &= \frac{1}{(\nRepLeft+1)(\nRepRight+1)} 
  \Bigg{(}(\fug + \nRepLeft + \nRepRight)\tilde{E}_\nRepLeft +\\
    &\quad + \bleftmone \Bigg(\sum_{i=0}^{\nRepRight-1} \prod_{j=0}^{i-1}
      P_{\nRepLeft+1+i-j} \Bigg) E_\nRepLeft P_{\nRepLeft-1} 
      P_{\nRepLeft+1} E_\nRepLeft 
     \Bigg(\sum_{i=0}^{\nRepLeft-1} \prod_{j=0}^{i-1} P_{\nRepLeft-2-j}\Bigg)
    \brightmone \Bigg{)} = \\
  \label{eq:ebbe_m_mbar}
  &= \frac{1}{(\nRepLeft+1)(\nRepRight+1)} 
  \left((\fug + \nRepLeft + \nRepRight)\tilde{E}_\nRepLeft 
    + \nRepLeft \nRepRight\;
    \bleftmone \brightmone 
    E_\nRepLeft P_{\nRepLeft-1} P_{\nRepLeft+1} E_\nRepLeft
    \bleftmone \brightmone \right)  \, ,
\end{align}
where the last equality holds thanks to the fact that $\bleftmone$ can
be written as itself times the Young symmetrizer related to the
tableau obtained by removing the box numbered $\nRepLeft-1$ in
$\hat{t}_1$, and the analogous one for $\bright$.

Now, we can generalize our reasoning done for the case
$\nRepLeft=2,\nRepRight=2$.  If we redefine the operator
\begin{equation}
  \label{eq:R_m_mbar}
  R := \left(\tilde{E}_\nRepLeft \bleft \bright - \fugLR^k\right) 
  \tilde{E}_\nRepLeft \, ,
\end{equation}
then every state of a sector with given $k$ will be in the kernel of
$R$ with the same value of $\fugLR^k$, a consequence of the projection
performed by $\tilde{E}_\nRepLeft$.  The values of $\fugLR^k$ in the
general case $\nRepLeft,\nRepRight$ are then found to be:
\begin{equation}
  \label{eq:n12k_m_mbar}
  \fugLR^k = 
  \left\{
    \begin{array}{rl}
     \displaystyle{
      \frac{(\nRepRight-k+1)(\nRepLeft+\fug+k)}{(\nRepLeft+1)(\nRepRight+1)}
       }
     & \text{if } \nRepLeft \ge \nRepRight,\\
     \displaystyle{
       \frac{(\nRepLeft-k+1)(\nRepRight+\fug+k)}{(\nRepLeft+1)(\nRepRight+1)}
     }
     & \text{if } \nRepLeft < \nRepRight.
    \end{array} \right.
\end{equation}
Note that $\fugLR^k$ is symmetric in exchanging
$\nRepLeft,\nRepRight$, as it should in our  problem.  The
computation of the weights $\fugLR^k$ in the model described by
$\tilde{H}$, equation \eqref{eq:Hallperm}, concludes the analysis of
the sector $j=0$.

As explained in section \ref{sec:newsec}, when we fix $j\ge 1$, we
have only to consider $k=\minlr$, because if $j=1$, the
other values of $k$ are related to corresponding sectors in the $j=0$
case, and when $j>1$, no contraction of strings is allowed.  So for
$j\ge 1$, the mapping onto the $2$BLM can be directly done without any
further issue.

The $(2j+2k+\ell)$-leg exponents $h_{\nRepLeft,\nRepRight}(j+k)$ in
the model \eqref{eq:Hallperm} are finally listed in table
\ref{tab:crit_exp} where we have used the parametrization of
$\fugL,\fugR,\fugLR^k$ in terms of $r_1,r_2,r_{12}^k$ introduced in
\eqref{eq:notation2BLM}.  Recall that $\fugL$ and $\fugR$ are simple
functions of $\fug$ and of respectively $\nRepLeft$ and $\nRepRight$,
equations \eqref{eq:n1_m} and \eqref{eq:n2_mbar}, and that $\minlr$ is
the minimum between $\nRepLeft$ and $\nRepRight$.
\begin{table}[h!c]
  \centering
  \begin{tabular}{|c|c|c|c|}
    \hline
    $j$ & $k$ & $\#(\mbox{legs})$ & $h^{\nRepLeft,\nRepRight}(j+k)$ \\
    \hline 
    $0$ & $0$ & $\ell$ & $h_{r_{12}^0,r_{12}^0}$ \\
    $0$ & $1$ & $\ell+2$ & $h_{r_{12}^1,r_{12}^1}$ \\
    $0$ & $2$ & $\ell+4$ & $h_{r_{12}^2,r_{12}^2}$ \\
    $\vdots$ & $\vdots$ & $\vdots$ & $\vdots$ \\
    $0$ & $\minlr$ & $\nRepLeft+\nRepRight$ & $h_{r_{12}^\minlr,r_{12}^\minlr}$ \\
    $1$ & $\minlr$ & $\nRepLeft+\nRepRight+2$ & $h_{r_1+r_2-1,r_1+r_2-1+2}$ \\
    $2$ & $\minlr$ & $\nRepLeft+\nRepRight+4$ & $h_{r_1+r_2-1,r_1+r_2-1+4}$ \\
    $3$ & $\minlr$ & $\nRepLeft+\nRepRight+6$ & $h_{r_1+r_2-1,r_1+r_2-1+6}$ \\
    $\vdots$ & $\vdots$ & $\vdots$ & $\vdots$ \\
    \hline 
  \end{tabular}
  \caption{Critical exponents for the two-boundary problem.}
  \label{tab:crit_exp}
\end{table}
When the number of legs is between $\ell$ and $\nRepLeft+\nRepRight$, we are
in the $j=0$ sector of a $2$BLM, and for $k=0,\dots,\minlr$, the
exponents $h_{r_{12}^k,r_{12}^k}$ are computed from our expression of
$\fugLR^k$, equation \eqref{eq:n12k_m_mbar}.  When we have instead
$\nRepLeft+\nRepRight+2j$ legs, with $j>0$, we are in the sector $j$
of a $2$BLM which is completely described by the two parameters
$\fugL$ and $\fugR$, equations \eqref{eq:n1_m} and \eqref{eq:n2_mbar},
and the exponents then follow from fusion of the boundary conformal
weights of the one-boundary case, the leading exponents in a sector
being $h_{r_1+r_2-1,r_1+r_2-1+2j}$.

Invoking the usual universality hypothesis---namely that the critical
behaviour is unchanged upon adding permutations other than the
generators to the Hamiltonian $H$---the above expressions completely
determine the critical exponents in our original model
\eqref{eq:hamiltonian_loop}.

Lastly, we have verified that as in the one-boundary case, the
version of the model with higher spins on the boundary, $\tilde{\tilde{H}} =
\bleftmone \brightmone H \bleftmone \brightmone$, is in the same 
universality class as well.

\section{Back to the spin chains spectrum}
\label{sec:back_spin}

In section \ref{sec:alg_quot} we have suggested that the
representation of the quotient algebra $\Quot(\diffBosFer)$ is
faithful if the number of states per site in the spin chain is greater
or equal to three. Therefore, every irreducible representation of the
diagram algebra studied previously should be present in the $\slnm$
spin chains, except when $\nFer=1$, $\diffBosFer=0$.  Whenever the
representation is faithful, the eigenvalues of the loop and of the
super spin representations are the same, the difference in the Hilbert
spaces manifesting itself only in the degeneracies.  We can thus use
the loop model---which is technically more convenient---to solve the
full spin chain problem. This requires understanding all exponents,
and calculating their degeneracies. While such a program can easily be
carried out for the pure Temperley-Lieb problem \cite{Read2001}, in
the present case, a lot is still lacking.

One first incomplete aspect of our study is that we have not
determined all the geometrical exponents: there are definitely plenty
of subleading eigenvalues in a given sector. When a direct mapping
onto a Boundary Loop Model is available, we can reinterpret the
$(j,k)$ sectors of the edge state model in terms of the
blobbed/unblobbed sectors of the BLM and we hence know exactly which
conformal weights one finds. Such an exact mapping is possible in the
one-boundary system with $\nRepLeft=1$ and $\nRepLeft=0$, for which
the conformal weights of primary operators in the sector $j=0$ are
$h_{r_1,r_1}$, and for $j>0$ are $h_{r_1,r_1+2j}$ (blobbed sector) and
$h_{r_1,r_1-2j}$ (unblobbed sector). It is also possible in the
two-boundary case with $\nRepLeft=\nRepRight=1$ if we fix $k=1$, where
for $j=0$ we find primaries with weight $h_{r_{12}^1-2n,r_{12}^1}$ and
$n\in\mathbb{Z}$, while for $j>0$ we find $h_{\epsilon_1 r_1 +
  \epsilon_2 r_2 - 1 - 2n, \epsilon_1 r_1 + \epsilon_2 r_2 - 1 + 2j}$
with $\epsilon_{1,2} = \pm 1$ and $n \in \mathbb{N}$. But already in
the sector $j=k=0$ when $\nRepLeft=\nRepRight=1$, our algebra gives an
unfaithful representation of the two-boundary Temperley-Lieb algebra,
since the mapping to a $2$BLM involves a quotient version of the
latter, and we do not find all the subleading eigenvalues
$h_{r_{12}^0-2n,r_{12}^0},n\in\mathbb{Z}$. In general we do not have
full control of which subleading eigenvalues appear. However the value
of possible eigenvalues we expect should be of the form given by the
$2$BLM conformal weights with some possible omissions and
blobbed/unblobbed sectors mixing, as we have indeed observed in the
systems studied numerically. Clearly a deeper algebraic understanding
of the problem is needed here to characterize the full spectrum.

The other incomplete aspect is that we lack a clear understanding of
the degeneracies. One way to understand these in general is to
think of the generating function of eigenvalues (for the super spin
chain), which could in principle be computed as the modified partition
function of a loop model where non-contractible loops are weighted
differently \cite{Read2001}. In algebraic terms, this generating
function would be written as a sum over irreducible representations of
the algebra to which the Hamiltonian belongs. The multiplicities of
summands would then be the dimensions of irreducibles of the commutant
algebra in the Hilbert space at hand; the problem is that very little is
known in general about this commutant.

Another way to attack the problem of degeneracies is to study the
combinatorial aspect of the loop representation of the model, along
the lines of \cite{LykkeJacobsen2008}. One would then need to develop
a theory of combined projectors of the symmetric group (on the
boundary) and of the Jones-Wenzl type (in the bulk) in which the
degeneracies would be identified with the Markov traces of these
projectors.  One way or the other, it is clear that further study is
needed to finish the program.

For some physical applications \cite{GruzbergObuse}, only the leading
exponents will matter, and the exact knowledge of degeneracies and
subleading corrections to scaling does not play a paramount role.
However, if these two aspects were completely elucidated, one could
construct exact continuum limit partition functions and compute
various kinds of exact crossing probabilities, presumably with
applications to transport properties in the spin quantum Hall effect.
We leave these developments for future work.

Let us finally recall that the $\ssl{1}{1}$ super spin chain---alias
symplectic fermions; see appendix \ref{sec:glone_case}---seems to be
the only case where the spin chain representation is not faithful.  In
particular this manifests itself in the fact that for
$\nRepLeft=\nRepRight$, the spectrum in this chain is equal to that
obtained for free ($\nRepLeft=\nRepRight=0$) boundary conditions, when
the Hamiltonian belongs to the Temperley-Lieb algebra and the critical
exponents are those of the conformal field theory of symplectic
fermions.  Instead, as explicitly showed in table
\ref{tab:exp_faith-unfaith}, in the case $\nRepLeft=\nRepRight\neq 0$,
the $\ssl{\nFer}{\nFer}$ spin chain when $\nFer>1$ possesses an
infinity of new exponents which are not contained in symplectic
fermion theory. Note that these exponents in general are not
rational---a rather unusual feature in conformal field theory.

\begin{table}[h!c]
  \centering
  \begin{tabular}{|p{6cm}|c|c|}
    \hline
   \centering Critical Exponents $\fug=0$
    &  $\ssl{\nFer}{\nFer}_{\nFer>1}$ &   $\ssl{1}{1}$\\
    \hline
    \begin{tikzpicture}[scale=0.5]
      \useasboundingbox (-2,-0.5) rectangle (3,1);
      \fill (0,0) circle (0.1);
      \node at (3.35,-0.5) {$ = h_{1,1} = 0$};
    \end{tikzpicture}
    & \ding{51} & \ding{51} \\
    \begin{tikzpicture}[scale=0.5]
      \useasboundingbox (-2,-0.5) rectangle (3,1);
      \foreach \x in {-1,1} { \draw (0,0)--(\x,-1); }  
      \fill (0,0) circle (0.1);
      \node at (3.35,-0.5) {$ = h_{1,1} = 0$};
    \end{tikzpicture}
    & \ding{51} & \ding{51} \\
    \begin{tikzpicture}[scale=0.5]
      \useasboundingbox (-2,-0.5) rectangle (3,1);
      \foreach \x in {-1,-0.34,0.34,1} { \draw (0,0)--(\x,-1); }  
      \fill (0,0) circle (0.1);
      \node at (5.5,-0.5) {$ \approx h_{0.410,0.410 + 2} \approx 0.191$};
    \end{tikzpicture}
    & \ding{51} & {\color{red}\ding{55}} \\
    \begin{tikzpicture}[scale=0.5]
      \useasboundingbox (-2,-0.5) rectangle (3,1);
      \foreach \x in {-1,-0.6,-0.2,0.2,0.6,1} { \draw (0,0)--(\x,-1); }  
      \fill (0,0) circle (0.1);
      \node at (5.5,-0.5) {$ \approx h_{0.410,0.410 + 4} \approx 1.486$};
    \end{tikzpicture}
    & \ding{51} & {\color{red}\ding{55}} \\
    \begin{tikzpicture}[scale=0.5]
      \useasboundingbox (-2,-0.5) rectangle (3,1);
      \foreach \x in {-1,-0.72,-0.44,-0.16,0.16,0.44,0.72,1} 
      { \draw (0,0)--(\x,-1); }  
      \fill (0,0) circle (0.1);
      \node at (5.5,-0.5) {$ \approx h_{0.410,0.410 + 6} \approx 3.781$};
    \end{tikzpicture}
    & \ding{51} & {\color{red}\ding{55}} \\
    \centering $\vdots$ & $\vdots$ & $\vdots$ \\    
    \hline
  \end{tabular}
  \caption{In the first column, watermelon exponents computed in the
    geometrical model when $\nRepLeft=\nRepRight=1$ and $\fug=0$.
    The number of legs is the coordination of the vertex.
    In the second and third column we note the presence or not of these 
    exponents in the faithful ($\nFer>1$) and unfaithful ($\nFer=1$) 
    representations of the algebra for the $\ssl{\nFer}{\nFer}$ spin chains.}
  \label{tab:exp_faith-unfaith}
\end{table}

\section{Conclusion}

In conclusion, we have seen that edge states in super spin chains
describe an infinity of conformal invariant boundary conditions for
conformal superprojective sigma models. The technology involved to
characterize these boundary conditions is definitely involved, and the
exponents very far from trivial. How these exponents can be organized
in a complete description of the boundary and bulk logarithmic CFT
describing the low energy physics of the sigma models remains an open
question.

Meanwhile, it is natural to expect that the boundary conditions we
have uncovered, together with their edge states description in the
quantum spin chains, will play an important role in the description of
transitions between plateaus in the spin quantum Hall effect, as well
as in the ordinary quantum Hall effect (though in this case a
different sigma model would have to be considered)
\cite{Xiong1997}. We hope to get back to this question soon.

\bigskip

\noindent {\bf Acknowledgments}: We thank J. Dubail for very useful
comments on the potential relations with boundary Temperley-Lieb
algebras.  H.S. thanks C. Candu, N. Read, and V. Schomerus for many
discussions and earlier collaborations on related matters.  This work
was supported in part by the Agence Nationale de la Recherche (grants
ANR-06-BLAN-0124 and ANR-10-BLAN-0414).


\appendix

%
%

\section{The $\gl{1}{1}$ spin chain}
\label{sec:glone_case}

Here we will work out in details the solutions of the model in the
simplest case of the $\gl{1}{1}$ super spin chain, where the
Hamiltonian reduces to free fermions and explicit computations can be
done.

\subsection{The lie superalgebra $\gl{1}{1}$}
\label{sec:glone_liesuper}
First we will briefly recall some facts about the Lie superalgebra
$\gl{1}{1}$ and its representation theory \cite{GOTZ2007}.

The $\gl{1}{1}$ Lie superalgebra is defined by two bosonic generators
$E,N$ and two fermionic ones $F^-, F^+$, and its superdimension is
zero. $E$ is central, and the other defining relations are
\begin{equation}
  \label{eq:glone_rel}
  [N,F^\pm] = \pm F^\pm, \quad \{F^-,F^+ \}=E \, .
\end{equation}
In particular $N$ counts the number of fermions.

When dealing with the irreducible representations of superalgebras, we
have to distinguish between typical or atypical representations. For
$\gl{1}{1}$ typical representations are the two-dimensional
representations $\langle e,n \rangle$, $e \neq 0$, $n\in \mathbb{R}$,
which read in matrix notation:
\begin{equation}
  \label{eq:gen_glone_fund}
  N = \left( \begin{array}{cc}
      n-1  & 0 \\
      0  & n  \end{array} \right)\, \quad 
  F^- = \left( \begin{array}{cc}
      0  & 1 \\
      0  & 0  \end{array} \right)\, \quad 
  F^+ = \left( \begin{array}{cc}
      0  & 0 \\
      e  & 0  \end{array} \right)\, \quad 
  E = e \id_2 \, .
\end{equation}
The state annihilated by $F^-$ is a boson.  Atypical representations
are the one-dimensional irreducible representations $\langle n
\rangle$, which have $F^+ = F^- = E = 0, N=n$.

Other finite-dimensional representations that occur are
indecomposable, not fully reducible. They are composites of atypical
constituents linked by the action of $F^\pm$.  We denote by
$\mathcal{P}_{n}$ the indecomposable four-dimensional representation
playing the role of the projective cover of $\langle n \rangle$, that
is it can never appear as a subrepresentation of a larger
indecomposable and contains $\langle n \rangle$.

The spin chain \eqref{eq:space_states} is built with the fundamental
$V = \langle 1,1/2 \rangle$ and the dual $V^\star = \langle -1,1/2
\rangle$.  The (graded) tensor product of representations is given by
the following formulas:
\begin{align} 
  \langle e_1,n_1 \rangle \otimes \langle e_2,n_2 \rangle
  &= \left\{ 
    \begin{array}{ll} 
      \mathcal{P}_{n_1+n_2-1} & \text{if } e_1+e_2 = 0,\\ 
      \langle e_1+e_2,n_1+n_2 \rangle \oplus
      \langle e_1+e_2,n_1+n_2-1 \rangle & \text{otherwise} .
    \end{array} \right. \nonumber
  \\
  \label{eq:typproj}
  \langle e,n \rangle \otimes \mathcal{P}_{m} 
  &= 
  \langle e,n+m+1
  \rangle \oplus 2 \cdot \langle e,n+m \rangle \oplus \langle e,n+m-1
  \rangle \, .
\end{align}
Then we have for $l_1 \neq l_2$:
\begin{equation}
  \label{eq:Vr1_V*r2}
  V^{\otimes l_1} \otimes (V^\star)^{\otimes l_2} 
  \simeq 
  \bigoplus_{p=0}^{1} 
  \bigoplus_{i=0}^{l_1-1} 
  \bigoplus_{j=0}^{l_2-1} 
  \binom{l_1-1}{i}
  \binom{l_2-1}{j}
  \left\langle l_1+l_2, i+j+2-\frac{l_1+l_2}{2} +p\right\rangle \, .
\end{equation}
This is the decomposition of the Hilbert space \eqref{eq:space_states}
when the number of extra representations on the two boundaries of the
spin chain is different. It involves $\gl{1}{1}$ $2$-dimensional
representations, since changing the order of representations in the
spin chain leads to isomorphic spaces.  When instead $l_1=l_2=l$
\begin{equation}
  \label{eq:VV*L}
  (V \otimes V^\star)^{\otimes l} 
  \simeq 
  \bigoplus_{i=1-l}^{l-1} 
  \binom{2l-2}{l-1+i}
  \mathcal{P}_{i} \, ,
\end{equation}
this corresponding to the decomposition of the Hilbert space in terms
of $4$-dimensional representations when we have the same number of
representations at the two boundaries of the chain.

\subsection{Free fermions formulation}
\label{sec:free_fermions}

Denote by $f_i, f^\dagger_i$ the fermionic operators of $\gl{1}{1}$
acting on $V$ at site $i$, and by $\bar{f}_i, \bar{f}^\dagger_i$ those
acting on $V^\star$. They satisfy the anti-commutation relations
$\{f_i, f_j^\dagger\} = \delta_{ij}$, $\{\bar{f}_i,
\bar{f}_j^\dagger\} = -\delta_{ij}$, where the presence of the
additional minus sign implies that the Hilbert space has indefinite
inner product.  To deal with ordinary fermions $c_i,c^\dagger_i$ that
satisfy $\{c_i,c_j^\dagger\} = \delta_{ij}$, we define $c_i = f_i$ and
$c^\dagger_i = f_i^\dagger$ if we have $V$ on site $i$; if we have
$V^\star$ on site $i$ define instead $c_i =
\bar{f}_i$ and $c^\dagger_i = -\bar{f}_i^\dagger$.

$P_i$ and $E_i$ defined in section \ref{sec:alg_quot} can be written
in terms of these fermions as:
\begin{align}
  \label{eq:brauer_fermions}
  P_{j,j+1} &= (1 - (c_j^\dagger c_j - c_{j+1}^\dagger c_j - c_j^\dagger c_{j+1} +
  c_{j+1}^\dagger c_{j+1})),  \quad j = 0,\dots,\nRepLeft-1 \\
  E_j &= (-1)^{j+m+1} (c_j^\dagger c_j - c_{j+1}^\dagger c_j + c_j^\dagger c_{j+1} -
  c_{j+1}^\dagger c_{j+1}), \quad j = \nRepLeft,\dots,2\nRepBulk+\nRepLeft-2\\
  P_{j,j+1} &= -(1 - (c_j^\dagger c_j - c_{j+1}^\dagger c_j - c_j^\dagger c_{j+1} +
  c_{j+1}^\dagger c_{j+1})), \quad j = 2\nRepBulk+\nRepLeft-1,\dots, 
  2\nRepBulk+\nRepLeft+\nRepRight-2 \, .
\end{align}
One can check by hand that the relations \eqref{eq:def_A1}--\eqref{eq:def_A4} are
satisfied. The Hamiltonian \eqref{eq:hamiltonian} in this
representation becomes quadratic in the fermionic operators:
\begin{equation}
  \label{eq:hamil_ferm}
\begin{split}
  H &= u \sum_{j=0}^{\nRepLeft-1}(c_j^\dagger c_j - c_{j+1}^\dagger c_j - 
  c_j^\dagger c_{j+1} + c_{j+1}^\dagger c_{j+1}) +\\ 
  &\quad +
  \sum_{j=\nRepLeft}^{2\nRepBulk+\nRepLeft-2} (-1)^{j+\nRepLeft}
  (c_j^\dagger c_j - c_{j+1}^\dagger c_j + c_j^\dagger c_{j+1} -
  c_{j+1}^\dagger c_{j+1}) + \\
  &\quad - v  \sum_{j=2\nRepBulk+\nRepLeft-1}^{2\nRepBulk+\nRepLeft+\nRepRight-2}
  (c_j^\dagger c_j - c_{j+1}^\dagger c_j - 
  c_j^\dagger c_{j+1} + c_{j+1}^\dagger c_{j+1})
  -\nRepLeft u + \nRepRight v \, .
\end{split}
\end{equation}

\subsubsection{Diagonalization procedure}
\label{sec:diag_proc}
Rewrite \eqref{eq:hamil_ferm} as
\begin{equation}
  \label{eq:ham_A}
  H = \sum_{i,j}c_i^\dagger A_{ij}c_j  -\nRepLeft u + \nRepRight v\, ,
\end{equation}
where $A$ is an appropriate $(2\nRepBulk+\nRepLeft+\nRepRight) \times
(2\nRepBulk+\nRepLeft+\nRepRight)$ real non-symmetric tridiagonal
matrix.

The problem of diagonalizing a Hamiltonian of the form
\eqref{eq:ham_A} is a well-known problem, solved by Lieb, Mattis and
Schultz in \cite{Lieb1961407}.  We introduce by a canonical
transformation the fermions $\eta_k,\eta^\dagger_k$, in terms of which
$H$ is diagonal:
\begin{equation}
  \label{eq:Heta}
  H = \sum_k \Lambda_k \left(\eta_k^\dagger \eta_k - \frac{1}{2}\right)\, .
\end{equation}

However the equations involved in the diagonalization procedure even
in the simplest case $\nRepLeft=1,\nRepRight=0$ are rather
complicated, and instead we will proceed diagonalizing the matrix
numerically. This task we can accomplish more efficiently by using a
closed formula for the characteristic polynomial of $A$ for generic
values of $\nRepLeft$ and $\nRepRight$.  Indeed we have found that
$P_\nRepBulk^{\nRepLeft,\nRepRight} (u,v,x) = \Det(A - x
\id_{2\nRepBulk+\nRepLeft+\nRepRight})$ has the following expression:
\begin{equation}
  \label{eq:char_poly} 
  \begin{split}
    &P_\nRepBulk^{\nRepLeft,\nRepRight} (u,v,x) = x
    (-1)^{\nRepBulk+\nRepRight} u^{\nRepLeft-1} v^{\nRepRight-1}
    \bigg\{ U_{\nRepBulk+1}\left(1-\frac{x^2}{2}\right) \cdot \\
    &\quad \cdot \bigg( U_{\nRepLeft+1}\left(1-\frac{x}{2 u}\right)
    \bigg( p_1(u,v,x) U_{\nRepRight}\left(1+\frac{x}{2 v}\right)
    + p_2(u,v,x) U_{\nRepRight+1}\left(1+\frac{x}{2 v}\right) \bigg) + \\
    &\quad +U_\nRepLeft\left(1-\frac{x}{2 u} \right) \bigg(
    p_3(u,v,x) U_{\nRepRight}\left(1+\frac{x}{2 v}\right)+
    p_1(-v,u,-x) U_{\nRepRight+1} 
    \left(1+\frac{x}{2 v}\right) \bigg) \bigg) +\\
    &+ U_\nRepBulk\left(1-\frac{x^2}{2}\right) \cdot\\
    &\quad \cdot \bigg( U_{\nRepLeft+1}\left(1-\frac{x}{2 u}\right)
    \bigg( q_1(u,v,x) U_{\nRepRight}\left(1+\frac{x}{2 v}\right)
    + q_2(u,v,x) U_{\nRepRight+1}\left(1+\frac{x}{2 v}\right) \bigg)+\\
    &\quad +U_\nRepLeft\left(1-\frac{x}{2 u} \right) \bigg(
    q_3(u,v,x) U_{\nRepRight}\left(1+\frac{x}{2 v}\right) +
    q_1(-v,u,-x)U_{\nRepRight+1} \left(1+\frac{x}{2 v}\right)
    \bigg) \bigg) \bigg\}
 \end{split}
\end{equation}
where $U_i(z)$ are the Chebyshev polynomials of the second kind,
\begin{equation}
  \label{eq:che_2nd}
  U_i(z)=\frac{\sin \left((i+1) \cos ^{-1}(z)\right)}{\sqrt{1-z^2}}\, ,
\end{equation}
and the coefficients $p$'s and $q$'s are simple polynomials 
depending only on $u,v$ and $x$:
\begin{align}
  \label{eq:alpha_beta}
  p_1(u,v,x) &= u \left(x^4+2 x^3 v-x^2 (v+2)-4 x v+v\right) \\
  p_2(u,v,x) &= -x \left(x^2-2\right) u v \\
  p_3(u,v,x) &= x \left(x^4+2 x^3 (v-u)-x^2 (4 u v+u+v+2)+4 x (u-v)+9 u v+u+v\right)\\
  q_1(u,v,x) &= -u \left(v-x \left(x^2-3\right) \left(x^3+2 x^2 v-x (v+1)-2 v\right)\right) \\
  q_2(u,v,x) &= -x \left(x^4-4 x^2+3\right) u v \\
  q_3(u,v,x) &= -x \big(v \left(x \left(x \left(x \left(4 u x-2
          x^2+x+8\right)-17 u-3\right)-6\right)
    +14 u\right) \nonumber \\
  &\qquad \qquad\qquad \qquad +x \left(x^2-3\right) \left(x \left(-2 u
      x+u+x^2-1\right)+2 u\right)+u+v\big)
\end{align}

Formula \eqref{eq:char_poly} seems to not help in the analytic
determination of the eigenvalues apart from the trivial case
$\nRepLeft=0,\nRepRight=0$, but it is remarkable that the dependence
on $\nRepBulk$, $\nRepLeft$, $\nRepRight$ is organized in terms of
Chebyshev polynomials, and further it allows efficient numerical
diagonalization of the Hamiltonian.  The presence of Chebyshev
polynomials does not come as a surprise since it is an old result that
these polynomials are related to the discrete Laplacian
\cite{Nash1986}, and our formula is in this sense a generalization of
that result.

We remark that the matrix $A$ introduced above is actually not
diagonalizable when $\nRepLeft=\nRepRight$, and it has Jordan cells of
rank 2. This is linked to the indecomposable nature of the conformal
field theory describing the $\gl{1}{1}$ chain \cite{Read2007,Creutzig2007}.

\subsection{Spectrum of the Hamiltonian and continuum limit}
\label{sec:spectrum_H_gl11}

We would like now to determine the spectrum of the Hamiltonian
\eqref{eq:Heta}. We define the vacuum $\ket{0}$ as the state with no
particles: $\eta_k \ket{0} = 0$ for every $k$, whose energy is
$-\smfrac{1}{2}\sum \Lambda_k$. Define $I^-$ and $I^+$ as the set of
values of $k$'s such that the real part of the corresponding
eigenvalues are respectively negative and positive.  The ground state
$\ket{\Omega}$ is obtained by filling the single particle states with
negative energy up to the Fermi surface, $\ket{\Omega} = \prod_{k \in
  I^{-}} \eta^\dagger_k \ket{0}$, and its energy is
\begin{equation}
  \label{eq:Egs}
  E_0 = \frac{1}{2} \left(\sum_{k\in I^-} \Lambda_k - \sum_{k\in I^+}\Lambda_k
  \right) \, .
\end{equation}

We can calculate the conformal weights $h_i$ from finite-size effects
using formula \eqref{eq:c_finite_size}.
Ultimately with the information gained from finite-size scaling, we
can compute for given boundary conditions $\nRepLeft,\nRepRight$:
\begin{equation}
  \label{eq:Zmmbar_generic}
  Z_{\nRepLeft,\nRepRight} = \lim_{\nRepBulk\to\infty} 
  \Tr e^{-\beta (H - E_0(\nRepBulk) \nRepBulk)} = \Tr q^{L_0 - c/24}
  = \sum_i q^{h_i - c/24} \, ,
\end{equation}
where $\Tr$ stands for an ordinary trace on the vector space, not a
supertrace, and $q$ is as usual the modular parameter. This corresponds
to a modified partition function, with antiperiodic boundary
conditions in the time direction \cite{Read2001}.

\subsubsection{Free boundary conditions}
\label{sec:free_bound}
We start by analyzing the simplest and already known case of the
Temperley-Lieb chain with open
boundary conditions ($m=n=0$). In this case indeed formula \eqref{eq:char_poly}
reduces to
\begin{equation}
  \label{eq:char_poly_m0mbar0}
  P_\nRepBulk^{0,0}(x)
  = (-1)^{\nRepBulk+1} x^2 U_{\nRepBulk-1}\left(1-\frac{x^2}{2}\right)\, ,
\end{equation}
and its roots are given by $x=0$ with multiplicity two, and the roots of 
$U_{\nRepBulk-1}\left(1-\frac{x^2}{2}\right)$:
\begin{equation}
  \label{eq:rootsm0mbar0}
  x = 2 \sin \left(\frac{\pi  k}{2 \nRepBulk}\right) \, ,
  \quad k=\pm 1,\dots,\pm (\nRepBulk-1)\, .
\end{equation}
The eigenvectors corresponding to these eigenvalues can also be easily
computed.

With these values of $\Lambda_k$ determined, we can use the
Euler-Maclaurin formula and equation \eqref{eq:c_finite_size} to
verify that the central charge of the model is $c=-2$ (using that
$v_s = 2$ for $p=1$):
\begin{equation}
  \label{eq:scaling_Egs_free}
  \begin{split}
    \frac{E_0}{2\nRepBulk} &= - \frac{1}{2\nRepBulk}
    \sum_{k=1}^{\nRepBulk-1}2 \sin\left(\pi \frac{k}{2\nRepBulk}
    \right) = \frac{1-\cot\left(\smfrac{\pi}{4\nRepBulk}\right)}{2}\\
    &\simeq -\frac{2}{\pi} -\frac{1}{2\nRepBulk} + \frac{\pi}{6
      (2\nRepBulk)^2} \, .
  \end{split}
\end{equation}

To compute the partition function in the free boundary case, we note
first that $h_0 = 0$. Then since $\nRepBulk/\pi \sum_{i=1}^p
\Lambda_{k_i} \sim \sum_{i=1}^p k_i$ for large $\nRepBulk$, $p \ll
\nRepBulk$, the exponents are all integers and their computation is
rather simple. Indeed it boils down to compute all the ways we can
build the integer $\sum_{i=1}^p k_i$ using at most two equal $k_i$'s
(one associated to the action of $c^\dagger_{k_i}$, the other of
$c_{-k_i}$).
If we write
\begin{equation}
  \label{eq:Zfree1}
  Z_{0,0} = q^{-c/24} \sum_{i = 0}^{\infty} a_i q^i \, ,
\end{equation}
and use the notation $\lambda=(\lambda_1^{n_1},\dots,\lambda_k^{n_k})
\vdash i$ for the partition of size $i$, $a_i$ is given by
\begin{equation}
  \label{eq:ahfree}
  a_i = 4 \sum_{\substack{
      \lambda \vdash i 
      \\ n_j =1,2 }}
  \,  \prod_{j=1}^k 2^{\delta_{n_j,1}} \, .
\end{equation}
The factor $4$ comes from having two zero $\Lambda_k$'s and parts
occurring only once have a weight two because they can come from the
action of $c_{k_i}^\dagger$ or $c_{-k_i}$. As customary when dealing
with the generating function of partitions, we are not able to give an
expression in closed form for $a_i$ but we can do this for the
generating function:
\begin{equation}
  \label{eq:Zfree2}
  Z_{0,0} = \prod_{h=0}^{\infty} \left(1 + 2q^h + q^{2h}\right) = 
  2 \frac{\theta_2(\tau)}{\eta(\tau)} = \det(-\mathcal{D}_{A,N})\, ,
\end{equation}
where $\mathcal{D}_{A,N}$ is the Laplacian on the strip with Neumann
boundary conditions in the space direction and antiperiodic boundary
conditions in the time direction.

Finally we note that this result is in agreement with the expression
of the partition function as sums of Virasoro
characters (recall however that the modules involved are not
  simple but indecomposable for this theory \cite{Read2007}):
\begin{align}
  \label{eq:Zfree3a}
  Z_{0,0} &= \sum_{j=0}^{\infty} (2j+1)
  \frac{q^{h_{1,1+2j}} - q^{h_{1,-1-2j}}}{\eta(\tau)} \\
  \label{eq:Zfree3b}
  &= 2 \frac{q^{h_{1,1}}}{\eta(\tau)} + 
  2 \sum_{j=1}^{\infty} \frac{q^{h_{1,1+2j}} + q^{h_{1,1-2j}}}
  {\eta(\tau)} \, .
\end{align}

\subsubsection{Generic boundary conditions}
\label{sec:gen_bound}
We now turn to the discussion of generic values of
$\nRepLeft,\nRepRight$. In this case we do not have an analytic
expression for the eigenvalues of the Hamiltonian and so we proceed
numerically using the results presented in \ref{sec:diag_proc}, and we
extract the finite-size scaling by fitting energies for different
sizes of the system. We have studied chains up to size $2\nRepBulk =
24$.

Define:
\begin{equation}
  \label{eq:rgl11}
  r = 
  1
  -
  \frac{2}{\pi} 
  \arctan
  \left(
    \frac{|\nRepLeft - \nRepRight|}
    {1 + \nRepLeft + \nRepRight + 2 \nRepLeft \nRepRight}
  \right) \, ,
\end{equation}
where the absolute value comes from the symmetry upon exchanging $m$
and $\nRepRight$.  We find that the partition function of the
$\gl{1}{1}$ chain is given by replacing the exponents $h_{1,1+k}$ by
$h_{r,r+k}$ in \eqref{eq:Zfree3b}:
\begin{equation}
  \label{eq:Zmmbar_gl11}
  Z_{\nRepLeft,\nRepRight} = 2 \frac{q^{h_{r,r}}}{\eta(\tau)} + 
  2 \sum_{j=1}^{\infty} \frac{q^{h_{r,r+2j}} + q^{h_{r,r-2j}}}
  {\eta(\tau)} \, .  
\end{equation}

Now we will discuss the relation between this spin chain and the
geometrical model at $\fug=0$. We have pointed out in section
\ref{sec:loop_repr} and \ref{sec:back_spin} that the $\gl{1}{1}$
representation is not faithful, so not all the eigenvalues of the
Hamiltonian in the geometrical formulation are present here. We now
wish to make this observation more quantitative.

In the $\gl{1}{1}$ spin chain the conformal weights appearing 
in (\ref{eq:Zmmbar_gl11}) are
$h_{r,r\pm 2j}, j=0,1,\dots$. If we compute explicitly $\fugLR^k$ from
our expression of exponents in the geometrical model (table
\ref{tab:crit_exp}) when $\fug=0$, we find that the function $r$
defined in \eqref{eq:rgl11} coincides with $r_{12}^0$, so we interpret
$h_{r,r}$ as the $\ell$-leg exponent. Further $r_{12}^1 = 2-r$ and
then $h_{r,r+2}$ is the $(\ell+2)$-leg exponent.  When we look at
other exponents the situation is different for the one-boundary and
the two-boundary cases.

For the one-boundary case $\nRepRight=0$, in the spin chain we always
find the eigenvalues of the Hamiltonian corresponding to the fully
symmetric (and also to the alternating) representation of the
symmetric group acting onto the strings, where the lowest eigenvalues
lie. So, $h_{r,r+2j}$ are the $(\nRepLeft+2j)$-legs critical exponents
we found in the loop representation.

When we look at the two-boundary problem, the lowest eigenvalues in
the geometrical model for $(j=0,k=0)$ and $(j=0,k=1)$ 
are present in the spin chain, as already noted. However, for the other
sectors this is no more the case, and the conformal weights $h_{r,r\pm
  2j},j>1$, appear as subleading eigenvalues in the sectors of the
geometrical model.

We remark that for $\nRepLeft = \nRepRight$ the spectrum in the
scaling limit of the $\gl{1}{1}$ spin chain is the same as that in the
free boundary case, and the multiplicity of the exponents $h_{r,r\pm
  2j}$ is $4$ while for $\nRepLeft \neq \nRepRight$ is $2$. This can
be interpreted from the analysis of the tensor product of
representations of $\gl{1}{1}$ given in
\eqref{sec:glone_liesuper}. These numbers correspond to the
multiplicities of modules of the commutant of $\gl{1}{1}$, the Walled
Brauer algebra, and are present also in our case.

As a last remark, we interpret this result in terms of the continuum
calculation presented in section \ref{sec:cpzero} for the \CPzero\
sigma model, which the spin chain introduced in this section should
discretize.  Indeed we note that when $\mu=0$ formula
(\ref{eq:charcpzero}) can be rewritten as equation \eqref{eq:Zmmbar_gl11},
where $r=1-2\lambda$.
%

\section{Young symmetrizers as blob operators}
\label{sec:YS_blob}

Irreducible representations of the symmetric group $\perm_{\nRepLeft+1}$,
the Specht modules, are obtained by a corresponding Young symmetrizer,
indexed by a standard Young tableau $t$ of size $\nRepLeft+1$, where
in each box of the diagram there are numbers from the set
$\{0,\dots,\nRepLeft\}$.  Young symmetrizers $b_t$ have the property
of being idempotent if properly normalized, $b_t^2 = b_t$.  More than
that, they satisfy also the other relation of the blob algebra:
\begin{equation}
  \label{eq:rel_b_t}
  E_\nRepLeft b_t E_\nRepLeft = \fugL^t E_\nRepLeft b_{\hat{t}} \, ,
\end{equation}
if we replace $E_\nRepLeft$ with $\tilde{E}_\nRepLeft = E_\nRepLeft
b_{\hat{t}}$. $\hat{t}$ is a standard Young tableau of size
$\nRepLeft$, obtained from $t$ by removing the box labelled
$\nRepLeft$.  So, one way to realize the blob algebra is taking a
Young symmetrizer acting on the $\nRepLeft$ boundary lines and the
leftmost bulk strand.  The weight of blobbed loops in this
construction will be indexed by the Young tableau and is given by:
\begin{equation}
  \label{eq:n1_t}
  \fugL^t = \frac{\prod_{x \in Y(\hat{t})}{\hook(x)}}
  {\prod_{x \in Y(t)}{\hook(x)}} (\fug + L_t(\nRepLeft) - A_t(\nRepLeft))
\end{equation}
$Y(t)$ is the Young diagrams to which the tableau $t$ is associated,
$\hook(x)$ is the hook length of the box $x$, $L_t(\nRepLeft)$ is the
number of boxes to the left of the box numbered $\nRepLeft$, and
$A_t(\nRepLeft)$ is the number of boxes above the box numbered
$\nRepLeft$.

One can verify formulas \eqref{eq:rel_b_t}--\eqref{eq:n1_t} by
computing the first cases, doable by hand. For example, for the case
of the Young symmetrizer of the symmetric representation, our formula
is consistent with the result of equation \eqref{eq:n1_m}.  If we take
the alternating representation instead, we have $\fugL =(\fug -
\nRepLeft) /(\nRepLeft+1)$.  When one looks at $1$BLM realizations of
this algebra, however the interpretation of the critical behaviour of
such a model can be difficult for negative $\fugL$, since
phenomena like level crossings appear, as has been discussed in the
context of the boundary chromatic polynomial \cite{Jacobsen2008a}.

We defer to future work further comments about the generalizations of
this construction to the case of two-boundary Temperley-Lieb, which is
definitely a more involved task.

\bibliographystyle{utphys}
%
%
\bibliography{Edgestates-v2.5}

\end{document}